\documentclass[12pt]{article}
\usepackage[latin1]{inputenc}
\usepackage[T1]{fontenc}
\usepackage[english]{babel}
\usepackage{epsfig}
\usepackage{amssymb}
\usepackage[centertags]{amsmath}
\usepackage{amsthm}
\usepackage{amsfonts}
\usepackage{makeidx}
\usepackage{graphics}
\usepackage{graphicx}
\usepackage{psfrag}
\usepackage{subfigure}
\usepackage{dsfont}
\usepackage{eucal}
\usepackage{eufrak}
\usepackage{color}
\usepackage{bm}
\newcommand{\bc}{\begin{center}}
\newcommand{\ec}{\end{center}}
\newcommand{\be}{\begin{equation}}
\newcommand{\ee}{\end{equation}}
\newcommand{\bea}{\begin{eqnarray}}
\newcommand{\eea}{\end{eqnarray}}
\newcommand{\ba}{\begin{array}}
\newcommand{\ea}{\end{array}}

\newcommand{\bfg}{\begin{figure}[htbp]}
\newcommand{\efg}{\end{figure}}

\def \de {\partial}

\def \a {\alpha}
\def \b {\beta}
\def \g {\gamma}
\def \G {\Gamma}
\def \d {\delta}

\def \l {\lambda}

\def \m {\mu}
\def \n {\nu}

\def \s {\sigma}

\def \ta {\tau}
\def \te {\theta}

\def \x {\xi}

\def \be {\begin{equation}}
\def \ee {\end{equation}}
\def \bea {\begin{eqnarray}}
\def \eea {\end{eqnarray}}
\def \non {\nonumber}

\def \ra {\rightarrow}
\def \Ra {\Rightarrow}

\def\laq{~\raise 0.4ex\hbox{$<$}\kern -0.8em\lower 0.62
ex\hbox{$\sim$}~}
\def\gaq{~\raise 0.4ex\hbox{$>$}\kern -0.7em\lower 0.62
ex\hbox{$\sim$}~}

\begin{document}

\vspace{0.5 cm} \bc {\large \textbf{Hadron potentials within the
gauge/string correspondence}} \vspace{1. cm}

Frédéric Jugeau$^{1,2}$\\
\vspace{0.2 cm}
\textit{$^1$ Institute of High Energy Physics, Chinese Academy of Sciences,\\
P. O. Box 918(4), Beijing 100049, China\\
$^2$ Theoretical Physics Center for Science Facilities, Chinese
Academy of Sciences,\\
Beijing 100049, China}\\
\vspace{0.3 cm}
jugeau@ihep.ac.cn\\
\ec
\par
\vspace{0.5 cm}

\begin{center}
{\large Abstract}
\end{center}

It is known, since the 70's, that the large $N$ 't Hooft limit of
gauge theories is related to string theories. In 1998, J. M.
Maldacena identified precisely such a relation: the so-called
\emph{AdS}/CFT correspondence which speculates a duality between a
large $N$ strongly-coupled supersymmetric and conformal Yang-Mills
theory in four dimensions and a weakly-coupled string theory
defined in a five-dimensional anti-de Sitter $AdS_5$ space-time.
This review aims at introducing concepts and methods used to
derive, in the framework of the gauge/string correspondence, the
interaction potentials of mesons and baryons at zero and finite
temperature. The dual string configurations associated with the
different kinds of hadrons are described and their behaviours at
short and large distances are understood. Although the application
of Maldacena's \emph{AdS}/CFT conjecture to QCD is not
straightforward, QCD being neither supersymmetric nor conformal,
the \emph{AdS}/QCD correspondence approach attempts to identify
the dual theory of QCD. Especially, the study of heavy
quark-antiquark bound-states leads to establish general dual
criteria for the confinement.

\par
\vspace{4.0 cm} PACS numbers: 11.25.Tq, 11.25.-w, 12.38.Aw,
12.38.Lg, 12.40.Yx.
\par
Keywords: \emph{AdS}/CFT correspondence, holographic models of
QCD,
Wilson loop,\\
\hspace*{2.65 cm}hadron potentials, confinement.
\newpage

\tableofcontents

\newpage
\section{Introduction}

A crucial breakthrough in the attempt to deal with
strongly-coupled Yang-Mills theories came with the \emph{AdS}/CFT
correspondence, proposed by J. M. Maldacena in 1998
\cite{Maldacena AdS/CFT}, that conjectures a duality between the
supergravity approximation of a superstring/M-theory living in a
$d$-dimensional anti-de Sitter space ($AdS_d$) times a compact
manifold and the 't Hooft limit of a maximally $\CMcal{N}=4$
superconformal $SU(N)$ gauge theory defined on the
$(d-1)$-dimensional boundary space ($\de AdS_d$). Shortly
afterwards, \cite{Witten, GKP} established the method for deriving
conformal dimensions of operators and correlators in conformal
field theory by means of dual superstring theory. In addition, J.
M. Maldacena considered the problem of calculating expectation
value of Wilson loop \cite{Maldacena wilson loop}. This issue is
of particular importance since the Wilson loop, through the area
law, consists of one of the most efficient tools for probing the
large distance properties of confining QCD-like gauge theories.
The purpose of the present paper is thus to review, within the
\emph{AdS}/CFT and the \emph{AdS}/QCD correspondences, the short
and large distance behaviours of the Wilson loop in the conformal
and non-conformal cases, and its use for deriving the interaction
potentials of bound-states of quarks.\\

First of all, let us motivate the gauge/string correspondence
which arises from low energy arguments when one considers energies
$E$ much smaller than the energy scale associated with the typical
string length $\ell_s=\sqrt{\a'}$:
\begin{equation}\label{low energy argument}
E\ll\frac{1}{\sqrt{\a'}}\;\;.
\end{equation}
As a matter of fact, we will be mostly interested, in the sequel,
in a stack of $N$ coincident D3-branes (A D$p$-brane is an
extended physical object with $p$ spatial dimensions. The capital
letter D in D$p$-brane stands for Dirichlet as the coordinates of
the open string endpoints normal to the brane must satisfy
Dirichlet boundary conditions). Therefore, it is worthwhile to
recall briefly some results of the open string spectroscopy in the
presence of D$p$-branes. In a $d$-dimensional flat space-time, the
(square) mass spectrum of an open string of which the endpoints
lie on a D$p$-brane reads
\begin{equation}
M^2=\frac{1}{\a'}\Big(N^{(i)}+N^{(a)}-1\Big)
\end{equation}
where $N^{(i)}\equiv\displaystyle\sum_{n=1}^{\infty}\sum_{i=2}^p
n\,{a_n^i}^{\dagger}a_n^i$ and
$N^{(a)}\equiv\displaystyle\sum_{m=1}^{\infty}\sum_{a=p+1}^d
m\,{a_m^a}^{\dagger}a_m^a$, expressed in terms of creation and
annihilation operators, count the number of modes along the
directions, respectively tangential ($i=2,\ldots,p$) and normal
($a=p+1,\ldots,d$) to the D$p$-brane. The low energy limit
\eqref{low energy argument} can then be understood as follows.
Whereas the energies are kept bounded (and the closed string
coupling constant $g_s$ is kept fixed) the typical length scale of
the string is put to zero \cite{Maldacena AdS/CFT}:
\begin{equation}\label{decoupling limit}
\a'\to0\;\;\;\;(E\;\;\textrm{bounded
and}\;\;g_s\;\;\textrm{fixed})\;\;.
\end{equation}
Hence, the massive modes of the open string decouple and only
remain the massless states. For this reason, the low energy limit
is also called the decoupling limit. The massless modes are of two
types. The first kind of massless states consists of the
oscillators ${a_m^a}^{\dagger}$'s normal to the brane acting on
the (tachyonic $M^2=-\frac{1}{\a'}$) ground-states
$|\Omega\rangle$: ${a_m^a}^{\dagger}|\Omega\rangle$. Since the
index $a$ lives in the space normal to the brane, it is not a
Lorentz index for the brane and the $(d-p)$ corresponding states
are massless scalar states (these states represent slight parallel
displacements of the brane. In fact, a space-filling brane does
not have massless scalar excitation because it has precisely
nowhere to turn). More interesting for us are the oscillators
${a_n^i}^{\dagger}$'s tangent to the brane. Since the index $i$
lives on the brane, they give rise to $(p+1)-2$ massless vector
states. When $p=3$, one recognizes the two polarization states of
the photon field living in the four-dimensional world-volume of
the D3-brane. What happens with a stack of $N$ coincident
D3-branes is the following. We know that a gauge field lives on
each of the branes. But now, the open strings are allowed to pass
from a brane to another without requiring additional amount of
energy (contrary to an open string stretched out between two
parallel non-coincident D$p$-branes where the mass-squared gains
then an extra positive contribution stemming from the classical
stretching energy \eqref{string-mass-squared}). Since the branes
are distinguishable, the orientations $[\a\b]$ and $[\b\a]$ of an
open string from the brane $\a$ to the brane $\b$ for instance are
not equivalent (the indices $\a,\b=1,\ldots,N$ are called
Chan-Paton indices). Therefore, with the $N$ gauge fields
associated with the $N$ D3-branes, we have altogether
$N+N(N-1)=N^2$ self-interacting gauge fields living in the
(p+1=4)-dimensional world-volume of the stack. In other words, an
$U(N)$ Yang-Mills theory in the Minkowski space-time. As a matter
of fact, the previous statement is not rigorously true: all the
gauge fields do not interact with each other. Provided that we
perform a change of basis in the space of states, a
non-interacting $U(1)$ gauge field can be identified which carries
indeed a zero charge as seen by the other eight self-interacting
gauge fields which define, in turn, a $SU(N)$ Yang-Mills theory.

We are now ready to establish the \emph{AdS}/CFT correspondence
\cite{Maldacena AdS/CFT,report}. First of all, let us consider the
$N$ coincident D3-branes described above plunged into a
ten-dimensional flat space-time (the bulk with $d=10$). Although
the branes carry charge and mass, their back-reaction on the bulk
is neglected here and they are just considered as boundary
conditions for the open strings. At low energies, the
four-dimensional physics is described by a maximally $\CMcal{N}=4$
$SU(N)$ supersymmetric Yang-Mills theory. Such a theory is also
invariant under the conformal $SO(2,4)$ group, \emph{i.e.} does
not contain any length scale. The number of generators in this
conformal field theory (or CFT) is
$\textrm{dim}\Big(SO(2,n)\Big)\;=\;\textrm{dim}\Big(
SO(n+2)\Big)\;=\;\frac{(n+2)(n+1)}{2}\underset{(n=4)}{=}15$, which
is also the dimension of the isometry group of the anti-de Sitter
space-time $AdS_5$: here is a first insight towards the
establishment of the $AdS$/CFT correspondence conjecture. As for
the closed strings, of which the massless spectrum contains the
graviton states\footnote{The square masses of the closed strings
are:
\begin{equation}
M^2=\frac{2}{\a'}\Big(N+\overline{N}-2\Big)
\end{equation}
with
$N=\displaystyle\sum_{n=1}^{\infty}\sum_{I=2}^{d}n\,{a_n^I}^{\dagger}a_n^I$
and
$\overline{N}=\displaystyle\sum_{n=1}^{\infty}\sum_{I=2}^{d}n\,{\overline{a}_n^I}^{\dagger}\overline{a}_n^I$
the number operators of respectively the left- and right-moving
modes which satisfy the constraint $N=\overline{N}$.}, they are
allowed to propagate throughout the bulk. The strength of the
gravitational interactions is governed by Newton's constant which,
at ten dimensions, is given by $G_{10}=8\pi^6g_s^2{\a'}^4$ (we
work in natural units where $\hbar=c=1$). In the decoupling limit
where $\a'\to0$ with $g_s$ fixed, Newton's constant vanishes and
the interactions in gravity become free in the \emph{infrared}
(IR): the closed strings do not interact with each other anymore
but also with the $SU(N)$ gauge fields living on the branes since
gravity couples universally to all forms of matter. Hence, we are
left with a $SU(N)$ super-Yang-Mills theory defined in the
four-dimensional Minkowski world-volume of the $N$ coincident
D3-branes and a \emph{decoupled} set of non-interacting closed
strings in ten-dimensional flat space-time.

In the second viewpoint, the $N$ D3-branes are solution of the
gravitational field equations of a ten-dimensional type IIB
(oriented closed) superstring theory. Since the branes carry
energy, they warp locally space-time geometry: a throat develops
in the depths of which are the branes while the space-time
infinitely far away the throat is asymptotically flat. This throat
can be visualized as an infinitely deep cylinder, the radius $R$
of which becoming constant (this so-called radius of the horizon
should not be confused with the horizon of a Schwarzschild black
hole since the geometry of the latter is quite different from the
throat geometry). Moreover, since the D3-branes define a
four-dimensional space-time on their world-volume, they appear as
a point source of energy with respect to the remaining six
space-like dimensions. Precisely, six coordinates are required in
order to define a compact 5-sphere $S^5$ and, thus, the point-like
throat is surrounded, in the space transverse to the branes, by
concentric $S^5$'s. As we move near the throat, the volume of the
5-spheres tends to a constant with also $R$ as the radius. Let us
write down the geometry of the $N$ coincident (extremal) D3-branes
\cite{Horowitz-Storminger}:
\begin{equation}\label{extremal geometry}
ds^2=\Big(1+\frac{R^4}{{\a'}^4u^4}\Big)^{-\frac{1}{2}}\eta_{\m\n}dx^\m
dx^\n+\Big(1+\frac{R^4}{{\a'}^4u^4}\Big)^{\frac{1}{2}}\Big({\a'}^2du^2+\a'^2u^2d\Omega_5^2\Big)
\end{equation}
where $\eta_{\m\n}=\textrm{diag}(-1,+1,+1,+1)$ is the Minkowski
metric tensor of the D3-brane world-volume $(\m,\n=0,\ldots,3)$
and $d\Omega_5^2$ is the line element of the unit sphere $S^5$.
The D3-branes are located infinitely deep in the throat
$\a'u\to0$. In the $AdS$/CFT correspondence, the "holographic"
radial coordinate $\a'u$ is dual to the energy scale under which
is observed the $SU(N)$ super-Yang-Mills theory ($u$ has indeed
the dimension of an energy). Being the $N$ D3-branes and the
horizon deep down the throat, they can never be reached.
Consequently, it appears two \emph{decoupled} regions: the
asymptotically flat space-time far away at infinity
($\a'u\to+\infty$) and the region near the horizon $(\a'u\to0)$.
Near the horizon, the metric \eqref{extremal geometry} becomes:
\begin{equation}\label{near-horizon approximation}
ds^2=\a'\Big(\frac{\a'u^2}{R^2}\eta_{\m\n}dx^\m
dx^\n+\frac{R^2}{\a'}\frac{du^2}{u^2}\Big)+R^2d\Omega_5^2
\end{equation}
which is nothing else than the $AdS_5\times S^5$ geometry. There
is an overall $\a'$ factor, so the metric remains constant in
$\a'$ units. The near-horizon approximation \eqref{near-horizon
approximation} could have also been obtained from the decoupling
limit \eqref{decoupling limit} as defined by \cite{Maldacena
AdS/CFT}: whereas $\a'\to0$, we keep the energies bounded which is
indeed consistent with the fact that the energy scale $u$ remains
constant when passing from \eqref{extremal geometry} to
\eqref{near-horizon approximation}.  As \emph{seen} by an observer
living in the asymptotically flat Minkowski space-time, two
distinctive low energy physics arise associated with the two
\emph{decoupled} space-time regions. On the one hand, there is
obviously the low energy physics which rules near the observer.
Without brane where attached open string endpoints, it consists,
according to the decoupling limit, of massless modes of
non-interacting closed strings. On the other hand, near the
horizon, any finite energy mode which attempts to reach the
observer need to overcome the gravitational well generated by the
throat. These modes are then perceived as massless by the observer
and deeper they are in the throat, higher can be their energies.
As a result, the whole tower of massive modes has to be taken into
account by the observer at asymptotic infinity. So, in the second
viewpoint, we have a type IIB closed superstring theory living,
near the horizon, in a ten-dimensional $AdS_5\times S^5$
space-time and, once again, a \emph{decoupled} set of
non-interacting closed strings in ten-dimensional flat space-time.
We see that the two descriptions share the same decoupled system,
namely, the set of non-interacting closed strings. We are thus led
to conjecture that, at low energies, a $SU(N)$ super-Yang-Mills
theory living in a four-dimensional Minkowski space-time and a
type IIB closed superstring theory on $AdS_5\times S^5$ describe
the same physics. This is the $AdS$/CFT correspondence
\cite{Maldacena AdS/CFT}.

As a matter of fact, although these two very different theoretical
frameworks are believed (and checked, see for instance
\cite{test}) to be dual to each other, their tractability domains
turn out to be opposite. To see that, let us focus on the
relations involving the two sets of parameters ($R$, $g_s$) and
($N$, $g_{YM}$) of the dual theories ($g_{YM}$ is the Yang-Mills
coupling constant). The radius of the horizon $R$ (the so-called
$AdS$ radius) is given in terms of the string length scale $\a'$
by
\begin{equation}\label{duality relation 1}
\frac{R^2}{\a'}\propto\sqrt{g_s N}
\end{equation}
while
\begin{equation}\label{duality relation 2}
g_s\propto g^2_{YM}\;\;.
\end{equation}
According to \eqref{duality relation 2}, the ratio \eqref{duality
relation 1} implies the so-called 't Hooft coupling constant:
\begin{equation}\label{'t hooft coupling}
\lambda\equiv g_{YM}^2N
\end{equation}
which is the relevant coupling constant when considering gauge
theory in the large $N$ limit. These two duality relations can be
derived, for the former, from general considerations on
D$p$-branes involving their number, their tension and Newton's
constant, and for the latter, by considering the weak field
expansion of the Dirac-Born-Infeld action \eqref{dirac-born-infeld
action} which governs the dynamics of D$p$-branes carrying
electromagnetic fields on their world-volumes. Especially, the
second relation tells us that to a weak string coupling constant
$g_s\ll1$ corresponds a weak Yang-Mills coupling constant
$g_{YM}\ll1$, which seems convenient since it is easier to deal
with any theory in the perturbative regime. Thus, the $AdS$/CFT
correspondence seems straightforward to check. Nevertheless, let
us consider the case where $N$ is \emph{finite}. The first
relation implies then $R^2\ll\a'$ \emph{i.e.} the $AdS_5$ scalar
curvature (proportional to $1/R^2$) is much larger than
$1/{\a'}^2$. As a result, all the stringy effects have to be
considered but we do not know yet how to solve tree-level
superstring theory on $AdS_5\times S^5$. So, let us see what
happens in the \emph{large} $N$ limit when the 't Hooft coupling
$\lambda$ is also \emph{large}. In this case $R^2\gg\a'$ and the
strings appear as point-like particles. The ten-dimensional free
superstring theory can then be approximated by its low energy
effective theory : a tractable ten-dimensional supergravity theory
where the stringy effects are corrections of order $O(\a'^3)$. On
the other side of the duality, we have a $SU(N)$ super-Yang-Mills
theory which is now strongly-coupled. Although QCD is neither
supersymmetric nor conformal, one sees the great interest in
establishing a gauge/string correspondence for such a confining
gauge theory. We already know that we have to break properly the
underlying supersymmetry nature of the $AdS$/CFT correspondence
and to deform the ten-dimensional $AdS_5\times S^5$ holographic
space-time since QCD has a mass gap and is asymptotically
conformal only at high energy (namely at $\a' u\to+\infty$ in
\eqref{extremal geometry}). The strong coupling regime could then
be investigated from a higher-dimensional dual supergravity theory
(in the large $N$ limit with $N$ identified to the number of
colours $N_c$). In practice, two complementary approaches exist
which aim at identifying the dual theory of QCD. In the so-called
\emph{top-down} approach, the \emph{AdS} geometry is deformed into
a Schwarzschild black hole$-AdS$ geometry where the horizon plays
the role of an IR brane, the location of which being given in
terms of the Beckenstein-Hawking temperature (\emph{i.e.} the
thermal temperature of the gauge theory) \cite{Witten conformal
breaking}. On the other hand, according to the more
phenomenological \emph{bottom-up} approach (or, roughly speaking,
in the \emph{AdS}/QCD correspondence though such a duality has not
been established yet), five-dimensional holographic models which
attempt to reproduce the main properties of QCD have been
proposed. In the IR Hard Wall Model \cite{PT,EKSS,pomarol}, a
confining gauge theory can be obtained considering a truncated
$AdS_5$ holographic space-time, the typical size of which
representing the IR cutoff associated to the QCD mass gap. The
light hadron spectroscopy \cite{hard wall model spectrum}, the
meson and nucleon form factors \cite{hard wall model form
factors}, the two-point correlation functions \cite{hard wall
model correlator}, the deep inelastic scattering structure
functions \cite{hard wall model DIS}, the chiral symmetry breaking
mechanism and the axial $U(1)$ anomaly \cite{hard wall model
chiral} have been investigated. In particular, it has been showed
that the spectrum satisfies a Kaluza-Klein behaviour $m_n^2\sim
n^2$ instead of the expected Regge behaviour $m_n^2\sim n$ ($n$ is
the radial excitation number). To remedy this shortcoming, the IR
Soft Wall Model has been proposed which consists in inserting a
dilaton field in the $AdS_5$ space-time \cite{KKSS}. The
phenomenological outcomes of this model have also been largely
studied \cite{soft wall model,soft wall scalar}.\\

The review is organized as follows: the main steps leading to the
confining linear static potential of a heavy quark-antiquark
bound-state through the four-point Green function and the area law
of the Wilson loop in QCD are described in the section 2. The
section 3 is devoted to Maldacena's calculation of the heavy quark
potential within the \emph{AdS}/CFT correspondence. Especially,
the underlying conformal invariance and the intrinsically
non-perturbative nature of the result is pointed out. In the
section 4, we consider the finite temperature version of the
gauge/string correspondence as described by E. Witten \cite{Witten
conformal breaking} and derive, in three and four dimensions,
linear quark-antiquark potentials of which the string tensions
depend on the temperature. The L\"{u}scher correction term at
large distance is also discussed. The section 5 focuses on
holographic models of QCD for which general dual criteria for the
confinement can be established considering the bulk dynamics of
dual string world-sheet. In the section 6, we review the
string/brane configuration dual to the baryons and stress, in
particular, the existence of \emph{AdS}/CFT "reduced" baryons made
of $k<N$ quarks. At the end of the review, an appendix is devoted
to the properties of the Wilson loop and to the loop space
formalism of QCD.
\newpage
\section{The static potential in QCD}

The static potential between an infinitely massive quark and
antiquark has been computed for a long time. The two-loop
perturbative calculation gives \cite{schroder-lepage}:
\begin{equation}
V(r)=-C_F\frac{\a(r)}{r}
\end{equation}
with
\begin{eqnarray}
\a(r)&=&\a_s\Big\{1+\Big(a_1+\beta_0L\Big)\frac{\a_s}{4\pi}\non\\
&&+\Big[a_2+\beta_0^2\Big(L^2+\frac{\pi^2}{3}\Big)+\Big(\beta_1+2\beta_0a_1\Big)L\Big]\Big(\frac{\a_s}{4\pi}\Big)^2+\ldots\Big\}\;\;.\label{Wilson-running
coupling constant}
\end{eqnarray}
$r$ is the distance between the quark and the antiquark and the
(renormalization scale $\m$-dependent) strong coupling constant
$\a_s$ is in the modified minimal subtraction scheme
$\overline{MS}$: $\a_s\equiv\a_s^{(\overline{MS})}(\mu^2)$. We
have defined $L\equiv2\g_E+\ln(\m^2r^2)$ with $\g_E\simeq0.5772$
the Euler-Mascheroni constant. The one-loop and two-loop constants
$a_1$ and $a_2$ write out respectively:
\begin{eqnarray}
a_1&=&\frac{31}{9}C_A-\frac{20}{9}T_F n_F\;\;,\\
a_2&=&\Big(\frac{4343}{162}+4\pi^2-\frac{\pi^2}{4}+\frac{22}{3}\zeta(3)\Big)C_A^2-\Big(\frac{1789}{81}+\frac{56}{3}\zeta(3)\Big)C_AT_Fn_F\non\\
&&-\Big(\frac{55}{3}-16\zeta(3)\Big)C_FT_Fn_F+\big(\frac{20}{9}T_Fn_F\big)^2
\end{eqnarray}
where $\zeta(s)$ is the Riemann zeta function and $n_F$ the number
of massless quarks. In \eqref{Wilson-running coupling constant},
the two first regularization scheme-independent coefficients of
the $\b$-function are:
\begin{eqnarray}
\b_0&=&\frac{11}{3}C_A-\frac{4}{3}T_Fn_F\;\;,\\
\b_1&=&\frac{34}{3}C_A^2-4C_FT_Fn_F-\frac{20}{3}C_AT_Fn_F\;\;.
\end{eqnarray}
$C_F$ and $C_A$ are the values of the Casimir operators in the
fundamental and the adjoint representations respectively
($T^aT^a=C_F\mathbb{I}_{N_c}$ with $T^a=\frac{\lambda^a}{2}$ the
Gell-Mann matrices and $f^{aij}f^{bij}=C_A\d^{ab}$ with
$T^a_{ij}=-if^{aij}$ the structure constants). The trace
normalization is $Tr(T^aT^b)=T_F\d^{ab}$ and, in the case of the
colour gauge group $SU(3)_c$ of QCD, the colour factors become:
\begin{eqnarray}
T_F&=&\frac{1}{2}\;\;,\\
C_F&=&T_F\Big(\frac{N_c^2-1}{N_c}\Big)=\frac{4}{3}\;\;,\\
C_A&=&N_c=3\;\;.
\end{eqnarray}
Especially, at the lowest perturbative order, one recovers the
one-gluon exchange contribution which has a Coulomb-like form:
\begin{equation}
V(r)=-\frac{4}{3}\frac{\a_s}{r}\;\;.
\end{equation}

More phenomenologically, E. Eichten, K. Gottfried, T. Kinoshita,
K. D. Lane and T.-M. Yan postulated, in the late 70s, that "many
of the gross features of the charmonium $c\overline{c}$ potential
can be simulated by the potential" (several phenomenological forms
of the static potential have been proposed. See, for instance, the
report \cite{potential models}):
\begin{equation}
V(r)=-\frac{\kappa}{r}+\s\,r+C\;\;.\label{Wilson-Cornell
potential}
\end{equation}
"This is chosen to give a simple interpolation between the known
Coulomb-like force at short distance and a linear growth of the
static potential" \cite{cornell potential}. In the so-called
\emph{Cornell potential} \eqref{Wilson-Cornell potential},
$\kappa$ represents the coulomb strength, $\s$ the string tension
and the constant $C$ fixes the origin of the potential. They are
regarded as free parameters to be fitted on the spectrum. On the
other hand, when one attempts to derive the potential on the
lattice \cite{bali}, a non-physical term in $1/r^2$ is also often
added to \eqref{Wilson-Cornell potential} in order to enhance the
fit to data by simulating, for instance, running coupling effects.

The confining linear potential can also be extracted, in the
static approximation, from the four-point, or two-particle
(meson), Green function. The formalism involves the Wilson loop, a
gauge invariant functional, which provides a physical observable
able to measure the heavy quark interaction potential (see
Eq.\eqref{Wilson-feynmann kac formula} below). The gauge invariant
state of a quarkonium, bound-state of a quark and an antiquark, is
defined by means of the so-called gauge line $U(y,x,\CMcal{C})$:
\begin{equation}
U(y,x;\CMcal{C})\equiv Pe^{-ig\int_{x}^{y}A_{\m}(x)dx^{\m}}
\end{equation}
which carries out the parallel displacement of the gauge
transformation from the point $x$ to the point $y$ along the curve
$\CMcal{C}$:
\begin{equation}\label{Wilson-quarkonium state}
|\phi(y,x)\rangle=\phi(y,x)|0\rangle=\overline{q}(y)U(y,x,\CMcal{C})q(x)|0\rangle\;\;.
\end{equation}
The corresponding state, hermitian conjugate of
\eqref{Wilson-quarkonium state}, is:
\begin{equation}
\langle\phi(y,x)|=\langle0|\phi^{\dagger}(y,x)=\langle0|\overline{q}(x)U(x,y,\CMcal{C}')q(y)
\end{equation}
where $\CMcal{C}$ and $\CMcal{C}'$ describe the same space-time
curve but oriented in opposite directions. Let us then consider
the evolution amplitude of this quark-antiquark state during the
lapse of time $T$. It consists of a four-point gauge invariant and
colour-singlet Green function:
\begin{eqnarray}
G^{(4)}(x_1,x_2,x'_1,x'_2)&\equiv&\frac{1}{N_c}\langle\phi(x_1,x_2)|\phi(x'_1,x'_2)\rangle\non\\
&\equiv&\frac{1}{N_c}\langle\,\CMcal{T}\,[\overline{q}(x_2)U(x_2,x_1)q(x_1)\overline{q}(x'_1)U(x'_1,x'_2)q(x'_2)]\rangle_{A,q,\overline{q}}\;\;,\non\\
\label{Wilson-green function}
\end{eqnarray}
the averaging being defined in the path-integral formalism. The
dependence of the two gauge lines $U(x_2,x_1)$ and $U(x'_1,x'_2)$
on the curves $\CMcal{C}_{[x_2\,x_1]}$ and
$\CMcal{C}_{[x'_1\,x'_2]}$ is implicit and the ordering
prescription in \eqref{Wilson-green function} is the usual one
defined by the chronological operator $\CMcal{T}$:
\begin{equation}
\CMcal{T}[q(x)\overline{q}(y)]=\te(x_0-y_0)q(x)\overline{q}(y)-\te(y_0-x_0)\overline{q}(y)q(x)\;\;.
\end{equation}
The four-point Green function is normalized, in the large $N_c$
limit, by a factor $1/N_c$. Since the action of QCD is quadratic
in the fermion fields, the path-integral over the quarks is
gaussian and gives according to the Wick theorem:
\begin{equation}\label{Wilson-green function integral}
G^{(4)}(x_1,x_2,x'_1,x'_2)=-\frac{1}{N_c}\langle\textrm{det}(i\g^\m
D_\m-m_q)U(x_2,x_1)S_1(x_1,x'_1;A_\m)U(x'_1,x'_2)S_2(x'_2,x_2;A_\m)\rangle_A
\end{equation}
were $D_\m$ is the covariant derivative and $m_q$ the quark mass.
It is usual to neglect, in the quenched approximation, the
fermionic determinant giving rise, in perturbation theory, to
quark-antiquark loops. In the large $N_c$ limit, this
approximation is exact and we have:
\begin{equation}
\textrm{det}(i\g^\m D_\m-m_q)=1\;\;.
\end{equation}
Moreover, \eqref{Wilson-green function integral} does not account
for the other contribution associated to the $q\overline{q}$
annihilation process, only possible if $q(x_1)$ and
$\overline{q}(x_2)$ (respectively $q(x'_2)$ and
$\overline{q}(x'_1)$) have the same flavor.

In order to derive the static potential, Brown and Weisberger
\cite{brown} wrote \eqref{Wilson-green function integral} in terms
of the static quark and antiquark propagators in the presence of
the external gluon field $A_0$ (formally, the static approximation
consists in neglecting, in the equations of motion of the
propagators, the spatial components of the covariant derivatives
with respect to the temporal ones):
\begin{eqnarray}
S_{1\,stat.}(x_1,x'_1;A_0)&=&S_1^{(0)}(x_1-x'_1)U(x_1,x'_1)\;\;,\label{Wilson-quark
propagator}\\
S_{2\,stat.}(x'_2,x_2;A_0)&=&S_2^{(0)}(x'_2-x_2)U(x'_2,x_2)\label{Wilson-antiquark
propagator}
\end{eqnarray}
were $S_1^{(0)}(x_1-x'_1)$ and $S_2^{(0)}(x'_2-x_2)$ are the free
static propagators (namely, in the absence of gluon) which verify
the following equations of motion:
\begin{eqnarray}
(i\g_1^0\de_{1\,0}-m_1)S_1^{(0)}(x_1-x'_1)=i\d^4(x_1-x'_1)\;\;,\\
S_2^{(0)}(x'_2-x_2)(-i\g_2^0\overleftarrow{\de}_{2\,0}-m_2)=i\d^4(x'_2-x_2)\;\;.
\end{eqnarray}
The expressions \eqref{Wilson-quark propagator} and
\eqref{Wilson-antiquark propagator} allow one to write the static
evolution amplitude into the form:
\begin{equation}\label{Wilson-static green function}
G^{(4)}_{stat.}(x_1,x_2,x'_1,x'_2)=-\langle
S_1^{(0)}(x_1-x'_1)S_2^{(0)}(x'_2-x_2)\frac{1}{N_c}Tr_c\,U(x_2,x_1)U(x_1,x'_1)U(x'_1,x'_2)U(x'_2,x_2)\rangle_A\;\;.
\end{equation}
Then, the path-ordered product of the four gauge lines generates
the trace on the colour space of a closed gauge line which is
nothing else than the Wilson loop:
\begin{equation}\label{wilson loop}
\phi(\CMcal{C})\equiv
Tr_c\,U(x_1,x_1,\CMcal{C})=Tr_c\,P\,e^{-ig\oint_{\CMcal{C}}
A_{\m}(x)dx^{\m}}\;\;.
\end{equation}
Let us pointing out that the manifestly gauge invariant Wilson
loop does not depend in \eqref{wilson loop} on the point $x_1$
from which is parametrized the loop $\CMcal{C}$ because of the
colour trace $Tr_c$. We have thus:
\begin{equation}\label{Wilson-static green function final}
G^{(4)}_{stat.}(x_1,x_2,x'_1,x'_2)=-S_1^{(0)}(x_1-x'_1)S_2^{(0)}(x'_2-x_2)W[\CMcal{C}]
\end{equation}
where $W[\CMcal{C}]$ is the gauge invariant one-loop functional:
\begin{equation}\label{Wilson-expectation value}
W[\CMcal{C}]\equiv\frac{1}{N_c}\langle\phi(\CMcal{C})\rangle_A=\frac{1}{Z}\int
[d A_\m(x)]\frac{1}{N_c}\phi(\CMcal{C})e^{iS_{YM}[A_\m]}\;\;.
\end{equation}
Without loss of generality, let us consider the specific case
where $x_1^0=x_2^0\equiv X^0$ and $x'_1{^0}=x'_2{^0}\equiv
X'{^0}$. The loop is then reduced to a rectangular contour
(symbolized by $\square$) with temporal extent $T$ and spatial
extent $r$. In the limit where $T\equiv(X^0-X'^0)\ra\infty$ (in
the static approximation, the distance between the fixed quark and
antiquark
$r\equiv|\overrightarrow{x}_1-\overrightarrow{x}_2|=|\overrightarrow{x}'_1-\overrightarrow{x}'_2|$
remains constant over time), the loop becomes infinitely stretched
out and one finds Wilson's confinement criterion or area law
\cite{Wilson1974}:
\begin{equation}\label{Wilson-area law}
W[\square]\underset{\CMcal{\square\,\ra\infty}}{=}e^{-i\s_tA[\square]}
\end{equation}
where $\s_t$ is the string tension and $A[\square]$ is the area of
the minimal surface with the rectangle as boundary. In the
relativistic flux tube model \cite{olsson} in which the
confinement results from the formation of a chromo-electric field
tube (the effective "QCD string") between the quark and the
antiquark, the constant $\s_t=\frac{1}{2\pi\a'}$, of the order of
0.18 GeV$^2$, stands for the linear energy density of the tube and
is related to the slope $\a'$ of the Regge trajectories:
$J(m^2)=\a_0+\a'm^2$ ($J$ and $m^2$ are the total spin and the
square mass of the hadrons while $\a_0$ is the intercept). As a
final result, we obtain:
\begin{equation}\label{Wilson-static green function final 2}
G^{(4)}_{stat.}(x_1,x_2,x'_1,x'_2)=-\d^3(\overrightarrow{x}_1-\overrightarrow{x}'_1)\d^3(\overrightarrow{x}'_2-\overrightarrow{x}_2)e^{-im_1T}e^{-im_2T}e^{-i\s_trT}\;\;.
\end{equation}

On the other hand, it is possible to extract explicitly the energy
dependence of the evolution amplitude of the quarkonium, as
defined in \eqref{Wilson-green function}. In the Heisenberg
representation, the operators depend on the time and their
dynamics are governed by the Hamiltonian of the system:
\begin{equation}
\phi(\overrightarrow{x}_1,\overrightarrow{x}_2,X^0)=e^{iHX^0}\phi(\overrightarrow{x}_1,\overrightarrow{x}_2,0)e^{-iHX^0}\;\;.
\end{equation}
This relation can also be understood as a particular case of the
space-time translation:
\begin{equation}
\phi(x)=e^{ip\cdot x}\phi(0)e^{-ip\cdot x}
\end{equation}
where $p=(H,\overrightarrow{0})$ is the energy-momentum operator.
Thus we have:
\begin{eqnarray}
G^{(4)}_{stat.}(x_1,x_2,x'_1,x'_2)&=&\frac{1}{N_c}\langle0|e^{iHX^0}\phi^{\dagger}(x_1,x_2)e^{-iHX^0}e^{iHX'{^0}}\phi(x'_1,x'_2)e^{-iHX'{^0}}|0\rangle\non\\
&=&\frac{1}{N_c}\langle0|\phi^{\dagger}(\overrightarrow{x}_1,\overrightarrow{x}_2,0)e^{-iH(X^0-X'{^0})}\phi(\overrightarrow{x}'_1,\overrightarrow{x}'_2,0)|0\rangle\non\\
&=&\frac{1}{N_c}\langle\phi(\overrightarrow{x}_1,\overrightarrow{x}_2,0)|e^{-iHT}|\phi(\overrightarrow{x}'_1,\overrightarrow{x}'_2,0)\rangle\;\;.\label{Wilson-heisenberg}\\
\end{eqnarray}
Let $\{|P_n\rangle\}$ be a complete set of eigenvectors of $H$
associated to the eigenvalues $\{E_n\}$:
\begin{eqnarray}
H|P_n\rangle=E_n|P_n\rangle\;\;,\\
\sum_{n}|P_n\rangle\langle P_n|=1\label{Wilson-closure
relation}\;\;.
\end{eqnarray}
If one puts the closure relation \eqref{Wilson-closure relation}
into \eqref{Wilson-heisenberg}, one obtains:
\begin{equation}
G^{(4)}_{stat.}(x_1,x_2,x'_1,x'_2)=\sum_{n}\frac{1}{N_c}\langle\phi(\overrightarrow{x}_1,\overrightarrow{x}_2,0)|P_n\rangle\langle
P_n|\phi(\overrightarrow{x}'_1,\overrightarrow{x}'_2,0)\rangle
e^{-iE_nT}\;\;.
\end{equation}
In the limit $T\ra\infty$, the leading contribution corresponds to
the quarkonium ground-state $|P_0\rangle$ with the energy $E_0$ in
the complex exponential function:
\begin{equation}
G^{(4)}_{stat.}(x_1,x_2,x'_1,x'_2)=\frac{1}{N_c}\langle\phi(\overrightarrow{x}_1,\overrightarrow{x}_2,0)|P_0\rangle\langle
P_0|\phi(\overrightarrow{x}'_1,\overrightarrow{x}'_2,0)\rangle
e^{-iE_0T}
\end{equation}
where
$\frac{1}{\sqrt{N_c}}\langle\phi(\overrightarrow{x}_1,\overrightarrow{x}_2,0)|P_0\rangle$
is the wave function of the bound-state. One can understand the
ground state of the quarkonium as the one in which the number of
quark-gluon and gluon-gluon interactions is the smallest. In the
static limit where the kinetic energies of the quark and antiquark
vanish, we find:
\begin{equation}
E_0=m_1+m_2+V(r)\;\;.
\end{equation}
In comparison with \eqref{Wilson-static green function final 2},
we infer the confining linear static potential $V(r)$ of an
infinitely massive quark-antiquark pair:
\begin{equation}\label{Wilson-feynmann kac formula}
V(r)=-\underset{T\to\infty}{\textrm{lim}}\frac{1}{iT}\ln
W[\square]=\s_tr\;\;.
\end{equation}
The only condition for the validity of the Feynman-Kac formula
\eqref{Wilson-feynmann kac formula} is that the states
$|\phi\rangle$'s have a non-vanishing component over the ground
state.

\section{The heavy quark potential in the \emph{AdS}/CFT correspondence}

The gauge/string duality aims at relating gauge theory observables
to calculations in higher-dimensional dual space-times. For
instance, the method for deriving conformal dimensions of
operators and correlators in conformal field theory via dual
string theory has been described in \cite{Witten, GKP}. On the
other hand, J. M. Maldacena considered the problem of calculating
the (expectation value of the) Wilson loop $W[\CMcal{C}]$
\cite{Maldacena wilson loop}. He suggested the following
\emph{AdS}/CFT duality relation:
\begin{equation}\label{wilson-nambu goto action}
W[\CMcal{C}]\sim Z_{string}[\CMcal{C}]
\end{equation}
where $Z_{string}[\CMcal{C}]$ is the full partition function of
the dual string theory. As previously discussed in the
Introduction, although the string theory is weakly coupled in the
't Hooft limit ($g_s$ and $g_{YM}$ are the closed string and the
Yang-Mills coupling constants respectively):
\begin{equation}
g_s=\frac{g_{YM}^2}{2\pi}=\frac{1}{2\pi}\frac{\lambda}{N}\underset{\underset{\lambda\;\textrm{\bf{fixed}}}{N\to\infty}}{\to}0\;\;,
\end{equation}
we do not known yet how to solve free string theory on
$AdS_5\times S^5$. Nevertheless, in the case where the 't Hooft
coupling is large $\lambda\equiv g_{YM}^2N\gg1$, the typical
length scale of the string $\ell_s=\sqrt{\a'}$ is small in
comparison with the $AdS_5$ radius $R$:
\begin{equation}\label{wilson-duality relation}
\frac{R^4}{\a'\,^2}=4\pi g_s N=2\lambda\gg1\;\;.
\end{equation}
This is the so-called supergravity limit where strings appear as
point-like particles and the stringy effects can be neglected. The
\emph{AdS}/CFT prescription consists then in calculating the
Wilson loop $W[\CMcal{C}]$ in terms of the proper area of a string
world-sheet which describes the closed loop $\CMcal{C}$ on the
boundary:
\begin{equation}
W[\CMcal{C}]\sim e^{-S[\CMcal{C}]}
\end{equation}
where $S[\CMcal{C}]$ is the classical (Euclidean) action of the
string world-sheet. As a matter of fact, the world-sheet does not
describe the Wilson loop \eqref{Wilson-expectation value} but its
supersymmetric generalization since $AdS_5\times S^5$ is the
actual holographic space of the $\CMcal{N}=4$ $SU(N)$
super-Yang-Mills theory. Especially, we do not expect to
necessarily recover the area law \eqref{Wilson-area law} of the
Wilson loop (which is indeed a four-dimensional space-time
result).

Since we consider an infinitely heavy $Q\overline{Q}$ pair
(non-dynamical external probes), the inter-quark distance $r$ is a
constant of time and the string connecting the quark to the
antiquark gives rise to a rectangular contour $\CMcal{C}$ on the
boundary with temporal extent $T$ and spatial extent $r$:
\begin{eqnarray}
-\frac{T}{2}\leq t\leq\frac{T}{2}\;\;,\label{wilson-parameter1}\\
\non\\
-\frac{r}{2}\leq x\leq\frac{r}{2}\;\;.\label{wilson-parameter2}
\end{eqnarray}
The quark $Q$ and the antiquark $\overline{Q}$ are put down at
$x=r/2$ and $x=-r/2$ respectively. Nevertheless, because the
string is now allowed to move along the fifth holographic
coordinate of $AdS_5$, the minimal area of the world-sheet, of
which the boundary is the loop $\CMcal{C}$, is no longer the
rectangle. Indeed, $AdS_5$ is a curved space-time (the gravity
effects are non-zero) and, as a result, the string world-sheet
will not span only the four-dimensional surface enclosed by the
contour $\CMcal{C}$ at the boundary $\de AdS_5$. Instead of that,
the string will move inside the bulk and, held back by its
tension, will reach an extremal value of the holographic
coordinate.

If we choose the Poincaré coordinates $x^M=(x^\m,z)$
$(\m=0,1,2,3)$, the $AdS_5\times S^5$ line element reads
(throughout this section, we will work with the Euclidean
signature):
\begin{equation}\label{wilson-line element}
ds^2=\frac{R^2}{z^2}\Big(\d_{\m\n}dx^{\m}dx^{\n}+dz^2\Big)+R^2d\Omega_5^2\;\;.
\end{equation}
$AdS_5$ is the domain $z>0$ and
$\d_{\m\n}=\textrm{diag}(+1,+1,+1,+1)$ is the Euclidean metric
tensor of the boundary space-time $\de AdS_5$ which can be defined
by multiplying \eqref{wilson-line element} by $z^2$ and setting
$z=0$. Let us then define the dimensionless $AdS$ radius
$\tilde{R}$ as:
\begin{equation}\label{Wilson-R tilde}
\frac{R^4}{\a'\,^2}\equiv\tilde{R}^4\;\;.
\end{equation}
We also introduce a new holographic coordinate:
\begin{equation}\label{Wilson-new holographic coordinate}
u=\frac{R^2}{\a'z}=\frac{\tilde{R}^2}{z}
\end{equation}
which has the dimension of an energy. Not surprisingly, the line
element of the bulk \eqref{wilson-line element} becomes
\eqref{near-horizon approximation}:
\begin{equation}\label{wilson-line element 2}
ds^2=g_{MN}(x)dx^Mdx^N+R^2d\Omega_5^2=\alpha'\Big(\frac{u^2}{\tilde{R}^2}\d_{\m\n}dx^{\m}dx^{\n}+\frac{\tilde{R}^2}{u^2}du^2\Big)+R^2d\Omega_5^2\;\;.
\end{equation}
To the high (low) energy region $z\to0$ $(z\to+\infty)$
corresponds $u\to+\infty$ ($u\to0$).

Let us consider the simplest action which describes the dynamics
of an open string, namely the so-called Nambu-Goto action:
\begin{equation}\label{wilson-nambu goto action}
S_{NG}[\CMcal{C}]=T_0\int d^2\xi\sqrt{det(\g_{ab})}\;\;.
\end{equation}
$T_0^{-1}=2\pi\a'$ is the (inverse of the) fundamental string
tension and $\g_{ab}(\xi)$ ($a,b=1,2$) is the induced metric
tensor on the two-dimensional world-sheet:
\begin{equation}\label{Wilson-induced metric tensor}
\g_{ab}(\xi)\equiv G_{MN}(X)\frac{\de X^M}{\de\xi^a}\frac{\de
X^N}{\de\xi^b}
\end{equation}
which requires two parameters $\xi^1\equiv\s$ and $\xi^2\equiv\ta$
(the measure is $d^2\xi=d\s d\ta$). The metric tensor $G_{MN}(X)$
of the bulk \eqref{wilson-line element 2} in written in terms of
the world-sheet coordinates $X^M(\xi^a)=(X^0,X^1,X^2,X^3,X^5\equiv
U)(\xi^a)$ ($M=\m,5$ and, following a standard convention in
string theory, these coordinates (and the metric) are denoted with
capital letters). In static configuration, an useful
parametrization of the string world-sheet is the so-called static
gauge where:
\begin{equation}\label{wilson-static gauge}
X^0(\ta,\s)\equiv t(\ta,\s)=\ta\;\;.
\end{equation}
The lines of constant $\ta$ are "static strings" in the chosen
Lorentz frame. On the other hand, according to the
reparametrization invariance of the Nambu-Goto action
\eqref{wilson-nambu goto action}, we can choose:
\begin{equation}\label{wilson-reparametrization}
\s=x\;\;.
\end{equation}

According to \eqref{Wilson-feynmann kac formula}, the energy of a
quark-antiquark pair is obtained in the limit $T\ra\infty$. The
string world-sheet is then invariant under a translation along the
time coordinate and symmetric under the mirror transformation
$x\leftrightarrow-x$. Consequently, the holographic coordinate of
the string $U(x)$, a function of $x$ only, presents a minimum
which, by symmetry, occurs at $x=0$: in the sequel, we will denote
$U_0\equiv U(0)$ this noteworthy value. Let us express explicitly
the Nambu-Goto action \eqref{wilson-nambu goto action} in terms of
the string coordinates. For this, we have to evaluate the
components of the induced metric tensor \eqref{Wilson-induced
metric tensor}:
\begin{eqnarray}
\g_{11}&=&G_{11}(X)\frac{\de X^1}{\de\s}\frac{\de X^1}{\de\s}+G_{55}(X)\frac{\de U}{\de\s}\frac{\de U}{\de\s}=\a'\frac{U^2}{\tilde{R}^2}+\a'\frac{\tilde{R}^2}{U^2}U'\,^2\;\;,\\
\non\\
\g_{12}&=&\g_{21}=G_{MN}(X)\frac{\de X^M}{\de\s}\frac{\de X^N}{\de\ta}=0\;\;,\\
\non\\
\g_{22}&=&G_{00}(X)\frac{\de X^0}{\de\ta}\frac{\de
X^0}{\de\ta}=\a'\frac{U^2}{\tilde{R}^2}
\end{eqnarray}
such that
\begin{equation}\label{Wilson-Nambu-Goto action}
S_{NG}[\CMcal{C}]=T_0\int_{-T/2}^{T/2}dt\int_{-r/2}^{r/2}dx\,\CMcal{L}\big(U(x),U'(x)\big)
\end{equation}
with the Lagrangian density
\begin{equation}\label{Wilson-Lagrangian density}
\CMcal{L}=\a'\sqrt{U'\,^2+\frac{U^4}{\tilde{R}^4}}
\end{equation}
and where the prime denotes the derivative with respect to the
spatial coordinate $dU=U'(x)dx$. The shape $U(x)$ of the string
world-sheet stretching into the bulk is governed by the
Euler-Lagrange equation:
\begin{equation}\label{wilson-equation of motion}
\d_U S_{NG}=0\;\;\Ra\;\;\frac{\de\CMcal{L}}{\de
U}-\frac{d}{dx}\frac{\de\CMcal{L}}{\de
U'}=0\;\;\Ra\;\;UU''-4U'\,^2-2\frac{U^4}{\tilde{R}^4}=0\;\;.
\end{equation}
Moreover, since the Lagrangian density \eqref{Wilson-Lagrangian
density} does not depend explicitly on $x$, we have the following
first integral:
\begin{equation}
\frac{d\CMcal{L}}{dx}=U'\frac{\de\CMcal{L}}{\de U}+U''\frac{\de
\CMcal{L}}{\de U'}=\frac{d}{dx}\Big(U'\frac{\de\CMcal{L}}{\de
U'}\Big)+U'\Big(\frac{\de \CMcal{L}}{\de
U}-\frac{d}{dx}\frac{\de\CMcal{L}}{\de U'}\Big)=0\;\;.
\end{equation}
The last term vanishes according to the equation of motion
\eqref{wilson-equation of motion} and it remains:
\begin{equation}\label{wilson-first integral definition}
\frac{d}{dx}\Big(U'\frac{\de\CMcal{L}}{\de U'}-\CMcal{L}\Big)=0
\end{equation}
which finally gives:
\begin{equation}\label{Wilson-first integral}
\frac{U^4}{\sqrt{U'\,^2+\frac{U^4}{\tilde{R}^4}}}=\tilde{R}^2U_0^2=\textrm{const.}
\end{equation}
The constant can be evaluated, for instance, at $x=0$ where
$U(0)=U_0$ and $U'(0)=0$ by symmetry. One can also derived a
useful relation for the derivative of the string coordinate (the
positive (negative) sign corresponds to $0<x<r/2$ ($-r/2<x<0$)):
\begin{equation}\label{Wilson-expression of the derivative}
U'(x)=\pm\frac{U^2}{\tilde{R}^2U_0^2}\sqrt{U^4-U_0^4}\;\;.
\end{equation}
At this stage, it is suitable to determine the dependence of the
quark-antiquark distance $r$ on the holographic string coordinate.
To do so, we start from the following integral:
\begin{equation}
x=\int_{0}^{x}dx'=\int_{U_0}^{U(x)}\frac{d
U}{U'(x')}=\int_{U_0}^{U(x)}\frac{dU}{U^2}\frac{\tilde{R}^2U_0^2}{\sqrt{U^4-U_0^4}}\underset{(v\equiv\frac{U}{U_0})}{=}\frac{\tilde{R}^2}{U_0}\int_{1}^{U(x)/U_0}\frac{dv}{v^2\sqrt{v^4-1}}\;\;.
\end{equation}
For $x=\frac{r}{2}$ (where one puts the heavy quark $Q$),
$U(\frac{r}{2})\to+\infty$ such that
\begin{equation}\label{wilson-Maldacena interquark distance}
r(U_0)=\frac{2\tilde{R}^2}{U_0}\int_1^\infty\frac{dv}{v^2\sqrt{v^4-1}}=\frac{\tilde{R}^2}{U_0}\frac{1}{\rho}
\end{equation}
where we have defined the numerical factor:
\begin{equation}
\rho=\frac{\Gamma(1/4)^2}{(2\pi)^{3/2}}\;\;.
\end{equation}
This relation could have also been guessed from the underlying
conformal nature of the theory since $r\ra\lambda\,r$ under a
dilatation $z\ra\lambda z$, \emph{i.e.} when
$U_0\ra\frac{U_0}{\lambda}$ (see Eq.\eqref{Wilson-new holographic
coordinate}). For later convenience, it is also straightforward to
express $U_0$ in terms of $r$:
\begin{equation}\label{Wilson-U0}
U_0(r)=\frac{\tilde{R}^2}{\rho}\frac{1}{r}\;\;.
\end{equation}
We are now ready to derive, at least naively, the static
$V_{Q\overline{Q}}(r)$ potential. The Euclidean version of the
Feynman-Kac formula \eqref{Wilson-feynmann kac formula} allows us
to write:
\begin{equation}
V_{Q\overline{Q}}(r)=\underset{T\to\infty}{\textrm{lim}}\,\frac{1}{T}S_{NG}[\CMcal{C}]=\frac{1}{2\pi}\int_{-r/2}^{r/2}dx\sqrt{U'\,^2+\frac{U^4}{\tilde{R}^4}}\;\;.
\end{equation}
Using successively the properties of the holographic coordinate
($U(-x)=U(x)$ and $U'(-x)=-U'(x)$), the first integral
\eqref{Wilson-first integral} and the expression of the derivative
$U'(x)$ \eqref{Wilson-expression of the derivative}, we have:
\begin{eqnarray}
V_{Q\overline{Q}}(r)&=&\frac{1}{\pi}\int_0^{r/2}dx\sqrt{U'\,^2+\frac{U^4}{\tilde{R}^4}}=\frac{1}{\pi}\int_{U_0}^{\infty}\frac{d
U}{U'}\frac{U^4}{\tilde{R}^2U_0^2}\non\\
&=&\frac{1}{\pi\tilde{R}^2U_0^2}\int_{U_0}^{\infty}dU
U^4\frac{\tilde{R}^2U_0^2}{U^2\sqrt{U^4-U_0^4}}\underset{(v\equiv\frac{U}{U_0})}{=}\frac{U_0}{\pi}\int_1^\infty
dv\frac{v^2}{\sqrt{v^4-1}}\;\;.\label{wilson-potential steps}
\end{eqnarray}
As a matter of fact, the last integral gives rise to an infinite
result: the potential indeed needs to be regularized according to
the prescription $U(x)\leq U_{max}$ (the limit $U_{max}\ra+\infty$
being taken at the end of the calculations). Actually, the final
recipe for computing the Wilson loop, as proposed by Maldacena
\cite{Maldacena wilson loop}, turns out to be:
\begin{equation}\label{wilson-static heavy quark antiquark potential
formula}
V^{(R)}_{Q\overline{Q}}(r)=\underset{\underset{M\to\infty}{T\to\infty}}{\textrm{lim}}\,\frac{1}{T}\big(S_{NG}-\ell
M\big)
\end{equation}
where $\ell$ is the perimeter of the loop $\CMcal{C}$ on the
boundary:
\begin{equation}
\ell=2T+2r\underset{T\gg r}{=}2T
\end{equation}
which amounts, in the limit $T\ra\infty$, to twice the temporal
extent $T$ of the contour. $M$ is the mass of the so-called
"W-boson string" associated to the quarks \cite{Maldacena wilson
loop}. For infinitely massive quark and antiquark, this open
string stretches all the way from $U=0$ (one brane which is far
away from the boundary) to $U=\infty$ (where $N$ coincident
D3-branes defines, at low energies, a $SU(N)$ Yang-Mills theory).
In a flat space-time, the mass-squared $M^2$ of a stretched string
reads as (the quantum fluctuations are neglected)\footnote{In a
$d$-dimensional flat space-time, an open string between two
parallel D$p$- and D$q$-branes, located respectively at $x_1$ and
$x_2$, has the following square mass spectrum ($p>q$):
\begin{equation}\label{string-mass-squared}
M^2=\Big(\frac{x_2^a-x_1^a}{2\pi\a'}\Big)^2+\frac{1}{\a'}\Big(N^{(i,r,a)}-1+\frac{1}{16}(p-q)\Big)
\end{equation}
with the number operator $\displaystyle
N^{(i,r,a)}=\sum_{n=1}^{\infty}\sum_{i=2}^q
n\,{a_n^i}^{\dagger}a_n^i+\sum_{k\in\mathbb{Z}_{odd}^+}\sum_{r=q+1}^p
\frac{k}{2}{a_{\frac{k}{2}}^r}^{\dagger}a_{\frac{k}{2}}^r+\sum_{m=1}^{\infty}\sum_{a=p+1}^d
m\,{a_m^a}^{\dagger}a_m^a$. The indices $i$ and $a$ refer
respectively to the tangential and to the normal directions to the
two branes while the index $r$ stands for the remaining tangential
coordinates for the D$p$-brane and, thus, additional normal
coordinates to the D$q$-brane.}:
\begin{equation}\label{Wilson-square mass}
M^2=\Big(\frac{x_2^a-x_1^a}{2\pi\a'}\Big)^2\;\;.
\end{equation}
The locations of the two parallel, separated D-branes, on which
lie each of the endpoints of the open string, are specified by the
values $x_1^a$ and $x_2^a$ of the coordinates normal,
respectively, to the first and to the second branes. In our case,
the bulk is $AdS_5$ and the superscript $a$ corresponds to the
only normal coordinate to the branes, \emph{i.e.} the fifth
holographic coordinate ($a=5$). Since the string tension is
$T_0=\frac{1}{2\pi\a'}$, one sees that \eqref{Wilson-square mass}
is simply the square of the energy of a classical static string
stretched between two D-branes. In short, the second term on the
\emph{r.h.s.} of \eqref{wilson-static heavy quark antiquark
potential formula} consists in subtracting in the Nambu-Goto
action $S_{NG}$ the contribution of the (infinitely stretched)
string associated to the (infinitely massive) quark and antiquark.
In \eqref{Wilson-square mass}, the ratio $\frac{x}{\a'}$ has the
dimension of an energy and, therefore, can be identified to the
holographic coordinate $U(x)$. The regularized mass of the
"W-boson string" (in string theory, such a stretched open string
corresponds to a massive vector field) is
\begin{equation}\label{wilson-string theory square mass}
M=\frac{U_{max}}{2\pi}
\end{equation}
such that
\begin{eqnarray}
V^{(R)}_{Q\overline{Q}}(r)&=&\frac{1}{2\pi}\int_{-r/2}^{r/2}dx\sqrt{U'\,^2+\frac{U^4}{\tilde{R}^4}}-\frac{U_{max}}{\pi}\non\\
\non\\
&=&\frac{U_0}{\pi}\int_{1}^{U_{max}/U_0}dv\frac{v^2}{\sqrt{v^4-1}}-\frac{U_{max}}{\pi}\non\\
\non\\
&=&\frac{U_0}{\pi}\int_{1}^{U_{max}/U_0}dv\Big(\frac{v^2}{\sqrt{v^4-1}}-1\Big)+\frac{U_0}{\pi}\int_1^{U_{max}/U_0}dv-\frac{U_{max}}{\pi}\non\\
\non\\
V^{(R)}_{Q\overline{Q}}(r)&=&\frac{U_0}{\pi}\Big[\int_{1}^{U_{max}/U_0}dv\Big(\frac{v^2}{\sqrt{v^4-1}}-1\Big)-1\Big]\;\;.\label{Wilson-regularized
potential}
\end{eqnarray}
The integral turns out to be finite in the limit
$U_{\max}\to+\infty$ with the result:
\begin{eqnarray}
V^{(R)}_{Q\overline{Q}}(r)&=&-\frac{U_0}{\pi}\Big(E(-1)-(2-i)K(-1)+K(2)\Big)\non\\
&=&-\frac{U_0}{\pi}\Big(E(-1)-K(-1)\Big)\;\;,\label{wilson-AdS/CFT
potential}
\end{eqnarray}
expressed in terms of the complete Elliptic Integrals $K(z)$ and
$E(z)$ of the first and second kind respectively and where we have
used the relation
\begin{equation}
K(1/z)=\sqrt{z}\Big(K(z)-\sqrt{-\frac{1}{z}}\sqrt{\frac{1}{1-z}}\sqrt{z(1-z)}K(1-z)\Big)
\end{equation}
with $z=-1$. According to \eqref{Wilson-U0} and to the numerical
values of $K(-1)$ and $E(-1)$:
\begin{eqnarray}
K(-1)&=&\frac{\G(1/4)^2}{4\sqrt{2\pi}}=\frac{\pi}{2}\rho\;\;,\\
E(-1)&=&\frac{2\G(3/4)^4+\pi^2}{2\sqrt{2\pi}\G(3/4)^2}=\frac{1}{2\rho}+\frac{\pi}{2}\rho\;\;,\label{E(-1)}
\end{eqnarray}
it is possible to rewrite the static potential as:
\begin{equation}
V^{(R)}_{Q\overline{Q}}(r)=-\frac{\tilde{R}^2}{2\pi\rho^2}\frac{1}{r}
\end{equation}
or, in terms of the 't Hooft coupling $\lambda$
\eqref{wilson-duality relation}:
\begin{equation}\label{wilson-potential}
V^{(R)}_{Q\overline{Q}}(r)=-\frac{4\pi^2\sqrt{2\lambda}}{\G(1/4)^4}\frac{1}{r}\;\;.
\end{equation}
This result is valid for all distances $r$ when
$\lambda=g_{YM}^2N$ is large independently of the value of
$g_{YM}$. Especially, we do not recover the area law. Moreover,
although the potential seems to have a short-distance Coulomb-like
behaviour in $1/r$ (a fact which is determined by conformal
invariance: the factor $1/\a'$ in the string tension disappears in
\eqref{Wilson-Nambu-Goto action} and, consequently, in the
expression of the potential \eqref{Wilson-regularized potential}),
it goes as $\sqrt{\lambda}$ instead of $\lambda$ which is the
perturbative one-loop result: actually, the potential
\eqref{wilson-potential} turns out to be intrinsically
non-perturbative.

\section{The static potential at finite temperature}

\subsection{The conformal behaviour of the Wilson loop at finite
temperature}

Following Hawking and Page's work on the thermodynamics of black
holes in the anti-de Sitter space-time $AdS_4$ \cite{Hawking}, a
gauge/string duality involving a gauge theory at finite
temperature was proposed by Witten \cite{Witten conformal
breaking}. In this framework, on the supergravity side, the
$AdS_5$ space-time \eqref{wilson-line element 2} turns out to
accommodate a Schwarzschild black hole (BH) with (the Euclidean
version of) the metric:
\begin{equation}\label{wilson-finite temperature line element}
ds_{BH}^2=\a'\Big\{\frac{u^2}{\tilde{R}^2}\big(f(u)dt^2+\sum_{i=1}^3dx_i^2\big)+\frac{\tilde{R}^2}{u^2}\frac{du^2}{f(u)}+\tilde{R}^2d\Omega_5^2\Big\}\;\;.
\end{equation}
We have defined:
\begin{equation}
f(u)=1-\frac{u_T^4}{u^4}
\end{equation}
and $\tilde{R}^2$ as in \eqref{Wilson-R tilde}. $d\Omega_5^2$ is
the line element of the unit radius compact 5-sphere $S^5$. There
is a curvature singularity in the IR at $u=0$ hidden behind an
event horizon at $u=u_T$. In particular, at zero temperature,
which corresponds to $u_T=0$ as we shall see below
\eqref{wilson-relation temperature-horizon}, we recover the metric
of the $AdS_5\times S^5$ space-time \eqref{wilson-line element 2}.
Such a solution (with an event horizon) is also called the
near-extremal D3-brane solution of the equations of motion for the
metric in type IIB superstring theory, the extremal case
corresponding to the absence of horizon $u_T=0$ and $f(u)=1$,
\emph{i.e.} to the zero temperature case.

The Euclidean time direction shrinks to a zero-size geometrical
point at the horizon since $f(u_T)=0$ and is thus compactified on
a circle with period $\b$: $t\sim t+\b$. This period
$\b=\frac{1}{T}$ is the inverse of the Beckenstein-Hawking
temperature (which corresponds to the thermal temperature of the
gauge theory) of the near-extremal solution \eqref{wilson-finite
temperature line element} and gives the location $u_T$ of the
horizon. We can reason as in \cite{GKT}. The region of interest
here is near $u_T$. So, let us define a new holographic coordinate
$\rho$ as $u(\rho)=u_T\big(1+\frac{\rho^2}{\a'\tilde{R}^2}\big)$.
The relevant two-dimensional part of the metric
\eqref{wilson-finite temperature line element} becomes:
\begin{eqnarray}
ds^2_{BH}&=&\a'\frac{u^2}{\tilde{R}^2}f(u)dt^2+\frac{\tilde{R}^2}{u^2}\frac{du^2}{f(u)}+\ldots\non\\
&=&\frac{\a'}{\tilde{R}^2}u_T^2\big(1+\frac{\rho^2}{\a'\tilde{R}^2}\big)^2\big[1-\big(1+\frac{\rho^2}{\a'\tilde{R}^2}\big)^{-4}\big]dt^2\non\\
&&+\frac{\a'\tilde{R}^2}{u_T^2}\big(1+\frac{\rho^2}{\a'\tilde{R}^2}\big)^{-2}\big[1-\big(1+\frac{\rho^2}{\a'\tilde{R}^2}\big)^{-4}\big]^{-1}\frac{4u_T^2}{{\a'}^2\tilde{R}^4}\rho^2d\rho^2+\ldots\non\\
ds^2_{BH}&=&d\rho^2+\rho^2d\big(\frac{2u_T}{\tilde{R}^2}t\big)^2+\ldots\label{wilson-temperature}
\end{eqnarray}
We then recognize on the \emph{r.h.s.} of
\eqref{wilson-temperature} the (dimensionless) angle
$\theta\equiv\frac{2u_T}{\tilde{R}^2}t$ with period $2\pi$ such
that $\frac{2u_T}{\tilde{R}^2}\b=2\pi$ or $u_T=\pi\tilde{R}^2T$.
The Beckenstein-Hawking temperature can also be derived from the
formula ($\d_{00}$ is the Euclidean metric component with time
indices):
\begin{eqnarray}
T&=&\frac{1}{4\pi\a'}\frac{\de\,\d_{00}}{\de u}\Big|_{u=u_T}\\
&=&\frac{1}{4\pi\a'}\frac{\de}{\de
u}\Big(\a'\frac{u^2}{\tilde{R}^2}f(u)\Big)\Big|_{u=u_T}\\
T&=&\frac{u_T}{\pi\tilde{R}^2}\;\;.\label{wilson-relation
temperature-horizon}
\end{eqnarray}

We still work with a \emph{space-time} Wilson loop (or
\emph{ordinary} Wilson loop, in contrast to the \emph{spatial}
Wilson loop that we shall consider in the non-conformal cases in
the next sections) for which the boundary temporal and spatial
extents are given by \eqref{wilson-parameter1} and
\eqref{wilson-parameter2}. In the static gauge $X^0(\tau,\s)\equiv
t=\tau$ \eqref{wilson-static gauge} and $\s=x$
\eqref{wilson-reparametrization}, the Nambu-Goto action is
\cite{BISY conformal}:
\begin{equation}\label{wilson-conformal action}
S_{NG}[\CMcal{C}]=\frac{1}{2\pi\a'}\int
d^2\xi\sqrt{det(\g_{ab})}=\frac{T}{2\pi}\int_{-r/2}^{r/2}dx\sqrt{{U'}^2+\frac{U^4-U_T^4}{\tilde{R}^4}}
\end{equation}
with
\begin{eqnarray}
\g_{11}&=&\a'\frac{U^2}{\tilde{R}^2}+\a'\frac{\tilde{R}^2}{U^2}\frac{{U'}^2}{f(U)}\;\;,\\
\g_{22}&=&\a'\frac{U^2}{\tilde{R}^2}f(U)
\end{eqnarray}
the non-vanishing components of the induced metric tensor on the
two-dimensional world-sheet (with our conventions, $\xi^1=\s$ and
$\xi^2=\tau$). $U'\equiv U'(x)$ is the derivative of the
holographic coordinate with respect to the spatial boundary
coordinate $x$. The Lagrangian density does not depend explicitly
on this $x$. Thus, the Hamiltonian in the $x$ direction is a
constant of motion and the first integral \eqref{wilson-first
integral definition} can be evaluated at $x=0$ where $U(0)=U_0$
and $U'(0)=0$ by symmetry. We obtain:
\begin{equation}\label{wilson-conformal first derivative}
\frac{U^4-U_T^4}{\sqrt{{U'}^2+\frac{U^4-U_T^4}{\tilde{R}^4}}}=\tilde{R}^2\sqrt{U_0^4-U_T^4}=\,\textrm{const.}
\end{equation}
from which can be derived an expression for the derivative
$U'(x)$:
\begin{equation}\label{wilson-conformal finite temp derivative}
U'(x)=\pm\frac{\sqrt{(U^4-U_T^4)(U^4-U_0^4)}}{\tilde{R}^2\sqrt{U_0^4-U_T^4}}
\end{equation}
where the positive (negative) square root corresponds to $0<
x\leq\frac{r}{2}$ ($-\frac{r}{2}\leq x<0$).

The integral expression for the distance $r$ between the quark and
the antiquark is derived as usual ($x\geq0$):
\begin{equation}
\frac{r}{2}-x=\int_{x}^{r/2}dx'=\int_{U(x)}^{\infty}\frac{dU}{U'}=\tilde{R}^2\sqrt{U_0^4-U_T^4}\int_{U(x)}^{\infty}\frac{dU}{\sqrt{(U^4-U_T^4)(U^4-U_0^4)}}\;\;.
\end{equation}
With respect to $v\equiv U/U_0$ and defining $\epsilon\equiv
f(U_0)=1-\frac{U_T^4}{U_0^4}\geq0$ (since $U_0\geq U_T$), we get:
\begin{equation}
\frac{r}{2}-x=\frac{\tilde{R}^2}{U_0}\sqrt{\epsilon}\int_{U(x)/U_0}^{\infty}\frac{dv}{\sqrt{(v^4-1+\epsilon)(v^4-1)}}\;\;.
\end{equation}
In particular, if $x=0$, then
\begin{equation}\label{wilson-distance temperature}
r(U_0,U_T)=\frac{2\tilde{R}^2}{U_0}\sqrt{\epsilon}\int_1^{\infty}\frac{dv}{\sqrt{(v^4-1+\epsilon)(v^4-1)}}\;\;.
\end{equation}
In the supergravity approach, the static potential stemming from
the \emph{space-time} Wilson loop with the background
\eqref{wilson-finite temperature line element} is obtained as
follows. Thanks to the first integral \eqref{wilson-conformal
first derivative}, we have derived an expression for $U'$
\eqref{wilson-conformal finite temp derivative}. It is then easy
to rewrite the action \eqref{wilson-conformal action} following
the same steps as in \eqref{wilson-potential steps}:
\begin{eqnarray}
S_{NG}[\CMcal{C}]&=&\frac{T}{\pi}\int_0^{r/2}dx\sqrt{{U'}^2+\frac{U^4-U_T^4}{\tilde{R}^2}}=\frac{T}{\pi}\int_{0}^{r/2}dx\frac{U^4-U_T^4}{\tilde{R}^2\sqrt{U_0^4-U_T^4}}\non\\
&=&\frac{T}{\pi}\int_{U_0}^{\infty}\frac{dU}{U'}\frac{U^4-U_T^4}{\tilde{R}^2\sqrt{U_0^4-U_T^4}}\underset{(v\equiv\frac{U}{U_0})}{=}T\frac{U_0}{\pi}\int_{1}^{\infty}dv\sqrt{\frac{v^4-1+\epsilon}{v^4-1}}\;\;.\label{wilson-conformal
divergence potential finite T}
\end{eqnarray}
The heavy quark potential is given by Maldacena's prescription
\eqref{wilson-static heavy quark antiquark potential formula}:
\begin{equation}\label{wilson-maldacena prescription}
V^{(R)}_{Q\overline{Q}}(r)=\underset{\underset{M\to\infty}{T\to\infty}}{\textrm{lim}}\,\frac{1}{T}\big(S_{NG}-\ell
M\big)
\end{equation}
where the regularization procedure introduces in the action
$S_{NG}$ an \emph{ultraviolet} (UV) cutoff $U\leq U_{max}$ ($v\leq
U_{max}/U_0$). The second contribution on the \emph{r.h.s.} of
\eqref{wilson-maldacena prescription} is the required counter-term
which subtracts the very massive quark and antiquark contributions
to the regularized potential and which takes in this case the
following form ($\ell=2r+2T\underset{T\gg r}{\simeq}2T$ and
$M=\frac{U_{max}-U_T}{2\pi}$):
\begin{equation}\label{wilson-conformal w-boson string}
V_{c.t.}=-\underset{U_{max}\to\infty}{\textrm{lim}}\frac{U_{max}-U_T}{\pi}\;\;.
\end{equation}
Indeed, the "W-boson string" corresponds here to an open string
stretched between the brane at $U_{max}$ and the Schwarzschild
horizon $U=U_T$ \cite{BISY conformal,RTY}. Hence, the counter-term
\eqref{wilson-conformal w-boson string} and the renormalized (or,
more properly, subtracted) static potential:
\begin{eqnarray}
V^{(R)}_{Q\overline{Q}}(U_0,U_T)&=&\frac{U_0}{\pi}\int_1^{U_{max}/U_0}dv\Big(\sqrt{\frac{v^4-1+\epsilon}{v^4-1}}-1\Big)+\frac{U_0}{\pi}\int_1^{U_{max}/U_0}dv-\frac{U_{max}-U_T}{\pi}\non\\
&=&\frac{U_0}{\pi}\int_1^{U_{max}/U_0}dv\Big(\sqrt{\frac{v^4-1+\epsilon}{v^4-1}}-1\Big)+\frac{U_T-U_0}{\pi}\;\;.\label{wilson-conformal-potential
temperature}
\end{eqnarray}

Let us focus on the limit case $U_0\gg U_T$ where the string
world-sheet is close to the boundary such that it does not feel
the presence of the horizon. In fact, this configuration
corresponds to the low temperature limit $r\ll\b$ or $rT\ll1$ (we
have $u_T=\pi\tilde{R}^2T$). Obviously, for small temperatures,
the potential behaves approximately as in the zero temperature
case \eqref{wilson-potential} $V\sim-\frac{1}{r}$ since we recover
for $\epsilon\simeq1$ the expressions \eqref{wilson-Maldacena
interquark distance} and \eqref{Wilson-regularized potential}.
Moreover, the leading non-zero temperature correction exhibits
scaling consistent with the conformal invariance of the boundary
theory. \cite{BISY conformal} obtained:
\begin{equation}
V\propto-\frac{1}{r}\Big(1+a(rT)^4\Big)
\end{equation}
with $a$ a positive numerical constant which does not depend on
$\tilde{R}$. Without length scale, it is indeed meaningless to
speak, at low temperature, of a large or small compactification
radius of the Euclidean temporal dimension.

The high temperature limit $r\gg\b$ or $rT\gg1$ when $U_0\simeq
U_T$ is more subtle. As shown in \cite{BISY conformal,RTY}, there
is a critical value of the inter-quark distance $r_c$ above which
the potential starts to be positive. At this point, the
bound-state equations \eqref{wilson-distance temperature} and
\eqref{wilson-conformal-potential temperature} are no longer valid
because the lowest energy configuration consists instead of two
straight strings ending at the horizon. In other words, the quarks
become free as screened by the effects of the temperature. Hence,
the potential exhibits a behaviour expected for the deconfinement
phase at high temperature when the meson decays into a
configuration of quarks without interaction.

\subsection{The area law in three-dimensional Yang-Mills theory}

Following \cite{Witten conformal breaking,greensite-olesen,BISY},
we consider a \emph{spatial} Wilson loop $W[\CMcal{C}]$ (along two
space-like dimensions) at \emph{fixed value} of the temperature
and take the spatial extent $Y$ to be large with respect to the
other spatial direction $Y\gg r$. We choose the following
parametrization for the string world-sheet:
\begin{eqnarray}
-\frac{r}{2}\leq x\leq\frac{r}{2}\;\;,\\
-\frac{Y}{2}\leq y\leq\frac{Y}{2}\;\;.
\end{eqnarray}
In the limit $Y\to\infty$, the world-sheet configuration is
invariant under translation in the $Y$ direction. It is then
straightforward to write out the classical action
\eqref{wilson-nambu goto action} of the space-like Nambu-Goto
string in the background \eqref{wilson-finite temperature line
element}. The relevant components of the induced metric tensor are
this time ($\xi^1=x$ and $\xi^2=y$):
\begin{eqnarray}
\g_{11}&=&\a'\frac{U^2}{\tilde{R}^2}+\a'\frac{\tilde{R}^2}{U^2}\frac{{U'}^2}{f(U)}\;\;,\\
\g_{22}&=&\a'\frac{U^2}{\tilde{R}^2}
\end{eqnarray}
such that
\begin{equation}\label{wilson-nambu-goto action temperature}
S_{NG}[\CMcal{C}]=\frac{1}{2\pi\a'}\int_{-Y/2}^{Y/2}dy\int_{-r/2}^{r/2}dx\sqrt{det(\g_{ab})}=\frac{Y}{2\pi}\int_{-r/2}^{r/2}dx\sqrt{\frac{U^4}{\tilde{R}^4}+\frac{U^4}{U^4-U_T^4}U'\,^2}\;\;.
\end{equation}
As always, since the Lagrangian density does not depend explicitly
on $x$, we find a first integral which here takes the following
form ($U(0)\equiv U_0$ and $U'(0)=0$ by symmetry):
\begin{equation}\label{wilson-first integral temperature}
\frac{U^4}{\sqrt{\frac{U^4}{\tilde{R}^4}+\frac{U^4}{U^4-U_T^4}U'\,^2}}=\tilde{R}^2U_0^2=\textrm{const.}
\end{equation}
or, in terms of the derivative $U'(x)$ of the holographic
coordinate:
\begin{equation}\label{wilson-expression of the derivative
temperature}
U'(x)=\pm\frac{U^2}{\tilde{R}^2}\sqrt{\Big(1-\frac{U_T^4}{U^4}\Big)\Big(\frac{U^4}{U_0^4}-1\Big)}\;\;.
\end{equation}

The next step consists in deriving the expressions of the
inter-quark distance and of the heavy quark potential as functions
of $U_0$ and $U_T$. As for the distance $r(U_0,U_T)$, we have:
\begin{eqnarray}\label{wilson-interquark distance}
r(U_0,U_T)&=&\int_{-r/2}^{r/2}dx=2\int_0^{r/2}dx=2\int_{U_0}^{\infty}\frac{d U}{U'}=2\tilde{R}^2\int_{U_0}^{\infty}\frac{dU}{U^2}\frac{1}{\sqrt{(1-\frac{U_T^4}{U^4})(\frac{U^4}{U_0^4}-1)}}\non\\
&\underset{(v\equiv\frac{U}{U_0})}{=}&\frac{2\tilde{R}^2}{U_0}\int_1^{\infty}\frac{dv}{\sqrt{(v^4-1+\epsilon)(v^4-1)}}
\end{eqnarray}
with $\epsilon\equiv f(U_0)=1-\frac{U_T^4}{U_0^4}$. In the limit
$U_0\simeq U_T$ ($\epsilon\ll1$) where the string world-sheet
reaches the horizon, the inter-quark distance diverges:
\begin{equation}\label{wilson-interquark distance divergence}
r(U_0)=\frac{2\tilde{R}^2}{U_0}\int_{a}^{\infty}dv\frac{1}{v^4-1}\underset{a\to1}{\sim}-\frac{\tilde{R}^2}{2U_0}\ln(a-1)\;\;.
\end{equation}
Thus, we see that the large distance limit (where the confinement
is expected to appear) consists then in taking the limit
$U_0\simeq U_T$. On the other hand, when $r\gg\b=\frac{1}{T}$, the
circle $S^1(\b)$ around the compactified Euclidean time direction
is small and, as a result, the number of dimensions of the gauge
theory on the boundary reduces to three. By choosing appropriate
boundary conditions along this circle (namely, by taking
anti-periodic fermions around $S^1(\b)$ in contrast to the
periodic bosons), the supersymmetry can also be broken
\cite{Witten conformal breaking}. Moreover, as both fermions and
scalars get masses related to the temperature (due to
renormalization for the latter), they decouple at high enough
temperature and the theory reduces to a pure non-conformal gauge
theory. We are thus considering, at large distances,
three-dimensional non-supersymmetric Yang-Mills theory at zero
temperature (hence the title of this section). On the contrary, at
small distances $r\ll\b$, the compactification radius of the
circle turns out to be sizeable. We deal therefore with
four-dimensional supersymmetric Yang-Mills theory at zero
temperature and, not surprisingly, we recover Maldacena's result
\eqref{wilson-Maldacena interquark distance} ($U_0\gg U_T$ or
$\epsilon\simeq1$):
\begin{equation}
r(U_0)=\frac{2\tilde{R}^2}{U_0}\int_{1}^{\infty}dv\frac{1}{v^2\sqrt{v^4-1}}=\frac{\tilde{R}^2}{U_0}\frac{1}{\rho}
\end{equation}
and the Coulomb-like behaviour of the potential
\eqref{wilson-potential}. We are now ready to treat, in the
supergravity approach, the static potential derived from a
\emph{spatial} Wilson loop with the background
\eqref{wilson-finite temperature line element}. Following the
standard procedure, the action can be rewritten as
\begin{eqnarray}
S_{NG}[\CMcal{C}]&=&\frac{Y}{\pi}\int_0^{r/2}dx\sqrt{\frac{U^4}{\tilde{R}^2}+\frac{U^4}{U^4-U_T^4}\,U'^2}=\frac{Y}{\pi}\int_{0}^{r/2}dx\frac{U^4}{\tilde{R}^2U_0^2}\non\\
&=&\frac{Y}{\pi}\int_{U_0}^{\infty}\frac{dU}{U'}\frac{U^4}{\tilde{R}^2U_0^2}\underset{(v\equiv\frac{U}{U_0})}{=}Y\frac{U_0}{\pi}\int_{1}^{\infty}dv\frac{v^4}{\sqrt{(v^4-1+\epsilon)(v^4-1)}}\;\;.\label{wilson-divergence
potential finite T}
\end{eqnarray}
The heavy quark potential is then given by Maldacena's
prescription \eqref{wilson-static heavy quark antiquark potential
formula}:
\begin{equation}
V^{(R)}_{Q\overline{Q}}(r)=\underset{\underset{M\to\infty}{Y\to\infty}}{\textrm{lim}}\,\frac{1}{Y}\big(S_{NG}-\ell
M\big)
\end{equation}
where $S_{NG}$ is the regularized action (with the UV cutoff
$U\leq U_{max}$ or $v\leq U_{max}/U_0$) and the counter-term is
similar to \eqref{wilson-conformal w-boson string}
($\ell=2r+2Y\underset{Y\gg r}{\simeq}2Y$ and
$M=\frac{U_{max}-U_T}{2\pi}$):
\begin{equation}
V_{c.t.}=-\underset{U_{max}\to\infty}{\textrm{lim}}\frac{U_{max}-U_T}{\pi}\;\;.
\end{equation}
We have \cite{greensite-olesen}:
\begin{eqnarray}
V_{Q\overline{Q}}^{(R)}(r)&=&\frac{U_0}{\pi}\int_{1}^{U_{max}/U_0}dv\Big(\frac{v^4}{\sqrt{(v^4-1+\epsilon)(v^4-1)}}-1+1\Big)-\frac{U_{max}-U_T}{\pi}\non\\
&=&\frac{U_0}{\pi}\int_{1}^{U_{max}/U_0}dv\Big(\frac{(v^4-1)+1}{\sqrt{(v^4-1+\epsilon)(v^4-1)}}-1\Big)+\frac{U_0}{\pi}\int_{1}^{U_{max}/U_0}dv-\frac{U_{max}-U_T}{\pi}\non\\
&=&\frac{U_0}{\pi}\int_{1}^{\infty}\frac{dv}{\sqrt{(v^4-1+\epsilon)(v^4-1)}}+\frac{U_0}{\pi}\int_{1}^{\infty}dv\Big(\sqrt{\frac{v^4-1}{v^4-1+\epsilon}}-1\Big)+\frac{U_T-U_0}{\pi}\non\\
V_{Q\overline{Q}}^{(R)}(r)&=&\frac{U_0^2}{2\pi\tilde{R}^2}r+\frac{U_0}{\pi}\int_{1}^{\infty}dv\Big(\sqrt{\frac{v^4-1}{v^4-1+\epsilon}}-1\Big)+\frac{U_T-U_0}{\pi}\;\;.\label{wilson-renormalized
potential}
\end{eqnarray}

We are interested in the leading and subleading terms in the
static potential at large quark separation, \emph{i.e.} when
$U_0\simeq U_T$ ($\epsilon\ll1$). If we remarks that ($i$ is the
imaginary unit):
\begin{equation}\label{wilson-relation 1}
(v-1)(v+1)(v-i)(v+i)=v^4-1\;\;,
\end{equation}
\begin{equation}\label{wilson-relation 2}
(v-1+\frac{\epsilon}{4})(v+1-\frac{\epsilon}{4})(v-i+i\frac{\epsilon}{4})(v+i-i\frac{\epsilon}{4})=v^4-(1-\frac{\epsilon}{4})^4\underset{\epsilon\ll1}{\simeq}v^4-1+\epsilon+O(\epsilon^2)\;\;,
\end{equation}
then the inter-quark distance $r$ \eqref{wilson-interquark
distance} can be rewritten as
\begin{equation}\label{wilson-interquark separation modification}
r(U_0)\simeq\frac{2\tilde{R}^2}{U_0}\int_1^{\infty}\frac{dv}{\sqrt{(v-1)(v-1+\frac{\epsilon}{4})}}\frac{1}{\sqrt{F_{\epsilon}(v)}}
\end{equation}
where we have defined a new function:
\begin{equation}
F_{\epsilon}(v)=(v+1)(v-i)(v+i)(v+1-\frac{\epsilon}{4})(v-i+i\frac{\epsilon}{4})(v+i-i\frac{\epsilon}{4})\;\;,
\end{equation}
regular in $y=1$ and/or $\epsilon=0$:
\begin{eqnarray}
F_{\epsilon}(1)&=&(8-\epsilon)(2-\frac{\epsilon}{2}+\frac{\epsilon^2}{16})\underset{\epsilon\ll1}{=}16-6\epsilon+O(\epsilon^2)\;\;,\\
F_0(v)&=&(v+1)^2(v^2+1)^2\underset{v=1}{=}16
\end{eqnarray}
and which behaves asymptotically as
$F_{\epsilon}(v)\underset{v\gg1}{\sim}v^6$. In this way, we focus
on the main contribution of the integral \eqref{wilson-interquark
separation modification} which comes from the region $v=1$ (see
Eq.\eqref{wilson-interquark distance divergence}). A partial
integration gives:
\begin{eqnarray}
r(U_0)&\simeq&\frac{4\tilde{R}^2}{U_0}\Big[\ln\Big(\sqrt{v-1}+\sqrt{v-1+\frac{\epsilon}{4}}\big)\frac{1}{\sqrt{F_{\epsilon}(v)}}\Big]_1^{\infty}\non\\
&&+\frac{4\tilde{R}^2}{U_0}\int_1^{\infty}dv\ln\Big(\sqrt{v-1}+\sqrt{v-1+\frac{\epsilon}{4}}\big)\frac{F'_{\epsilon}(v)}{F_{\epsilon}(v)^{3/2}}\non\\
r(U_0)&\simeq&-\frac{\tilde{R}^2}{2U_0}\ln\,\epsilon+O(\epsilon\ln\epsilon)
\end{eqnarray}
or ($U_0\simeq U_T$)
\begin{equation}\label{wilson-epsilon in terms of r}
\epsilon\simeq e^{-\frac{2U_T}{\tilde{R}^2}r}\;\;.
\end{equation}
As for the heavy quark potential \eqref{wilson-renormalized
potential}, it is convenient to defined the function
$J(\epsilon)$:
\begin{eqnarray}
J(\epsilon)&\equiv&\int_{1}^{\infty}dv\Big(\sqrt{\frac{v^4-1}{v^4-1+\epsilon}}-1\Big)\;\;\textrm{with}\;\;J(0)=0\;\;,\\
\frac{\de
J(\epsilon)}{\de\epsilon}&=&-\frac{1}{2}\int_{1}^{\infty}dv\frac{\sqrt{v^4-1}}{(v^4-1+\epsilon)^{3/2}}\label{wilson-derivative
function}\;\;.
\end{eqnarray}
Because of \eqref{wilson-relation 1} and \eqref{wilson-relation
2}, the integral \eqref{wilson-derivative function} can be
rewritten, in the limit $\epsilon\ll1$, as
\begin{equation}
\frac{\de
J(\epsilon)}{\de\epsilon}\simeq-\frac{1}{2}\int_{1}^{\infty}dv\frac{\sqrt{v-1}}{(v-1+\frac{\epsilon}{4})^{3/2}}\phi_{\epsilon}(v)
\end{equation}
where the function:
\begin{equation}
\phi_{\epsilon}(v)=\frac{\sqrt{(v+1)(v-i)(v+i)}}{(v+1-\frac{\epsilon}{4})^{3/2}(v-i+i\frac{\epsilon}{4})^{3/2}(v+i-i\frac{\epsilon}{4})^{3/2}}
\end{equation}
is regular in $y=1$ and/or $\epsilon=0$:
\begin{eqnarray}
\phi_{\epsilon}(1)&=&\frac{2}{(2-\frac{\epsilon}{4})^3}\underset{\epsilon\ll1}{=}\frac{1}{4}+\frac{3}{16}\epsilon+O(\epsilon^2)\;\;,\\
\phi_{0}(v)&=&\frac{1}{(v+1)(v^2+1)}\underset{v=1}{=}\frac{1}{4}
\end{eqnarray}
with the asymptotic behaviour
$\phi_{\epsilon}(v)\underset{v\gg1}{\sim}\frac{1}{v^3}$. By
partial integration, we obtain:
\begin{eqnarray}
\frac{\de J(\epsilon)}{\de\epsilon}&\simeq&\Big\{\Big[\sqrt{\frac{v-1}{v-1+\frac{\epsilon}{4}}}-\ln\Big(\sqrt{v-1}+\sqrt{v-1+\frac{\epsilon}{4}}\Big)\Big]\phi_{\epsilon}(v)\Big\}_{1}^{\infty}\non\\
&&+\int_1^{\infty}dv\Big[-\sqrt{\frac{v-1}{v-1+\frac{\epsilon}{4}}}+\ln\Big(\sqrt{v-1}+\sqrt{v-1+\frac{\epsilon}{4}}\Big)\Big]\phi'_{\epsilon}(v)\\
\frac{\de
J(\epsilon)}{\de\epsilon}&\simeq&\frac{1}{8}\ln\,\epsilon+I(\epsilon)+O(\epsilon^0)\;\;.
\end{eqnarray}
The integral $I(\epsilon)$ can then be treated analogously:
\begin{eqnarray}
I(\epsilon)&\equiv&\int_1^{\infty} dv\Big[-\sqrt{\frac{v-1}{v-1+\frac{\epsilon}{4}}}+\ln\Big(\sqrt{v-1}+\sqrt{v-1+\frac{\epsilon}{4}}\Big)\Big]\phi'_{\epsilon}(v)\\
&=&\Big\{\Big[-\frac{3}{2}\sqrt{(v-1)(v-1+\frac{\epsilon}{4})}+\big(v-1+\frac{3\epsilon}{8}\big)\ln\Big(\sqrt{v-1}+\sqrt{v-1+\frac{\epsilon}{4}}\Big)\Big]\phi'_{\epsilon}(v)\Big\}_1^{\infty}\non\\
&&+\int_1^{\infty}
dv\Big[\frac{3}{2}\sqrt{(v-1)(v-1+\frac{\epsilon}{4})}-\big(v-1+\frac{3\epsilon}{8}\big)\ln\Big(\sqrt{v-1}+\sqrt{v-1+\frac{\epsilon}{4}}\Big)\Big]\phi''_{\epsilon}(v)\non\\
\end{eqnarray}
which shows, with
$\phi'_{\epsilon}(1)\underset{\epsilon\ll1}{=}-\frac{3}{8}-\frac{21}{64}\epsilon+O(\epsilon^2)$
and $\phi'_{\epsilon}(v)\underset{v\gg1}{\sim}\frac{1}{v^5}$, that
$I(\epsilon)$ is of order $O(\epsilon\ln\,\epsilon)$:
\begin{eqnarray}
I(\epsilon)&=&-\frac{3\epsilon}{16}\ln\Big(\frac{\epsilon}{4}\Big)\phi'_{\epsilon}(1)\non\\
&&+\int_1^\infty
dv\Big[\frac{3}{2}\sqrt{(v-1)(v-1+\frac{\epsilon}{4})}-\big(v-1+\frac{3\epsilon}{8}\big)\ln\Big(\sqrt{v-1}+\sqrt{v-1+\frac{\epsilon}{4}}\Big)\Big]\phi''_{\epsilon}(v)\non\\
I(\epsilon)&=&\frac{9}{128}\epsilon \ln\,\epsilon+O(\epsilon^2
\ln\,\epsilon)\;\;.
\end{eqnarray}
At the end of the day, the main contribution of
\eqref{wilson-derivative function} in the limit $\epsilon\ll1$
turns out to be
\begin{equation}
\frac{\de
J(\epsilon)}{\de\epsilon}=\frac{1}{8}\ln\,\epsilon+O(\epsilon\ln\,\epsilon)\;\;,
\end{equation}
that is \begin{equation}
J(\epsilon)=\int_0^{\epsilon}\de\epsilon'\frac{\de
J(\epsilon')}{d\epsilon'}=\frac{1}{8}\epsilon\ln(\epsilon)+O(\epsilon^2\ln\epsilon)\;\;.
\end{equation}
Thus, the heavy quark potential ($U_0\simeq U_T$):
\begin{equation}
V_{Q\overline{Q}}^{(R)}(r,U_T)=\frac{U_T^2}{2\pi\tilde{R}^2}r+\frac{U_T}{\pi}\frac{1}{8}\epsilon\ln(\epsilon)+O(\epsilon^2\ln\epsilon)\;\;.
\end{equation}
In terms of the distance between quarks \eqref{wilson-epsilon in
terms of r}, we find a leading correction to the linear potential
exponentially small for $rT\gg1$ \cite{greensite-olesen}:
\begin{equation}
V_{Q\overline{Q}}^{(R)}(r,U_T)\simeq\frac{U_T^2}{2\pi\tilde{R}^2}r\Big(1-\frac{1}{2}e^{-\frac{2U_T}{\tilde{R}^2}r}\Big)\;\;.
\end{equation}
On the other hand, as expected, the string tension is proportional
to (the square of) the temperature since it is our only
dimensionful parameter at hand:
\begin{equation}\label{wilson-string tension temperature}
\s=\frac{U_T^2}{2\pi\tilde{R}^2}=\frac{1}{2}\pi\tilde{R}^2T^2=\sqrt{\pi^3
g_s N}\,T^2\;\;.
\end{equation}

The subleading term in the static potential at large quark
separation is not in $1/r$ which is, at first sight, in
contradiction with predictions from effective string models and
Lattice QCD \cite{teper}. Instead, these latter tend to confirm a
subleading attractive Coulomb-like contribution to the linear
potential, the so-called L\"{u}scher term $-c/r$ where $c$ is a
universal numerical constant \cite{luscher}. Nevertheless, this
result is not so surprising since the limits at work in the
supergravity approach are the large $N$ and the large 't Hooft
coupling constant limits and it is known that there is no
L\"{u}scher term in the strong coupling regime on the lattice
whereas it appears in the weak coupling phase.
\cite{greensite-olesen,gross ooguri} stressed the fact that such a
phase transition could also occur in the supergravity approach as
the 't Hooft coupling is reduced and argued that the L\"{u}scher
term could arise from quantum fluctuations of the classical
world-sheet approximation \cite{greensite-olesen luscher}.

Finally, it is worth pointing out that the linear behaviour of the
static potential is not spoilt by the leading stringy corrections
$O({\a'}^3)$ (which consists also of an expansion in $1/\sqrt{N}$
according to \eqref{wilson-duality relation}) of the Schwarzschild
black hole$-AdS_5$ metric \eqref{wilson-finite temperature line
element}. The line element was found to be \cite{GKT}:
\begin{equation}
ds^2_{BH}=\a'\Big\{(1+\d_2)\frac{u^2}{\tilde{R}^2}f(u)dt^2+\frac{u^2}{\tilde{R}^2}\sum_{i=1}^3dx_i^2+(1+\d_1)\frac{\tilde{R}^2}{u^2}\frac{du^2}{f(u)}+\tilde{R}^2d\Omega_5\Big\}
\end{equation}
with the correction coefficients:
\begin{equation}
\begin{array}{lll}
\d_1&=&-\frac{15}{8}\zeta(3){\a'}^{3}\Big[5\Big(\frac{u_T}{u}\Big)^4+5\Big(\frac{u_T}{u}\Big)^8-3\Big(\frac{u_T}{u}\Big)^{12}\Big]\;\;,\\
\non\\
\d_2&=&\frac{15}{8}\zeta(3){\a'}^{3}\Big[5\Big(\frac{u_T}{u}\Big)^4+5\Big(\frac{u_T}{u}\Big)^8-19\Big(\frac{u_T}{u}\Big)^{12}\Big]\;\;.
\end{array}
\end{equation}
The classical action of the space-like Nambu-Goto string takes
then the following form:
\begin{equation}
S_{NG}[\CMcal{C}]=\frac{Y}{2\pi}\int_{-r/2}^{r/2}dx\sqrt{\frac{U^4}{\tilde{R}^4}+(1+\d_1)\frac{U^4}{U^4-U_T^4}U'^2}
\end{equation}
from which can be derived the inter-quark separation $r$ and the
(renormalized) static potential $V^{(R)}_{Q\overline{Q}}(r)$:
\begin{eqnarray}
r(U_0,U_T)&=&\frac{2\tilde{R}^2}{U_0}\int_0^{\infty}dv\frac{\sqrt{1+\d_1}}{\sqrt{(v^4-1+\epsilon)(v^4-1)}}\;\;,\\
V_{Q\overline{Q}}^{(R)}(r)&=&\frac{U_0}{2\pi\tilde{R}^2}r+\frac{U_0}{\pi}\int_1^{\infty}dv\Big(\sqrt{1+\d_1}\sqrt{\frac{v^4-1}{v^4-1+\epsilon}}-1\Big)+\frac{U_T-U_0}{\pi}\;\;.
\end{eqnarray}
The integrals are modified only by terms in $1/v$ which do not
rule out the logarithmic singularity in $v=1$ in the limit
$U_0\simeq U_T$ (see Eq.\eqref{wilson-interquark distance
divergence}).

\subsection{The area law in four-dimensional Yang-Mills theory}

We have seen previously that the Schwarzschild black hole$-AdS_5$
geometry was required in order to deal with a three-dimensional
gauge theory (after compactification of the Euclidean time
direction). If we are interested in studying higher-dimensional
gauge theories, it is then necessary to consider the general case
of a stack of $N$ coincident (extremal \emph{i.e.} without
horizon) $Dp$-branes in the decoupling limit. We are therefore led
to the (Euclidean) metric \cite{IMSY}:
\begin{equation}\label{wilson-metric general}
ds^2=\a'\Big\{\frac{u^{\frac{(7-p)}{2}}}{g_{YM}^{(p+1)}\sqrt{d_pN}}\Big(dt^2+\sum_{i=1}^pdx_i^2\Big)+\frac{g_{YM}^{(p+1)}\sqrt{d_pN}}{u^{\frac{7-p}{2}}}du^2+g_{YM}^{(p+1)}\sqrt{d_pN}u^{\frac{(p-3)}{2}}d\Omega_{8-p}^2\Big\}
\end{equation}
with $d_p\equiv 2^{7-2p}\pi^{\frac{9-3p}{2}}\G(\frac{7-p}{2})$.
The coupling constant $g_{YM}^{(p+1)}$ of the $(p+1)$-dimensional
$SU(N)$ super-Yang-Mills theory defined on the world-volume of the
$N$ $Dp$-branes is related to the closed string coupling constant
$g_s$ as follows:
\begin{equation}
{g_{YM}^{(p+1)}}^2=(2\pi)^{p-2}g_s\,\a'^{\frac{(p-3)}{2}}\;\;.
\end{equation}
The case of interest here consists of $p=4$ for which
${g_{YM}^{(5)}}^2=4\pi^2g_s\sqrt{\a'}$ (hence, $g_{YM}^{(5)}$ has
the dimension of a ($\emph{length})^{1/2}$) and the metric
\eqref{wilson-metric general} becomes:
\begin{equation}
ds^2=\a'\Big\{\frac{u^{3/2}}{R_4^{3/2}}\Big(dt^2+\sum_{i=1}^4dx_i^2\Big)+\frac{R_4^{3/2}}{u^{3/2}}du^2+R_4^{3/2}\sqrt{u}\,d\Omega_4^2\Big\}\;\;.
\end{equation}
We have defined $R_4^{3/2}\equiv
g_{YM}^{(5)}\sqrt{d_4N}=g_{YM}^{(5)}\sqrt{\frac{N}{4\pi}}$ such
that $R_4$ has the dimension of a (\emph{length})$^{1/3}$. When
one turns on the temperature, the metric is solution of the
equations of motion for a stack of $N$ coincident (non-extremal
\emph{i.e} in the presence of a horizon) $Dp$-branes in the
decoupling limit:
\begin{equation}
ds^2_{BH}=\a'\Big\{\frac{u^{3/2}}{R_4^{3/2}}\Big(g(u)dt^2+\sum_{i=1}^4dx_i^2\Big)+\frac{R_4^{3/2}}{u^{3/2}}\frac{du^2}{g(u)}+R_4^{3/2}\sqrt{u}\,d\Omega_4^2\Big\}
\end{equation}
where
\begin{equation}
g(u)=1-\frac{u_T^3}{u^3}\;\;.
\end{equation}
The event horizon at $u_T$ is given in terms of the
Beckenstein-Hawking temperature $T$:
\begin{equation}
T=\frac{1}{4\pi\a'}\frac{\de\,\d_{00}}{\de
u}\Big|_{u=u_T}=\frac{3}{4\pi R_4^{3/2}}\sqrt{u_T}
\end{equation}
which gives:
\begin{equation}
u_T=\frac{16}{9}\pi^2R_4^3T^2=\frac{4}{9}\pi
{g_{YM}^{(5)}}^2N\,T^2\;\;.
\end{equation}
 The Nambu-Goto action \eqref{wilson-nambu goto
action} of the space-like string world-sheet is this time:
\begin{equation}
S_{NG}=\frac{Y}{2\pi}\int_{-r/2}^{r/2}dx\sqrt{\frac{U^3}{R_4^3}+\frac{U^3}{U^3-U_T^3}U'^2}
\end{equation}
and the integral expressions for the distance $r$ between the
quarks and the static potential are ($\epsilon\equiv
g(U_0)=1-\frac{U_T^3}{U_0^3}$):
\begin{eqnarray}
r(U_0,U_T)&=&\frac{2R_4^{3/2}}{U_0^{1/2}}\int_1^{\infty}\frac{dv}{\sqrt{(v^3-1+\epsilon)(v^3-1)}}\;\;,\\
V_{Q\overline{Q}}^{(R)}(r)&=&\frac{U_0^{3/2}}{2\pi
R_4^{3/2}}r+\frac{U_0}{\pi}\int_1^{\infty}dv\Big(\sqrt{\frac{v^3-1}{v^3-1+\epsilon}}-1\Big)+\frac{U_T-U_0}{\pi}\;\;.
\end{eqnarray}
Also, we find that the potential presents an area law behaviour in
the case of a four-dimensional non-supersymmetric gauge theory
with a string tension ($U_0\simeq U_T$):
\begin{equation}
\s=\frac{U_T^{3/2}}{2\pi R_4^{3/2}}=\frac{8}{27}\pi
{g_{YM}^{(4)}}^2N\, T^2
\end{equation}
expressed in terms of the dimensionless coupling constant
$g_{YM}^{(4)}$ of the four-dimensional gauge theory \cite{BISY}.
This latter is obtained from $g_{YM}^{(5)}$ after compactification
of the Euclidean time direction along a circle $S^1(\b)$ of
circumference $\b$. We have indeed:
\begin{equation}
\int d^5x\frac{1}{{g_{YM}^{(5)}}^2}=\int
d^4x\frac{\b}{{g_{YM}^{(5)}}^2}=\int
d^4x\frac{1}{{g_{YM}^{(4)}}^2}\;\;,
\end{equation}
namely, ${g_{YM}^{(4)}}^2={g_{YM}^{(5)}}^2T$.

To summarize, we observe an area law for \emph{spatial} Wilson
loops in four- and five-dimensional supersymmetric Yang-Mills
theories at finite temperature. This can be interpreted as the
area law of \emph{ordinary} Wilson loops (after having identified
one of the spatial coordinates of the higher-dimensional theory as
the non-compactified Euclidean time) in three- and
four-dimensional non-supersymmetric Yang-Mills theories at zero
temperature which indicates confinement in these theories.

\section{The heavy quark potential in holographic models of QCD}

\subsection{Andreev and Zakharov's model}

The holographic models of QCD imply to introduce a dimensionful
parameter related in some way to the QCD mass gap. This can be the
cutoff $z_m$ where is located the IR brane in the Hard Wall Model
\cite{PT,EKSS,pomarol} or the dilaton parameter in the Soft Wall
Model \cite{KKSS}. In \cite{andreev,Andreev Zakharov confinement},
the authors chose to break the isometry group of the holographic
space-time $AdS_5$ (\emph{i.e.} the conformal invariance of the
boundary field theory) by means of a warp factor $h(z)$ in the
Euclidean metric:
\begin{equation}\label{wilson-AZ line element}
ds^2=g_{MN}(x)dx^Mdx^N=\frac{R^2}{z^2}h(z)\,\d_{MN}dx^Mdx^N\;\;.
\end{equation}
The bulk coordinates are $x^M=(x^\m,z)$ with $x^\m$
$(\m=0,\ldots,3)$ the boundary coordinates and $z>0$ the
holographic coordinate. $\d_{MN}=\textrm{diag}(+1,+1,+1,+1,+1)$ is
the Euclidean flat metric tensor. In this model, the warp factor
$h(z)\equiv e^{\frac{1}{2}c\,z^2}$ introduces the conformal
symmetry breaking parameter $c$ and we recover the $AdS_5$ metric
\eqref{wilson-line element} near the UV brane $z\to0$ where
$h(0)=1$. It is worth pointing out that this $c$ does not have to
be identified with the (square of the) dilaton parameter
$\Phi(z)=c^2_{\Phi}z^2$ in the IR Soft Wall approximation. For
example, even if the equivalence holds for the vector meson case
\cite{KKSS} (and then $c^2_{\Phi}=\frac{c}{4}$), it is not true in
general (see, for instance, the effective action for the scalar
mesons \cite{soft wall scalar}).

As usual, we start from the Nambu-Goto action \eqref{wilson-nambu
goto action} in the static gauge \eqref{wilson-static
gauge}-\eqref{wilson-reparametrization} $X^0(\tau,\s)=\tau$ with
$\s=x$. The non-vanishing components of the induced metric
$\g_{ab}$ ($a,b=1,2$) on the world-sheet \eqref{Wilson-induced
metric tensor} are:
\begin{equation}
\begin{array}{lll}
\g_{11}&=&\frac{R^2}{z^2}h(z)\;\;,\\
\g_{22}&=&\frac{R^2}{z^2}h(z)(1+{z'}^2)
\end{array}
\end{equation}
where the holographic coordinate of the string $z(x)$ is a
function only of $x$ in the limit $T\to\infty$ (in order to not
overweight the notation, we give up the convention of writing the
string coordinates with capital letters). The Nambu-Goto action of
the string is then (the notation $g$ instead of $\tilde{R}^2$ is
used in \cite{Andreev Zakharov confinement}):
\begin{equation}
S_{NG}[\CMcal{C}]=\frac{\tilde{R}^2}{2\pi}T\int_{-\frac{r}{2}}^{\frac{r}{2}}dx\frac{h}{z^2}\sqrt{1+{z'}^2}
\end{equation}
where $\CMcal{C}$ is the rectangular loop already considered in
\eqref{wilson-parameter1}-\eqref{wilson-parameter2}. The equation
of motion for $z(x)$ and the first integral read respectively as
\begin{equation}\label{wilson-AZ equation of motion}
\d_z S_{NG}=0\;\;\;\Ra\;\;\;zz''+(2-c\,z^2)(1+z'^2)=0
\end{equation}
and
\begin{equation}\label{wilson-AZ first integral}
\frac{h}{z^2\sqrt{1+(z')^2}}=C\;\;.
\end{equation}
The integration constant $C$ is positive and can be evaluated for
any value of $z(x)$. Especially, at $x=0$, we have $z(0)\equiv
z_0$ and $z'(0)=0$ by symmetry such that
$C=\frac{e^{\frac{1}{2}\lambda}}{z_0^2}$ where we have defined:
\begin{equation}
\lambda\equiv c\,z_0^2\;\;.
\end{equation}

The two parametric expressions for the inter-quark distance
$r(z_0,c)$ and the interaction potential $V(z_0,c)$ take the
following forms:
\begin{eqnarray}
r(z_0,c)&=&2\int_0^{\frac{r}{2}}dx=2\int_{z_0}^0\frac{dz}{z'}=2\int_0^{z_0}dz\frac{C\,z^2}{h}\Big(1-\frac{C^2z^4}{h^2}\Big)^{-\frac{1}{2}}\non\\
&=&2C\int_0^{z_0}dz\,z^2e^{-\frac{1}{2}c\,z^2}\Big(1-\frac{z^4}{z_0^4}\,e^{\lambda-c\,z^2}\Big)^{-\frac{1}{2}}\non\\
r(\lambda,c)&\underset{(v=\frac{z}{z_0})}{=}&2\sqrt{\frac{\lambda}{c}}\int_0^1dv\,v^2\,e^{\frac{1}{2}\lambda(1-v^2)}\Big(1-v^4e^{\lambda(1-v^2)}\Big)^{-\frac{1}{2}}\label{wilson-AZ
interquark distance}
\end{eqnarray}
and
\begin{eqnarray}
V(z_0,c)&=&\underset{T\to\infty}{\textrm{lim}}\frac{1}{T}S_{NG}[\CMcal{C}]=\frac{\tilde{R}^2}{2\pi}\int_{-\frac{r}{2}}^{\frac{r}{2}}dx\frac{h}{z^2}\sqrt{1+{z'}^2}\non\\
&=&\frac{\tilde{R}^2}{\pi}\int_{z_0}^0\frac{dz}{z'}\frac{h}{z^2}\sqrt{1+{z'}^2}=\frac{\tilde{R}^2}{\pi}\int_0^{z_0}dz\frac{h}{z^2}\Big(1-\frac{C^2\,z^4}{h^2}\Big)^{-\frac{1}{2}}\non\\
V(\lambda,c)&\underset{(v=\frac{z}{z_0})}{=}&\frac{\tilde{R}^2}{\pi}\sqrt{\frac{c}{\lambda}}\int_0^1
dv\frac{e^{\frac{1}{2}\lambda\,v^2}}{v^2}\Big(1-v^4\,e^{\lambda(1-v^2)}\Big)^{-\frac{1}{2}}\label{wilson-AZ
divergent integral potential}
\end{eqnarray}
where we have made use of the expression of $z'(x)$ derived from
the first integral \eqref{wilson-AZ first integral}:
\begin{equation}
z'(x)=\pm\frac{h}{C\,z^2}\Big(1-\frac{C^2\,z^4}{h^2}\Big)^{\frac{1}{2}}\;\;.
\end{equation}
The plus (minus) sign corresponds to $-\frac{r}{2}<x<0$
($0<x<\frac{r}{2}$). As expected from our previous studies (see,
\emph{e.g.}, the Eqs.\eqref{wilson-potential steps} and
\eqref{wilson-divergence potential finite T}), the integral
\eqref{wilson-AZ divergent integral potential} does not converge
when $v\to0$ and require an UV cutoff $z(x)\geq z_{min}$:
\begin{eqnarray}
V^{(reg.)}(\lambda,c,z_{min})&=&\frac{\tilde{R}^2}{\pi}\sqrt{\frac{c}{\lambda}}\int_0^1\frac{dv}{v^2}\Big[e^{\frac{1}{2}\lambda\,v^2}\Big(1-v^4\,e^{\lambda(1-v^2)}\Big)^{-\frac{1}{2}}-1\Big]+\frac{\tilde{R}^2}{\pi}\sqrt{\frac{c}{\lambda}}\int_{z_{min}/z_0}^1\frac{dv}{v^2}\non\\
\non\\
&=&\frac{\tilde{R}^2}{\pi}\sqrt{\frac{c}{\lambda}}\Big\{-1+\int_0^1\frac{dv}{v^2}\Big[e^{\frac{1}{2}\lambda\,v^2}\Big(1-v^4\,e^{\lambda(1-v^2)}\Big)^{-\frac{1}{2}}-1\Big]\Big\}+\frac{\tilde{R}^2}{\pi}\frac{1}{z_{min}}\;\;.\non\\
\label{wilson-AZ regularized potential}
\end{eqnarray}
If one remembers the relation $U=\frac{\tilde{R}^2}{z}$
\eqref{Wilson-new holographic coordinate} between the two
holographic coordinates $z$ and $U$, then the last term on the
\emph{r.h.s.} of \eqref{wilson-AZ regularized potential} is
canceled out, according to Maldacena's prescription, by the same
counter-term present in \eqref{Wilson-regularized potential}. At
the end of the day, in a holographic space-time with the
background metric \eqref{wilson-AZ line element}, the renormalized
(or subtracted) potential is:
\begin{equation}\label{wilson-AZ renormalized potential}
V^{(R)}(\lambda,c)=\frac{\tilde{R}^2}{\pi}\sqrt{\frac{c}{\lambda}}\Big\{-1+\int_0^1\frac{dv}{v^2}\Big[e^{\frac{1}{2}\lambda\,v^2}\Big(1-v^4\,e^{\lambda(1-v^2)}\Big)^{-\frac{1}{2}}-1\Big]\Big\}\;\;.
\end{equation}

\subsubsection{The heavy quark potential at large distances}

As a matter of fact, the expression \eqref{wilson-AZ interquark
distance} has a logarithmic singularity when $\lambda=c\,z_0^2=2$.
This peculiar finite value of $z_0=\sqrt{\frac{2}{c}}$ corresponds
to the maximal extent reached by the string world-sheet along the
holographic coordinate. There, the inter-quark distance
$r(\lambda,c)$ explodes, which mimics the confinement mechanism.
On the contrary, if the conformal symmetry breaking parameter
$c=0$, then $z_0$ is allowed to run over all the holographic
dimension ($0<z_0<\infty$) and we do not have confinement anymore.
Let us identify this logarithmic singularity. Since $r(2,c)$ does
not converge, it is not allowed to expand \eqref{wilson-AZ
interquark distance} in powers of $(2-\lambda)$. Nevertheless, as
we are interested in the region $z\sim z_0$, \emph{i.e.} $v\sim1$,
the integral in $r(\lambda,c)$ can be approximately replaced by
its main contribution:
\begin{equation}
r(\lambda,c)\simeq2\sqrt{\frac{\lambda}{c}}\int_0^1\frac{dv}{\sqrt{2(2-\lambda)(1-v)+(-2\lambda^2+9\lambda-6)(1-v)^2}}\;\;.
\end{equation}
The quark separation has clearly a logarithmic singularity at
$\lambda=2$:
\begin{equation}
r(\lambda,c)\underset{\underset{v\to1}{\lambda=2}}{\sim}-\sqrt{\frac{2}{c}}\ln(1-v)\;\;.
\end{equation}
The static potential \eqref{wilson-AZ renormalized potential}
develops the same singularity when $v\to1$ at $\lambda=2$. Indeed,
we can write:
\begin{equation}
V^{(R)}(\lambda,c)\simeq\frac{\tilde{R}^2}{\pi}\sqrt{\frac{c}{\lambda}}\Big\{-1+\int_0^1dv\Big[\frac{e}{\sqrt{2(2-\lambda)(1-v)+(-2\lambda^2+9\lambda-6)(1-v)^2}}-1\Big]\Big\}\non\\
\end{equation}
where $e$ is the exponential function of the unit. At large
distances, we have then:
\begin{equation}
V^{(R)}(\lambda,c)\underset{\underset{v\to1}{\lambda=2}}{\sim}-\frac{\tilde{R}^2}{2\pi}\sqrt{\frac{c}{2}}e\ln(1-v)
\end{equation}
which gives, in terms of $r$, a linear confining potential:
\begin{equation}
V^{(R)}(r,c)=\s r
\end{equation}
where we have defined the large-distance string tension:
\begin{equation}\label{wilson-AZ long distance string tension}
\s=\tilde{R}^2\frac{e}{4\pi}c\;\;.
\end{equation}

\subsubsection{The heavy quark potential at short distances}

The behaviours of $r(\lambda,c)$ and $V^{(R)}(\lambda,c)$ at short
distances correspond to a string configuration with $z_0\sim0$,
namely to the limit $\lambda\to0$ (since then the string
world-sheet does not go far away along the fifth holographic
coordinate, it mainly feels the UV geometry of the background
metric. As a consequence, there is no IR correction to the
potential \cite{Andreev Zakharov confinement}). The expansion of
the inter-quark distance in power series up to the order
$O(\lambda^2)$ yields:
\begin{eqnarray}
r(\lambda,c)&=&2\sqrt{\frac{\lambda}{c}}\int_0^1dv\frac{v^2}{\sqrt{1-v^4}}\Big(1+\frac{\lambda}{2}\frac{(1-v^2)}{(1-v^4)}+O(\lambda^2)\Big)\non\\
&=&2\sqrt{\frac{\lambda}{c}}\int_0^1dv\frac{v^2}{\sqrt{1-v^4}}+\lambda\sqrt{\frac{\lambda}{c}}\int_0^1dv\Big(\frac{v^2}{(1-v^4)^{3/2}}-\frac{v^4}{(1-v^4)^{3/2}}\Big)+O(\lambda^{5/2})\;\;.\non\\
\label{wilson-AZ short distance}
\end{eqnarray}
The first integral on the \emph{r.h.s.} gives the well-known
\emph{AdS}/CFT result \eqref{wilson-Maldacena interquark
distance}:
\begin{equation}
r(\lambda,c)=\sqrt{\frac{\lambda}{c}}\frac{1}{\rho}+O(\lambda^{3/2})
\end{equation}
with $\rho=\frac{\G(1/4)^2}{(2\pi)^{3/2}}$ the usual numerical
factor and $\sqrt{\frac{\lambda}{c}}=z_0=\frac{\tilde{R}^2}{U_0}$.
Although the second and the third integrals are singular when
$v\to1$, in fact their divergences $\sim1/\sqrt{1-v^4}$ cancel out
each other. This can be easily seen as follows:
\begin{eqnarray}
r(\lambda,c)&=&\sqrt{\frac{\lambda}{c}}\Big[\frac{1}{\rho}+\lambda\int_0^1dv\Big(-\frac{(1-v^2)-1}{(1-v^4)^{3/2}}+\frac{(1-v^4)-1}{(1-v^4)^{3/2}}\Big)+O(\lambda^2)\Big]\non\\
&=&\sqrt{\frac{\lambda}{c}}\Big[\frac{1}{\rho}+\lambda\int_0^1dv\Big(-\frac{1}{\sqrt{1-v^2}(1+v^2)^{3/2}}+\frac{1}{\sqrt{1-v^4}}\Big)+O(\lambda^2)\Big]\non\\
r(\lambda,c)&=&\sqrt{\frac{\lambda}{c}}\Big[\frac{1}{\rho}+\lambda\Big(-\frac{1}{2}E(-1)+\frac{\sqrt{\pi}\G(5/4)}{\G(3/4)}\Big)+O(\lambda^2)\Big]
\end{eqnarray}
where $E(-1)$ is the complete Elliptic integral of second kind
\eqref{E(-1)}. By standard handling of the Gamma functions (such
that the formulae $x\G(x)=\G(x+1)$ and
$\G(x)\G(1-x)=\frac{\pi}{\sin(\pi x)}$), we finally obtain:
\begin{equation}
r(\lambda,c)=\sqrt{\frac{\lambda}{c}}\frac{1}{\rho}\Big(1-\frac{\lambda}{4}(1-\pi\rho^2)+O(\lambda^2)\Big)\;\;.
\end{equation}
Furthermore, it will be worthwhile to express $\lambda$ in terms
of $r$ when we will attempt to write the potential $V^{(R)}(r,c)$.
Successive iterations give then:
\begin{equation}
\sqrt{\frac{\lambda}{c}}=\rho\,r\Big(1+\frac{\lambda}{4}(1-\pi\rho^2)+O(\lambda^2)\Big)=\rho\,r\Big(1+\frac{c}{4}\rho^2r^2(1-\pi\rho^2)+O(r^4)\Big)
\end{equation}
since
\begin{equation}
\lambda=c\rho^2r^2\Big(1+\frac{\lambda}{2}(1-\pi\rho^2)+O(\lambda^2)\Big)=c\rho^2r^2\Big(1+O(r^2)\Big)\;\;.
\end{equation}
The heavy quark potential \eqref{wilson-AZ renormalized potential}
is treated in the same way as the inter-quark distance. We get:
\begin{eqnarray}
V^{(R)}(\lambda,c)&=&\frac{\tilde{R}^2}{\pi}\sqrt{\frac{c}{\lambda}}\Big\{-1+\int_0^1\frac{dv}{v^2}\Big[\frac{1}{\sqrt{1-v^4}}-1\Big]+\frac{\lambda}{2}\int_0^1dv\frac{1+v^2-2v^4}{(1-v^4)^{3/2}}+O(\lambda^2)\Big\}\non\\
&=&\frac{\tilde{R}^2}{\pi}\sqrt{\frac{c}{\lambda}}\Big\{-\frac{1}{2\rho}+\frac{\lambda}{16\sqrt{2\pi}}\Big[12\G(1/4)\G(5/4)+\G(3/4)\G(-1/4)\Big]+O(\lambda^2)\Big\}\non\\
V^{(R)}(\lambda,c)&=&-\frac{\tilde{R}^2}{\pi}\sqrt{\frac{c}{\lambda}}\frac{1}{2\rho}\Big(1+\frac{\lambda}{4}(1-3\pi\rho^2)+O(\lambda^2)\Big)\label{wilson-AZ
final potential}
\end{eqnarray}
where the two first contributions in the first line consist of the
renormalized expression \eqref{Wilson-regularized potential} of
the \emph{AdS}/CFT potential (after appropriate variable changes).
We are now able to write the potential $V^{(R)}(r,c)$ as a
function of the distance $r$ between the quarks:
\begin{eqnarray}
V^{(R)}(r)&=&-\frac{\tilde{R}^2}{2\pi\rho^2}\frac{1}{r}\Big(1-\frac{c}{4}\rho^2
r^2(1-\pi\rho^2)+O(r^4)\Big)-\frac{\tilde{R}^2}{8\pi}(1-3\pi\rho^2)c\,r\Big(1+O(r^2)\Big)\non\\
\\
&=&-\frac{\kappa_0}{r}+\s_0\,r+O(r^3)\label{wilson-AZ potential}
\end{eqnarray}
with
\begin{equation}
\left\{
\begin{array}{lll}
\kappa_0&=&\frac{\tilde{R}^2}{2\pi\rho^2}=\frac{\sqrt{4\pi g_s N}}{2\pi\rho^2}\;\;,\\
\s_0&=&\tilde{R}^2\frac{c\,\rho^2}{4}\;\;.\label{wilson-AZ short
distance string tension}
\end{array}
\right.
\end{equation}
Although the linear term in the Cornell potential
\eqref{Wilson-Cornell potential} has only one string tension for
any length scale, it appears, in the supergravity side, two
tensions $\s$ \eqref{wilson-AZ long distance string tension} and
$\s_0$ \eqref{wilson-AZ short distance string tension}
corresponding respectively to the large and short distance
regimes. Nevertheless, their ratio turns out to be rather closed
to one:
\begin{equation}
\frac{\s}{\s_0}=\frac{e}{\pi\rho^2}=\frac{8\pi^2e}{\G(1/4)^4}\simeq1.24\;\;.
\end{equation}
Without being obviously conclusive, this estimate is satisfactory
at the accuracy level usually associated with holographic models
of QCD. As for the Coulomb-like term in \eqref{wilson-AZ
potential}, it does not have to be identified with the
perturbative part of the Cornell potential. This is reminiscent of
what happens in the \emph{AdS}/CFT correspondence where the
potential in $1/r$ \eqref{wilson-potential} behaves not as a power
of the 't Hooft coupling but as the square root thereof.
Nevertheless, it is hard, in the string picture used here, to
disentangle the contributions, if any, of the large distance
L\"{u}scher term in $1/r$ from the perturbative Coulomb term at
short distances.\\

It is worth pointing out that Andreev and Zakharov's model has
also been used to explore finite temperature features of a heavy
quark-antiquark pair as, for instance, the spatial string tension
\cite{spatial string tension} or the free energy \cite{free
energy}. The line element \eqref{wilson-AZ line element} has also
been considered in \cite{AZ baryon} in order to study the baryon
potential and the Y-ansatz of the baryonic area law.

\subsection{The heavy quark potential from general geometry in
\emph{AdS}/QCD}

In the following, we will consider a general form of the metric
which respects Poincar\'e symmetry on the boundary
\cite{Shock,white}:
\begin{equation}\label{wilson-background metric}
ds^2=\a'\tilde{R}^2\Big(f(z)\d_{\m\n}dx^\m
dx^\n+\frac{dz^2}{z^2}\Big)
\end{equation}
with $\d_{\m\n}=\textrm{diag}(+1,+1,+1,+1)$ the four-dimensional
Euclidean flat metric tensor. The warp factor $f(z)>0$ is assumed
to be positive. In particular, $f(z)=\frac{1}{z^2}$ corresponds to
the Euclidean $AdS_5$ line element. The Nambu-Goto action in the
static gauge $X^0(\tau,\s)=\tau$ and $\s=x$ reads
\begin{equation}
S_{NG}[\CMcal{C}]=\frac{\tilde{R}^{2}}{2\pi}\int_{-T/2}^{T/2}dt\int_{-r/2}^{r/2}dx\,f(z)\sqrt{1+\frac{{z'}^2}{f(z)z^2}}
\end{equation}
where $z(x)$ is the holographic coordinate of the string. The
Lagrangian density does not depend explicitly on $x$ which gives
us the first integral:
\begin{equation}
\frac{f(z)}{\sqrt{1+\frac{{z'}^2}{f(z)z^2}}}=f_0
\end{equation}
where $z_0$ is the value of $z(x)$ at $x=0$, that is, the maximal
extent of the string world-sheet along the holographic dimension
where $z'(0)=0$ by symmetry and $f_0\equiv f(z_0)$. We derive the
expression of the derivative $z'(x)$ ($-r/2<x<0$ for the plus sign
and $0<x<r/2$ for the minus sign):
\begin{equation}
z'(x)=\pm z\sqrt{f(z)}\sqrt{\frac{f^2}{f_0^2}-1}\;\;.
\end{equation}
The equation of motion is:
\begin{equation}
zz''-{z'}^2-z^3f'(z)-\frac{3}{2}\frac{f'(z)}{f(z)}z{z'}^2=0
\end{equation}
where $z'(x)\equiv\frac{dz}{dx}$ and $f'(z)\equiv\frac{df}{dz}$
are the derivatives with respect to the arguments. In the anti-de
Sitter case where $f(z)=\frac{1}{z^2}$, we recover the equation of
motion \eqref{wilson-AZ equation of motion} (with $c=0$) which
describes the behaviour of the string world-sheet spreading into
the $AdS_5$ holographic space-time.

The inter-quark distance takes the general form:
\begin{eqnarray}
r(z_0)&=&\int_{-r/2}^{r/2}dx=2\int_{z_0}^0\frac{dz}{z'}=2\int_0^{z_0}\frac{dz}{z}\frac{1}{\sqrt{f}}\Big(\frac{f^2}{f_0^2}-1\Big)^{-\frac{1}{2}}\non\\
&=&2\int_0^{z_0}dz\frac{1}{\sqrt{\tilde{f}}}\Big(\frac{z_0^4\tilde{f}^2}{z^4\tilde{f}_0^2}-1\Big)^{-\frac{1}{2}}\;\;.\label{wilson-white-eq1}
\end{eqnarray}
Following \cite{white}, we have defined $\tilde{f}(z)=z^2\,f(z)$
such that $\tilde{f}(0)=1$. As for the interaction potential, we
find successively:
\begin{eqnarray}
V(z_0)&=&\frac{\tilde{R}^2}{\pi}\int_0^{z_0}\frac{dz}{z^2}\sqrt{\tilde{f}}\Big(1-\frac{z^4\tilde{f}_0^2}{z_0^4\tilde{f}^2}\Big)^{-\frac{1}{2}}\non\\
V^{(reg.)}(z_0,z_{min})&=&\frac{\tilde{R}^2}{\pi}\Big\{-\frac{1}{z_0}+\int_0^{z_0}\frac{dz}{z^2}\Big[\sqrt{\tilde{f}}\Big(1-\frac{z^4\tilde{f}_0^2}{z_0^4\tilde{f}^2}\Big)^{-\frac{1}{2}}-1\Big]\Big\}+\frac{\tilde{R}^2}{\pi}\frac{1}{z_{min}}\non\\
V^{(R)}(z_0)&=&\frac{\tilde{R}^2}{\pi}\Big\{-\frac{1}{z_0}+\int_0^{z_0}\frac{dz}{z^2}\Big[\sqrt{\tilde{f}}\Big(1-\frac{z^4\tilde{f}_0^2}{z_0^4\tilde{f}^2}\Big)^{-\frac{1}{2}}-1\Big]\Big\}\;\;.\label{wilson-white-eq2}
\end{eqnarray}
The last expression of the energy is obtained as usual, once the
infinite contribution $\frac{\tilde{R}^2}{\pi}\frac{1}{z_{min}}$
($z_{min}\to0$) stemming from the "W-boson string" associated with
the very massive quarks is subtracted.

On the one hand, at short distances \emph{i.e.} when the string
world-sheet is close enough to the boundary space-time, the bulk
geometry felt by this latter is nearly $AdS_5$. Not surprisingly,
the limit $z_0\to0$ (and then $z\to0$ and $\tilde{f}(z)\to1$ in
\eqref{wilson-white-eq1} and \eqref{wilson-white-eq2}) gives the
famous \emph{AdS}/CFT results \eqref{wilson-Maldacena interquark
distance} and \eqref{wilson-AdS/CFT potential}:
\begin{eqnarray}
r(z_0)&\underset{z_0\to0}{\simeq}&2\int_0^{z_0}dz\frac{z^2}{\sqrt{z_0^4-z^4}}=\frac{z_0}{\rho}\;\;,\\
V^{(R)}(z_0)&\underset{z_0\to0}{\simeq}&\frac{\tilde{R}^2}{\pi}\Big\{-\frac{1}{z_0}+\int_0^{z_0}\frac{dz}{z^2}\Big[\frac{1}{\sqrt{1-\frac{z^4}{z_0^4}}}-1\Big]\Big\}=-\frac{\tilde{R}^2}{2\pi\rho}\frac{1}{z_0}
\end{eqnarray}
such that
\begin{equation}
V^{(R)}(r)=-\frac{\tilde{R}^2}{2\pi\rho^2}\frac{1}{r}\;\;.
\end{equation}

On the other hand, in the case of mesonic bound-states, the
confinement criterion can be stated as follows: there exists a
finite value $z_0^{\ast}$ of the maximal extent of the world-sheet
along the holographic coordinate such that the distance
$r(z_0^{\ast})$ between quarks diverges. This peculiar value
$z_0^{\ast}$ is related to the QCD mass gap. In particular, it
enters the expression of the string tension in the confining
linear potential. Moreover, this divergence is logarithmic. By
expanding around $z_0^{\ast}$, we have indeed (where
$\tilde{f}(z_0^{\ast})\equiv\tilde{f}_0^{\ast}$):
\begin{eqnarray}
r(z_0^{\ast})&=&2\int_0^{z_0^{\ast}}dz\frac{1}{\sqrt{\tilde{f}(z)}}\Big(\frac{{z_0^{\ast}}^4\tilde{f}^2(z)}{z^4\tilde{f}_0^{\ast\,2}}-1\Big)^{-\frac{1}{2}}\\
&\simeq&\frac{2}{\sqrt{\tilde{f}(z_0^{\ast})}}\int_0^{z_0^{\ast}}dz\Big\{\Big[\frac{4}{z_0^{\ast}}-\frac{2}{\tilde{f}_0^{\ast}}\frac{d\tilde{f}}{dz}\Big|_{z_0^{\ast}}\Big](z_0^{\ast}-z)\non\\
&&+\Big[\frac{10}{{z_0^{\ast}}^2}-\frac{8}{z_0^{\ast}\tilde{f}_0^{\ast}}\frac{d\tilde{f}}{dz}\Big|_{z_0^{\ast}}+\frac{1}{\tilde{f}_0^{\ast}}\frac{d^2\tilde{f}}{dz^2}\Big|_{z_0^{\ast}}+\frac{1}{\tilde{f}_0^{\ast\,2}}\Big(\frac{d\tilde{f}}{dz}\Big|_{z_0^{\ast}}\Big)^2\Big](z_0^{\ast}-z)^2+\ldots\Big\}^{-\frac{1}{2}}\;\;.\non\\
\end{eqnarray}
With the background metric \eqref{wilson-background metric}, the
confinement criterion is then \cite{white}:
\begin{equation}
z_0^{\ast}\frac{d\tilde{f}}{dz}\Big|_{z_0^{\ast}}=2\tilde{f}_0^{\ast}
\end{equation}
such that $r(z_0^{\ast})$ diverges logarithmically:
\begin{equation}
r(z_0^{\ast})\underset{z\to
z_0^{\ast}}{\sim}-\ln(1-\frac{z}{z_0^{\ast}})\;\;.
\end{equation}
We are also interested in the asymptotic behaviour of the
potential. From our previous studies, we expect the same kind of
singularity than for $r(z_0^{\ast})$. We find indeed:
\begin{eqnarray}
V^{(R)}(z_0^{\ast})&=&\frac{\tilde{R}^2}{\pi}\Big\{-\frac{1}{z_0^{\ast}}+\int_0^{z_0^{\ast}}\frac{dz}{z^2}\Big[\sqrt{\tilde{f}(z)}\Big(1-\frac{z^4\tilde{f}_0^{\ast\,2}}{z_0^{\ast\,4}\tilde{f}^2(z)}\Big)^{-\frac{1}{2}}-1\Big]\Big\}\non\\
&\underset{z\to
z_0^{\ast}}{\simeq}&\frac{\tilde{R}^2}{\pi}\frac{\sqrt{\tilde{f}_0^{\ast}}}{z_0^{\ast\,2}}\int_0^{z_0^{\ast}}dz\Big(\frac{z_0^{\ast\,4}\tilde{f}^2(z)}{z^4\tilde{f}_0^{\ast\,2}}-1\Big)^{-\frac{1}{2}}\;\;.
\end{eqnarray}
The integral is the same that enters the expression of the
inter-quark distance such that
\begin{equation}
V^{(R)}(r,z_0^{\ast})=\s(z_0^{\ast})\,r
\end{equation}
with the string tension:
\begin{equation}
\s(z_0^{\ast})=\frac{\tilde{R}^2}{2\pi}\frac{\tilde{f}_0^{\ast}}{z_0^{\ast\,2}}\;\;.
\end{equation}

To conclude this section, let us mention that the heavy quark
potential has also been investigated in a realization of the hard
wall approximation: the used framework is the Randall-Sundrum
model \cite{randall-sundrum} which consists of an $AdS_5$ slice
between two D3-branes with the fields of the Standard Model living
on the four-dimensional world-volume of one of these branes
\cite{braga-static potential}. Thermal effects have been studied
by means of the Schwarzschild black hole$-AdS$ metric
\cite{braga-static potential finite temperature}. The issue of
finding general criteria for the confinement has also been
considered in \cite{kiritsis}.

\section{The supergravity description of baryons}

\subsection{The baryon potential within the \emph{AdS}/CFT
correspondence}

In a $SU(N)$ Yang-Mills theory, a colour-singlet baryon must be
made of $N$ quarks. As described in the supergravity dual, such a
baryon consists of $N$ quarks living on the boundary of a
holographic space-time. On each of these quarks ends a string with
the other endpoint attached to a D5-brane wrapped around the
5-sphere $S^5$: the so-called baryon vertex located at the
holographic coordinate $u_0$ \cite{gross ooguri,witten-baryon}.
The typical \emph{radius} of the baryon is denoted $r$. Moreover,
the configuration of the $N$ quarks on the boundary is symmetric
with respect to the boundary dimensions such that the resulting
force acting on the baryon vertex is zero along these directions.
In the following, we will consider only the induced metric
contribution of the Dirac-Born-Infeld action of the D5-brane:
\begin{equation}\label{wilson-baryons-brane action}
S_{\textrm{D5}}=T_5\int d^6x\sqrt{det\,g_{\textrm{D5}}}
\end{equation}
with $T_5^{-1}=(2\pi)^5{\a'}^3g_s$ the (inverse of the) tension of
the brane\footnote{In general, a D$p$-brane carries on its
$(p+1)$-dimensional world-volume electromagnetic fields of which
the dynamics is governed by the so-called Dirac-Born-Infeld
action:
\begin{equation}\label{dirac-born-infeld action}
S_{\textrm{Dp}}=T_p\int d^{p+1}x\sqrt{-det(\eta_{MN}+2\pi\a'
F_{MN})}
\end{equation}
with $T_p=\frac{2\pi}{(2\pi\ell_s)^{p+1}g_s}$ the brane tension
and $M,N=0,1,\ldots,p$ the space-time indices of the (flat)
world-volume of the D$p$-brane.}. The line element
$ds^2_{\textrm{D5}}$ which measures invariant distances between
two events located on the brane at $u_0$ can be derived
straightforwardly from the $AdS_5\times S^5$ line element
\eqref{wilson-line element 2}:
\begin{equation}
ds_{\textrm{D5}}^2=\a'\frac{u_0^2}{\tilde{R}^2}dt^2+\a'\tilde{R}^2d\Omega_5^2=\a'\frac{u_0^2}{\tilde{R}^2}dt^2+\a'\tilde{R}^2g_{ij}d\theta^id\theta^j=\a'\frac{u_0^2}{\tilde{R}^2}dt^2+g_{ij}d\omega^id\omega^j
\end{equation}
where we have defined the dimensionful coordinate $w^i\equiv
(\sqrt{\a'}\tilde{R})\,\theta^i$ $(i,j=1,\ldots,5)$. Being the
D5-brane static, the square root of the determinant of the induced
metric in \eqref{wilson-baryons-brane action} does not depend on
the time. The integral over the time coordinate $-T/2\geq t\geq
T/2$ gives rise to an overall factor $T$ in the action. The
remaining integrals involves five coordinates describing the
$S^5$. We have \cite{BISY baryons}:
\begin{eqnarray}
S_{\textrm{D5}}&=&\frac{1}{(2\pi)^5{\a'}^3g_s}\int_{-T/2}^{T/2}dt\int_{S^5}d^5x\sqrt{\a'\frac{U_0^2}{\tilde{R}^2}}\sqrt{det\,g_{S^5}}=\frac{T\,U_0}{(2\pi)^5({\a'})^{\frac{5}{2}}g_s\tilde{R}}\int_{S^5}d^5x\sqrt{det\,g_{S^5}}\non\\
&=&\frac{T\,U_0}{(2\pi)^5({\a'})^{\frac{5}{2}}g_s\tilde{R}}(\sqrt{\a'}\tilde{R})^5V(S^5)=\frac{T\,N\,U_0}{8\pi}\label{wilson-baryon
vertex action}
\end{eqnarray}
where $V(S^5)=\pi^3$ is the volume of the unit 5-sphere and
$\tilde{R}^4=4\pi g_sN$. The Nambu-Goto action of a string
world-sheet in the $AdS_5\times S^5$ background has already been
widely studied \eqref{Wilson-Nambu-Goto
action}-\eqref{Wilson-Lagrangian density}. However, the baryonic
system involves the additional contribution of the D5-brane. The
total action is thus (with $U(x=0)\equiv U_0$):
\begin{equation}
S_{total}=S_{\textrm{D5}}+\sum_{i=1}^N
S^{(i)}_{string}=\frac{T\,N\,U_0}{8\pi}+\frac{T\,N}{2\pi}\int_0^rdx\sqrt{{U'}^2+\frac{U^4}{\tilde{R}^4}}\;\;.
\end{equation}
Let us remark that the integral above over the boundary spatial
coordinate $x$ runs from $0$ to the typical radius $r$ of the
baryon. This latter should not be confused with the inter-quark
distance, also denoted $r$, in $Q\overline{Q}$ bound-states.
(Besides, the dual string configurations associated to baryons and
mesons are quite dissimilar.)

First of all, let us derive the stability condition of the baryon
vertex along the holographic coordinate. Since the endpoint $U_0$
of the $N$ strings is free to vary $\d U_0\neq0$ and the
variational principle gives:
\begin{eqnarray}
\d S_{total}&=&\frac{NT}{2\pi}\Big\{\frac{\d U_0}{4}+\int_0^r
dx\frac{1}{\sqrt{{U'}^2+\frac{U^4}{\tilde{R}^4}}}\Big(\frac{2U^3}{\tilde{R}^4}\d U+U'\frac{d}{dx}\d U\Big)\Big\}\non\\
&=&\frac{NT}{2\pi}\Big[\frac{\d
U_0}{4}+\int_0^rdx\frac{d}{dx}\Big(\frac{U'\d
U}{\sqrt{{U'}^2+\frac{U^4}{\tilde{R}^4}}}\Big)\Big]+\frac{NT}{2\pi}\int_0^rdx\,\d
U\Big[\frac{2U^3}{\tilde{R}^4\sqrt{{U'}^2+\frac{U^4}{\tilde{R}^4}}}-\frac{d}{dx}\Big(\frac{U'}{\sqrt{{U'}^2+\frac{U^4}{\tilde{R}^4}}}\Big)\Big]\non\\
\end{eqnarray}
The first contribution in square brackets stands for the surface
term at $U_0$ (since $\d U(r)=0$ on the boundary space) and must
vanishes for any $\d U_0$ which gives therefore the stability (or
no-force) condition of the D5-brane along the holographic
coordinate:
\begin{equation}\label{wilson-stability condition}
\frac{U'_0}{\sqrt{{U'_0}^2+\frac{U_0^4}{\tilde{R}^4}}}=\frac{1}{4}
\end{equation}
where $U'_0\equiv\frac{dU}{dx}\Big|_{x=0}$ is the slope of the $N$
strings at the baryon vertex. In particular, we have
${U_0'}^2=\frac{1}{15}\frac{U_0^4}{\tilde{R}^4}$ which allow us to
determine the value of the first integral \eqref{Wilson-first
integral} at $x=0$ in the baryon case:
\begin{equation}\label{wilson-baryon first integral}
\frac{U^4}{\sqrt{{U'}^2+\frac{U^4}{\tilde{R}^4}}}=\sqrt{\frac{15}{16}}\tilde{R}^2U_0^2=\textrm{const.}
\end{equation}
or, if we are interested in an expression of the derivative
(always positive since $0\leq x\leq r$):
\begin{equation}\label{wilson-baryon derivative}
U'(x)=\frac{U^2}{\tilde{R}^2}\sqrt{\b^2\frac{U^4}{U_0^4}-1}\;\;.
\end{equation}
We have defined $\b=\sqrt{\frac{16}{15}}$. The second contribution
in square brackets must also vanish for any interior $\d U$ and,
thus, gives the equation of motion for $U(x)$.

The typical radius of the baryon has the expression:
\begin{equation}\label{wilson-baryon radius}
r(U_0)=\int_0^rdx=\int_{U_0}^{\infty}\frac{dU}{U'}\underset{(v\equiv\frac{U}{U_0})}{=}\frac{\tilde{R}^2}{U_0}\int_1^{\infty}\frac{dv}{v^2\sqrt{\b^2v^4-1}}
\end{equation}
or
\begin{equation}
U_0(r)=\frac{\tilde{R}^2}{r}\int_1^{\infty}\frac{dv}{v^2\sqrt{\b^2v^4-1}}\;\;.
\end{equation}
The contribution of one string $(i)$ to the energy of the baryon
is obtained in a way similar to \eqref{Wilson-regularized
potential}:
\begin{eqnarray}
V^{(i)}_{string}(U_0)&=&\underset{T\to\infty}{\lim}\frac{1}{T}S^{(i)}_{string}=\frac{1}{2\pi}\int_0^r
dx\sqrt{{U'}^2+\frac{U^4}{\tilde{R}^4}}\non\\
&=&\frac{1}{2\pi}\int_{U_0}^{\infty}dU\b\frac{U^2}{U_0^2}\Big(\b^2\frac{U^4}{U_0^4}-1\Big)^{-\frac{1}{2}}\non\\
V^{(reg.)(i)}_{string}(U_0,U_{max})&\underset{(v\equiv\frac{U}{U_0})}{=}&\frac{U_0}{2\pi}\int_1^{U_{max}/U_0}dv\Big[\frac{\b
v^2}{\sqrt{\b^2v^4-1}}-1\Big]+\frac{U_0}{2\pi}\int_1^{U_{max}/U_0}dv\non\\
V^{(R)(i)}_{string}(U_0)&=&\frac{U_0}{2\pi}\Big\{\int_1^{\infty}dv\Big[\frac{\b
v^2}{\sqrt{\b^2v^4-1}}-1\Big]-1\Big\}\;\;.
\end{eqnarray}
So, the energy of the baryon is:
\begin{equation}\label{wilson-baryon potential}
V_B^{(R)}(U_0)=\frac{N\,U_0}{8\pi}+\frac{N\,U_0}{2\pi}\Big\{\int_1^{\infty}dv\Big[\frac{\b
v^2}{\sqrt{\b^2v^4-1}}-1\Big]-1\Big\}\;\;.
\end{equation}
In terms of the typical radius $r$, we obtain a potential which is
proportional to $N$ times the potential of a quark-antiquark
bound-state \eqref{wilson-potential} ($\lambda$ is the 't Hooft
coupling constant):
\begin{equation}
V_B(r)=-N\a_B\frac{\sqrt{2\lambda}}{r}
\end{equation}
with
\begin{equation}
\a_B=\frac{1}{2\pi}\int_1^{\infty}\frac{du}{u^2\sqrt{\b^2u^4-1}}\Big\{\frac{3}{4}-\int_1^{\infty}dv\Big[\frac{\b
v^2}{\sqrt{\b^2v^4-1}}-1\Big]\Big\}\simeq0.036\;\;.
\end{equation}
The behaviour in $1/r$ of the baryon potential is obviously
dictated by the conformal invariance of the field theory at the
boundary.

\subsection{Existence of \emph{AdS}/CFT baryons made of $k<N$ quarks}

Remarkably, another string configuration has been identified which
allows, on the supergravity side, to account for baryons made of a
smaller number of quark constituents $k<N$ \cite{BISY baryons}. In
that case, to the baryon vertex at $u_0$ are attached $k$ strings,
quite analogous to those studied above, which end, at the boundary
($u\to+\infty$), on the $k$ quarks. However, there are $N-k$
remaining strings which stretch out from the baryon vertex to the
brane at $u=0$. These strings are radial straight strings and are
described by the action ($j=k+1,\ldots,N-k$):
\begin{equation}
S^{(j)}_{string}=\frac{1}{2\pi\a'}\int
d^2\xi\sqrt{det(\g_{ab})}=\frac{1}{2\pi\a'}\int_{-T/2}^{T/2}dt\int_{0}^{U_0}dU\sqrt{{\a'}^2}=\frac{T\,U_0}{2\pi}
\end{equation}
since the non-vanishing components of the induced metric tensor on
the world-sheet are here ($\x^1=U$ and $\x^2=t$):
\begin{eqnarray}
\g_{11}&=&\a'\frac{\tilde{R}^2}{U^2}\;\;,\\
\g_{22}&=&\a'\frac{U^2}{\tilde{R}^2}\;\;.
\end{eqnarray}
Hence, the total action governing the dynamics of the baryon:
\begin{eqnarray}\label{wilson-CFT baryon action}
S_{total}&=&S_{\textrm{D5}}+\sum_{i=1}^{k}S^{(i)}_{string}+\sum_{j=1}^{N-k}S^{(j)}_{string}\non\\
&=&\frac{T\,N\,U_0}{8\pi}+\frac{k\,T}{2\pi}\int_0^{r}dx
\sqrt{{U'}^2+\frac{U^4}{\tilde{R}^4}}+\frac{T(N-k)U_0}{2\pi}\;\;.
\end{eqnarray}
The variational principle gives then the following stability
condition for the baryon vertex along the holographic coordinate:
\begin{equation}
\d S_{total}|_{\underset{\textrm{term at }U_0}{\textrm{surface
}}}=0\;\;\Ra\;\;\frac{U_0'}{\sqrt{\frac{U_0^4}{\tilde{R}^4}+{U_0'}^2}}=\frac{5N-4k}{4k}\equiv
A\;\;\Ra\;\;{U_0'}^2=\frac{A^2}{1-A^2}\frac{U_0^4}{\tilde{R}^4}\;\;.
\end{equation}
If $k=N$, then $A=\frac{1}{4}$ and we recover
\eqref{wilson-stability condition}. If the baryon has less quarks
$k\leq N$ then $A\geq\frac{1}{4}$. On the other hand, the upper
bound for $A$ (which corresponds to the lower bound for $k$) is
obtained for radial straight $k$-type strings ending on the baryon
vertex such that $U_0'\to\infty$. Then, $A=1$ and
$k=\frac{5N}{8}$. To summarize, the condition for having a stable
string/brane system into the bulk demands $\frac{5N}{8}\leq k\leq
N$.

The Lagrangian density in \eqref{wilson-CFT baryon action} depends
on $x$ only through $U(x)$. It results the first integral:
\begin{equation}\label{wilson-CFT baryon first integral}
\frac{U^4}{\sqrt{{U'}^2+\frac{U^4}{\tilde{R}^4}}}=\sqrt{1-A^2}\tilde{R}^2U_0^2
\end{equation}
which can be put into the form:
\begin{equation}\label{wilson-CFT baryon derivative}
U'(x)=\frac{U^2}{\tilde{R}^2\sqrt{1-A^2}}\sqrt{\frac{U^4}{U_0^4}-(1-A^2)}\;\;.
\end{equation}
The radius and the potential of these "reduced" baryons are then:
\begin{equation}\label{wilson-CFT baryon radius}
r(U_0)=\frac{\tilde{R}^2}{U_0}\sqrt{1-A^2}\int_1^{\infty}\frac{dv}{v^2\sqrt{v^4-(1-A^2)}}
\end{equation}
and
\begin{eqnarray}
V_B(U_0)&=&\frac{N\,U_0}{8\pi}+\frac{(N-k)U_0}{2\pi}+\frac{k}{2\pi}\int_0^{r}dx
\sqrt{{U'}^2+\frac{U^4}{\tilde{R}^4}}\non\\
&\underset{(v=\frac{U}{U_0})}{=}&\frac{N\,U_0}{8\pi}+\frac{(N-k)U_0}{2\pi}+\frac{k\,U_0}{2\pi}\int_1^{\infty}dv\frac{v^2}{\sqrt{v^4-(1-A^2)}}\non
\end{eqnarray}
\begin{equation}
V^{(reg.)}_B(U_0,U_{max})=\frac{N\,U_0}{8\pi}+\frac{(N-k)U_0}{2\pi}+\frac{k\,U_0}{2\pi}\int_1^{U_{max}/U_0}dv\Big[\frac{v^2}{\sqrt{v^4-(1-A^2)}}-1\Big]+\frac{k}{2\pi}(U_{max}-U_0)\non
\end{equation}
\begin{equation}
V^{(R)}_B(U_0)=\frac{N\,U_0}{8\pi}+\frac{(N-k)U_0}{2\pi}+\frac{k\,U_0}{2\pi}\Big\{\int_1^{\infty}dv\Big[\frac{v^2}{\sqrt{v^4-(1-A^2)}}-1\Big]-1\Big\}\label{wilson-CFT
baryon potential}
\end{equation}
where the counter-term required in order to absorb the UV
singularity consists here of $k$ radial straight strings stretched
out from the boundary at $u_{max}$ to the brane at $u=0$ (the
contribution of the baryon vertex vanishes according to
\eqref{wilson-baryon vertex action} with $U_0=0$):
\begin{equation}
V_{c.t.}=-\underset{U_{max}\to\infty}{\lim}k\frac{U_{max}}{2\pi}\;\;.
\end{equation}
When $k=N$, \eqref{wilson-CFT baryon first
integral}-\eqref{wilson-CFT baryon potential} reduces to
\eqref{wilson-baryon first integral}-\eqref{wilson-baryon radius}
and \eqref{wilson-baryon potential} respectively. If
$k=\frac{5N}{8}$ then $A=1$ which implies $r(U_0)=0$ (the baryon
size vanishes) and $V^{(R)}_B(U_0)=0$ independently of the
location $U_0$ of the D5-brane along the holographic coordinate.
If $\frac{5N}{8}<k\leq N$ then $A<1$ and the baryon energy
$V^{(R)}_B(r)=-\a\,U_0(r)$ can be written as the product of a
negative constant $-\a$ ($\a>0$) with $U_0$ (expressed in terms of
$r$) \cite{BISY baryons}.

\subsection{Baryons in three-dimensional Yang-Mills theory}

As largely discussed in preceding sections, we consider a spatial
string/brane configuration in the Schwarzschild black hole$-AdS_5$
background \eqref{wilson-finite temperature line element}. The
Nambu-Goto action of one space-like string world-sheet reads
$(i=1,\ldots,N)$
\begin{equation}
S^{(i)}_{string}=\frac{Y}{2\pi}\int_0^rdx\sqrt{\frac{U^4}{\tilde{R}^4}+\frac{U^4}{U^4-U_T^4}{U'}^2}\;\;,
\end{equation}
which is obviously reminiscent of \eqref{wilson-nambu-goto action
temperature}, and the total action is:
\begin{equation}\label{wilson-non susy baryon action}
S_{total}=S_{\textrm{D5}}+\sum_{i=1}^N
S^{(i)}_{string}=\frac{YNU_0}{8\pi}+\frac{N\,Y}{2\pi}\int_0^rdx\sqrt{\frac{U^4}{\tilde{R}^4}+\frac{U^4}{U^4-U_T^4}{U'}^2}\;\;.
\end{equation}
Apart from the equation of motion for the string coordinate
$U(x)$, the variational principle gives the following surface term
at $U_0$ where the baryon vertex wraps the 5-sphere $S^5$ ($\d
U_0\neq0$ while $\d U(r)=0$):
\begin{equation}\label{wilson-non susy baryon stability cond}
\d S_{total}|_{\underset{\textrm{term at }U_0}{\textrm{surface
}}}=0\;\;\Ra\;\;\frac{U_0'}{(1-\frac{U_T^4}{U_0^4})\sqrt{\frac{U_0^4}{\tilde{R}^4}+\frac{{U_0'}^2}{1-\frac{U_T^4}{U_0^4}}}}=\frac{1}{4}\;\;.
\end{equation}
At zero temperature where $u_T=\pi\tilde{R}^2T=0$, we recover the
stability condition \eqref{wilson-stability condition}.

We are interested in the large distance regime where the typical
radius of the Yang-Mills baryons is large. That corresponds to the
situation where the D5-brane reaches the horizon $(U_0\to U_T)$.
Then, according to \eqref{wilson-non susy baryon stability cond},
the slope of the strings must vanish at the horizon $(U_0'\to0)$:
the $N$ strings, attached to the quarks on the boundary, become
radial straight strings but, contrary to the \emph{AdS}/CFT
baryons made of $k=\frac{5N}{8}$ quarks considered previously, the
radius of the Yang-Mills baryons remains sizeable. Indeed, once
they hit the event horizon at $u_T$, the strings spread along the
transverse directions to the holographic dimension up to the
baryon vertex. In this limit case, the first integral derived from
\eqref{wilson-non susy baryon action} reduced to
\eqref{wilson-first integral temperature}:
\begin{equation}
\frac{U^4}{\sqrt{\frac{U^4}{\tilde{R}^4}+\frac{U^4}{U^4-U_T^4}{U'}^2}}=\frac{U_0^4}{\sqrt{\frac{U_0^4}{\tilde{R}^4}+\frac{U_0^4}{U_0^4-U_T^4}{U_0'}^2}}\underset{(U'_0\to0)}{\simeq}\tilde{R}^2U_0^2=\textrm{const.}
\end{equation}
It sheds light to express the radius (denoted $r_B$ here in order
to distinguish it from the meson inter-quark distance) and the
energy of the baryons in terms of the quark separation
\eqref{wilson-interquark distance} and of the potential
\eqref{wilson-renormalized potential} of the $Q\overline{Q}$
bound-states in three-dimensional Yang-Mills theory. We have at
large distances ($U_0\simeq U_T$):
\begin{eqnarray}
r_B(U_0,U_T)&\simeq&\frac{1}{2}r(U_0,U_T)\;\;,\\
V_B^{(R)}(U_0,U_T)&\simeq&\frac{N\,U_T}{8\pi}+\frac{N}{2}V_{Q\overline{Q}}^{(R)}(U_0,U_T)\;\;.
\end{eqnarray}
The integrals in $r_B(U_0,U_T)$ and $V_B^{(R)}(U_0,U_T)$ diverge
which gives rise, by identifying their singular contributions, to
a confining linear potential with a string tension equals to $N$
times the mesonic string tension \eqref{wilson-string tension
temperature}:
\begin{equation}
V_B^{(R)}(r)=N\Big(\frac{1}{2}\pi\tilde{R}^2T^2\Big)r\;\;.
\end{equation}

\section{Conclusion}

The expectation value of the Wilson loop $W[\CMcal{C}]$ provides,
through the area law at large distances, a criterion for the
confinement. According to the \emph{AdS}/CFT prescription
\cite{Maldacena wilson loop}, it can also be evaluated, on the
supergravity side, from the classical Nambu-Goto action of a
string world-sheet lying on the closed loop $\CMcal{C}$ at the
boundary. Since the world-sheet is no longer forced to span only
the four-dimensional boundary space-time but can spread out along
the fifth holographic coordinate, we do not expect to necessarily
recover the area law of the Wilson loop (which is indeed a
four-dimensional space-time result). In the absence of any length
scale, the interaction potentials of hadrons exhibit a
(non-perturbative) Coulomb-like behaviour $V(r)\propto-1/r$ in
agreement with the underlying conformal invariance of the boundary
theory. On the contrary, provided that a dimensionful parameter is
introduced in the formalism (which can be the Beckenstein-Hawking
temperature or by means of a warp factor in the $AdS_5$ metric),
the linear confinement $V(r)\propto r$ arises corresponding to the
situation where string world-sheet reaches a stationary point
along the holographic coordinate for which the inter-quark
distance $r$ explodes.\\
\\

\noindent{\bf Acknowledgement}

Most of the material exposed herehas been compiled during my INFN
fellowship in Bari, Italy while I was working on the phenomenology
of the IR Soft Wall Model. Especially, I am grateful to P.
Colangelo for having made my stay there so fruitful and exciting.
This work was partially supported by the Theoretical Physics
Center for Science Facilities (TPCSF), Institute of High Energy
Physics (IHEP), Chinese Academy of Sciences (CAS). \vspace{1cm}

\appendix
\section{Brief review of the Wilson loop in QCD}

When one attempts to formulate QCD in a discretized space-time,
one is naturally led to introduce the so-called Wilson loop
$W[\CMcal{C}]$ (especially, when one tries to build a gauge
invariant action for the gluon fields) of which the large distance
behaviour provides a confinement criterion \cite{Wilson1974}. The
(Euclidean version of the) area law:
\begin{equation}\label{App-area law}
W[\CMcal{C}]=e^{-\sigma_t\,r\,T}\;\;,
\end{equation}
where the contour $\CMcal{C}$ is taken as a rectangle with
time-like and space-like sides of length $T$ and $r$ respectively,
is equivalent to a confining interaction potential ($r$ is the
inter-quark distance):
\begin{equation}
V(r)=\sigma_t\,r\;\;.
\end{equation}
The Wilson loop is consequently a key ingredient of lattice QCD
and plays a fundamental role for the study of non-perturbative
properties of QCD.

We might wonder whether the Wilson loop plays such an important
role in the continuous theory. As a matter of fact, it turns out
to be at the basis of a formulation of QCD where all the
references to the gauge invariance of the theory (gauge
transformations, gauge-fixing terms, etc\ldots) are discarded.
Within this framework, QCD equations become functional equations
of the Wilson loop (or rather, of its multi-loop generalizations):
their resolving would then bring us valuable information on the
behaviour of QCD.

\subsection{The gauge line}

QCD is a non-Abelian gauge theory: the Lagrangian density
$\mathcal{L}_{QCD}$ is invariant under local transformations of
the quark field phases, which requires the presence of
self-interacting gluon fields in the theory. Being local, the
transformations involve space-time dependent parameters. That
explains why the Dirac mass term $-m\overline{q}(x)q(x)$ is
allowed in $\CMcal{L}_{QCD}$ whereas the non-gauge invariant
bilocal term $\overline{q}(y)q(x)$ is not for instance. So, which
sense to give to the partial derivative of the quark field
$\de_{\m}q(x)$ in a direction $\hat{\m}$:
\begin{equation}
\de_{\mu}q(x)\equiv\lim_{\epsilon\rightarrow0}\frac{q(x+\epsilon\hat{\mu})-q(x)}{\epsilon}
\end{equation}
which indeed involves two different space-time events. The gauge
transformation of $\de_{\mu}q(x)$ seems intricate since to each
point $x$ and $x+\epsilon\hat{\mu}$ corresponds a different
transformation law. To resolve this issue, one introduces a
non-local object, namely a phase factor or gauge line $U$ such
that $U(x+\epsilon\hat{\mu})q(x)$ and $q(x+\epsilon\hat{\mu})$
satisfy the same transformation law. In other words,
$U(x+\epsilon\hat{\mu})$ brings the gauge transformation from the
point $x$ to the point $x+\epsilon\hat{\mu}$. The new derivative
is then defined as:
\begin{equation}
D_{\mu}q(x)\equiv\lim_{\epsilon\rightarrow0}\frac{q(x+\epsilon\hat{\mu})-U(x+\epsilon\hat{\mu})q(x)}{\epsilon}
\end{equation}
which is nothing else than the standard covariant derivative:
\begin{equation}
D_{\mu}q(x)=(\de_{\mu}+igA_{\m}(x))q(x)
\end{equation}
such that
\begin{equation}
\begin{array}{rcl}
q(x)&\rightarrow&q'(x)=\Omega(x)q(x)\;\;,\\
D_{\mu}q(x)&\rightarrow&{D'}_{\mu}{q'}(x)=\Omega(x)D_{\mu}q(x)\;\;.
\end{array}
\end{equation}
$A_{\mu}(x)$ are the gauge fields, $\Omega(x)=e^{ig\omega(x)}$ is
an element of the gauge group $SU(N_c)$ and $g$ is the strong
coupling constant. In this appendix, the number of colours $N_c$
is regarded as a free parameter allowed to take all the possible
positive integers. As a result, we have $N_c^2-1$ (the dimension
of the group) real gauge parameters $\omega^i(x)$: the hermitian
parameter matrix reads
$\omega(x)\equiv\omega^i(x)\frac{\lambda^i}{2}$ and the
$\frac{\lambda^i}{2}$'s are the infinitesimal generators of the
algebra $su(N_c)$ ($i=1,2,\ldots,N_c^2-1$).

The definition of the gauge line is the following:
\begin{equation}\label{App_gauge line}
U(y,x;\CMcal{C})\equiv Pe^{-ig\int_{x}^{y}A_{\mu}(x)dx^{\mu}}
\end{equation}
which consists of a line integral of the gauge fields
$A_{\mu}(x)$: thus, $U(y,x;\CMcal{C})$ depends on the path
$\CMcal{C}$ oriented from $x$ to $y$. In the differential geometry
framework, one says that $U(y,x;\CMcal{C})$ performs a parallel
transport from $x$ to $y$ and that $A_{\mu}(x)$ is the
corresponding connection. The prescription $P$ in \eqref{App_gauge
line} is the path ordering operator, required in order to take
into account the non-Abelian nature of QCD. Indeed, the gauge
fields $A_{\mu}(x)\equiv A_{\mu}^{i}(x)\frac{\lambda^i}{2}$ are
anti-commuting square matrices of order $N_{c}$.

A path $\CMcal{C}(\sigma)$ (so, a mapping function) can be
parametrized as follows:
\begin{equation}
\begin{array}{lll}
\CMcal{C}&:&[0,1]\rightarrow\CMcal{M}^4\\
&&\sigma\rightarrow x^{\mu}(\sigma)
\end{array}
\end{equation}
where $\CMcal{M}^{4}$ is the $(3+1)-$dimensional Minkowski
space-time with the flat metric tensor
$\eta_{\mu\nu}=\textrm{diag}(+1,-1,-1,-1)$ and the parameter
$\sigma\in[0,1]$ such that $x^{\mu}(\sigma=0)\equiv x^{\mu}$ and
$x^{\mu}(\sigma=1)\equiv y^{\mu}$. Moreover,
$dx^{\mu}={\dot{x}}^{\mu}(\sigma)d\sigma$ where
${\dot{x}}^{\mu}=\frac{dx^{\mu}}{d\sigma}$ is the $\mu^{th}$
component of the four-vector tangent to $\CMcal{C}$ at
$x(\sigma)$. The gauge line \eqref{App_gauge line} has then the
following parametric representation:
\begin{equation}\label{App_parametrization}
U(y,x;\CMcal{C})=P\,e^{-ig\int_0^1
d\sigma{\dot{x}}^{\mu}(\sigma)A_{\mu}(x(\sigma))}
\end{equation}
and is invariant:

$\bullet$ by reparametrization: $x(\sigma)\rightarrow x(\sigma')$
where $\sigma'=f(\sigma)$ with $f'(\sigma)>0$ in order to keep
unchanged the point ordering along $\CMcal{C}$.

$\bullet$ under transformations of Poincar\'e's group (the flat
space-time isometry group):
$x^{\mu}(\sigma)\rightarrow{x'}^{\mu}(\sigma)={\Lambda^{\mu}}_{\nu}x^{\nu}(\sigma)+a^{\mu}$
where the Lorentz matrix ${\Lambda^{\mu}}_{\nu}$ and the
translation parameter $a^{\mu}$ are constant.

One sees that to each generator $\frac{\lambda^i}{2}$ is
associated, via the gauge field $A_{\mu}^i(x(\sigma))$, one and
only one value of $\sigma$: the operator $P$ puts in order,
through the parameter $\sigma$, the $su(N_c)$ generators
$\frac{\lambda^i}{2}$'s along $\CMcal{C}$. Let us now divide the
path $\CMcal{C}$ into infinitesimal straight lines
$\delta\CMcal{C}$'s, it is then possible to express the phase
factor $U(y,x;\CMcal{C})$ as an infinite product of elementary
gauge lines, each of them being associated with its own contour
$\delta\CMcal{C}$ and ordered along $\CMcal{C}$ according to the
prescription $P$:
\begin{equation}
U(y,x;\CMcal{C})=\lim_{n\rightarrow\infty}\prod_{n}U(x_n,x_{n-1};\delta\CMcal{C})\;.
\end{equation}
Another expression for $U(y,x;\CMcal{C})$ makes use of the
parametric representation \eqref{App_parametrization}:
\begin{eqnarray}\label{App_parametric heaviside expansion}
U^a_b(y,x;\CMcal{C})&=&\sum_{n=0}^{\infty}\big(-\frac{i
g}{\sqrt2}\big)^n\int_0^1d\sigma_1\ldots\int_0^1d\sigma_n\;\theta(\sigma_1-\sigma_2)\ldots\theta(\sigma_{n-1}-\sigma_n)\non\\
&&\times\dot{x}^{\mu_1}(\sigma_1)\ldots\dot{x}^{\mu_n}(\sigma_n){A_{\mu_1}}^a_{c_1}(\sigma_1)\ldots
{A_{\mu_n}}^{c_{n-1}}_b(\sigma_n)
\end{eqnarray}
where we have written $A_{\mu}(x(\sigma))\equiv A_{\mu}(\sigma)$
for the sake of simplicity and defined the matrix element
$(A_\mu)^a_b=(A_{\mu}^i\frac{\lambda^i}{2})^a_b\equiv\frac{1}{\sqrt2}{A_{\mu}}^a_b$
$(a,b=1,2,\ldots,N_c)$. At each order in the expansion, the
operator $P$ puts in order the gauge field product from the right
to the left by increasing value of $\sigma$. The Heaviside
functions play the role of the prescription $P$ and compensate for
the factors of $1/n!$ usually present when expanding the
exponential function. (The expansion \eqref{App_parametric
heaviside expansion} is similar to the well-known QFT expansion of
the evolution operator in interaction representation.)

Because the gauge line $U(y,x;\CMcal{C})$ carries out a parallel
transport from $x$ to $y$ along the path $\CMcal{C}(\sigma)$, the
fields $q(y)$ and $U(y,x;\CMcal{C})q(x)$ satisfy the same
transformation law (thus, the operator
$\overline{q}(y)U(y,x;\CMcal{C})q(x)$ is manifestly gauge
invariant) which implies the following gauge transformation for
$U(y,x;\CMcal{C})$:
\begin{equation}\label{App-gauge transformation}
U(y,x;\CMcal{C})\rightarrow{U'}(y,x;\CMcal{C})=\Omega(y)U(y,x;\CMcal{C})\Omega^{\dagger}(x)\;\;.
\end{equation}
Then, the corresponding matrix element $U^a_b(y,x;\CMcal{C})$
transforms as follows:
\begin{equation}
U^a_b(y,x;\CMcal{C})\rightarrow{U'}^a_b(y,x;\CMcal{C})=\Omega^a_c(y)U^c_d(y,x;\CMcal{C})\Omega^d_b(x)^{\dagger}
\end{equation}
from which can be easily obtained the infinitesimal gauge
transformation:
\begin{equation}
\delta_GU^a_b(y,x;\CMcal{C})=ig\big[w^a_c(y)U^c_b(y,x;\CMcal{C})-U^a_c(y,x;\CMcal{C})w^c_b(x)\big]\;\;.
\end{equation}

\subsection{Mandelstam's formula}

Mandelstam's formula \cite{Mandelstam} is one of the most
important equation when one attempts to rewrite Yang-Mills theory
within the loop space formalism (defined as the set of all the
continuous closed curves):
\begin{equation}
\frac{\delta}{\delta\sigma^{\mu\nu}(z)}U(y,x;\CMcal{C})=-igU(y,z;\CMcal{C})G_{\mu\nu}(z)U(z,x;\CMcal{C})
\end{equation}
with
$G_{\mu\nu}(z)=\de_{\mu}A_{\nu}(z)-\de_{\nu}A_{\mu}(z)+ig[A_{\mu},A_{\nu}](z)$
the non-Abelian strength field tensor. This relation can be
derived as follows. Thanks to the path ordering prescription $P$,
the gauge line \eqref{App_gauge line} can be written as the
product of three phase factors:
\begin{equation}
U(y,x;\CMcal{C})=U(y,x_2;\CMcal{C})U(x_2,x_1;\CMcal{C})U(x_1,x;\CMcal{C})\;\;,
\end{equation}
where each phase factor corresponds to one of the three stretches
$(x,x_1)$, $(x_1,x_2)$ and $(x_2,y)$ of $\CMcal{C}$. The two
points $x_1$ and $x_2$ are located anywhere between the endpoints
$x$ and $y$ of the curve. Let us then perform an infinitesimal
variation $\delta\CMcal{C}$ of the section $(x_1,x_2)$:
\begin{equation}
\left\{
\begin{array}{rll}
\CMcal{C}&\rightarrow&\CMcal{C}'=\CMcal{C}+\delta\CMcal{C}\\
U(x_2,x_1;\CMcal{C})&\rightarrow&U(x_2,x_1;\CMcal{C}')
\end{array}
\right.\;\;.
\end{equation}
The gauge line corresponding to the new curve $\CMcal{C}'$ reads:
\begin{equation}
U(y,x;\CMcal{C}')=U(y,x_2;\CMcal{C})U(x_2,x_1;\CMcal{C}+\delta\CMcal{C})U(x_1,x;\CMcal{C})\;\;.
\end{equation}
One defines the variation of the gauge line when passing from the
curves $\CMcal{C}$ to $\CMcal{C}'$ by the difference:
\begin{eqnarray}
\delta_{\CMcal{C}}
U(y,x;\CMcal{C})&=&U(y,x;\CMcal{C}')-U(y,x;\CMcal{C})\non\\
&=&U(y,x_2;\CMcal{C})\big[U(x_2,x_1;\CMcal{C}+\delta\CMcal{C})-U(x_2,x_1;\CMcal{C})\big]U(x_1,x;\CMcal{C})\non\\
\delta_{\CMcal{C}}
U(y,x;\CMcal{C})&=&U(y,x_2;\CMcal{C})\big[U(x_2,x_1;\CMcal{C}+\delta\CMcal{C})U^{-1}(x_2,x_1;\CMcal{C})-1\big]U(x_2,x_1;\CMcal{C})U(x_1,x;\CMcal{C})\;\;.\non\\
\label{App_change gauge line}
\end{eqnarray}
The phase factor $U^{-1}(x_2,x_1;\CMcal{C})$ corresponds to the
section $(x_1,x_2)$ of the contour $\CMcal{C}$ but oriented from
$x_2$ to $x_1$: $U^{-1}(x_2,x_1;\CMcal{C})=U(x_1,x_2;\CMcal{C})$.
Consequently, the product
$U(x_2,x_1;\CMcal{C}+\delta\CMcal{C})U^{-1}(x_2,x_1;\CMcal{C})$ in
\eqref{App_change gauge line} stands for a closed curve or loop,
denoted $C$. If the modification $\delta\CMcal{C}$ of the contour
$\CMcal{C}$ is infinitesimal, this loop $C$ gives rise to a
surface of infinitesimal area $\delta S$. We can then apply the
non-Abelian Stockes theorem valid for infinitesimal loops:
\begin{equation}\label{App_non Abelian Stockes theorem}
e^{\oint_{C}A_{\mu}dx^{\mu}}=e^{\int\int_{\delta
S}d\sigma_{\mu\nu}(z)G^{\mu\nu}(z)}\;\;,
\end{equation}
such that
\begin{eqnarray}
U(x_2,x_1;\CMcal{C}+\delta\CMcal{C})U^{-1}(x_2,x_1;\CMcal{C})&=&Pe^{-ig\int_{\CMcal{C}+\delta\CMcal{C}}A_{\mu}dx^{\mu}}Pe^{ig\int_{\CMcal{C}}A_{\mu}dx^{\mu}}\non\\
&=&Pe^{-ig\oint_{C}A_{\mu}dx^{\mu}}\non\\
U(x_2,x_1;\CMcal{C}+\delta\CMcal{C})U^{-1}(x_2,x_1;\CMcal{C})&=&Pe^{-ig\int\int_{\delta
S}d\sigma_{\mu\nu}(z)G^{\mu\nu}(z)}\label{App_flux}
\end{eqnarray}
where $d\sigma_{\mu\nu}(z)=dz_{\mu}\wedge dz_{\nu}$ is the area
element ($d\sigma_{\mu\nu}=-d\sigma_{\nu\mu}$) with the internal
point $z$ located between $x_1$ and $x_2$ along $\CMcal{C}$.

We see that
$U(x_2,x_1;\CMcal{C}+\delta\CMcal{C})U^{-1}(x_2,x_1;\CMcal{C})$
performs in \eqref{App_change gauge line} a rotation of the gauge
field $A_{\m}$ along the loop $C$. According to the non-Abelian
Stockes theorem \eqref{App_non Abelian Stockes theorem}, this
rotation is related to the flux \eqref{App_flux} of the strength
field tensor $G_{\mu\nu}$ through a surface with the loop $C$ as
boundary (in the differential geometry formalism, $G_{\mu\nu}$ is
the curvature tensor of the internal colour space). Then,
Mandelstam's formula can be obtained by expanding
\eqref{App_change gauge line} at the leading order in $\delta S$:
\begin{equation}
\delta_{\CMcal{C}}
U(y,x;\CMcal{C})=U(y,x_2;\CMcal{C})\big[1-ig\int\int_{\delta
S}d\sigma_{\mu\nu}(z)G^{\mu\nu}(z)-1\big]U(x_2,x_1;\CMcal{C})U(x_1,x;\CMcal{C})+O(\delta
S^2)
\end{equation}
so that
\begin{equation}\label{App_functional surface derivation}
\frac{\delta}{\delta\sigma^{\mu\nu}(z)}U(y,x;\CMcal{C})=-igU(y,z;\CMcal{C})G_{\mu\nu}(z)U(z,x;\CMcal{C})
\end{equation}
by deriving with respect to the area element at any point $z$
between $x_1$ and $x_2$. In brief, the area derivative
$\delta/\delta\sigma^{\mu\nu}(z)$ consists in inserting, at the
point $z$ along the contour $\CMcal{C}$, the non-Abelian strength
field tensor $G_{\mu\nu}(z)$.

Although elegant, this derivation presents a certain number of
shortcomings: first of all, the demonstration is geometrical and
depends on the form of the curves $\CMcal{C}$ and $\CMcal{C}'$.
Secondly, the functional derivation \eqref{App_functional surface
derivation} is not mathematically well-defined because arbitrary.
Indeed, there are several possible definitions
\cite{Polyakov,Makeenko} such as:
\begin{equation}
\frac{\delta}{\delta\sigma_{\mu\nu}(z)}U(y,x;\CMcal{C})\equiv\lim_{|\delta\sigma_{\mu\nu}|\rightarrow0}\frac{1}{|\delta\sigma_{\mu\nu}|}\Big(U(y,x;\CMcal{C}+\delta\CMcal{C}_{\mu\nu}(z))-U(y,x;\CMcal{C})\Big)
\end{equation}
as used by Mandelstam where $\delta\CMcal{C}_{\mu\nu}(z)$ is the
infinitesimal loop oriented along the plane $(\mu,\nu)$  at the
point $z$ and $|\delta\sigma_{\mu\nu}|$ is the area of the
associated minimal surface. In order to avoid these difficulties,
it is therefore interesting to find another way to derive
Mandelstam's formula \eqref{App_functional surface derivation}.
The solution consists in working with the (geometry independent)
parametric representation of the gauge line
\eqref{App_parametrization}. Let us write the variation of
$U(y,x;\CMcal{C})$ under an arbitrary transformation of
$\CMcal{C}$ following the notations $U(x(\sigma),x(\sigma'))\equiv
U(\sigma,\sigma')$, $A_{\alpha}(x(\sigma))\equiv
A_{\alpha}(\sigma)$ and $G_{\beta\alpha}(x(\sigma))\equiv
G_{\beta\alpha}(\sigma)$ \cite{jugeau}:
\begin{equation}\label{App_variation parametrization}
\begin{array}{lll}
\delta_{\CMcal{C}} U(y,x;\CMcal{C})&=&-ig\delta
x^{\alpha}(1)A_{\alpha}(1)U(1,0)+igU(1,0)A_{\alpha}(0)\delta x^{\alpha}(0)\\
&&+ig\int_0^1d\sigma
U(1,\sigma)\dot{x}^{\beta}(\sigma)G_{\beta\alpha}(\sigma)\delta
x^{\alpha}(\sigma)U(\sigma,0)\;\;.
\end{array}
\end{equation}
The first two terms on the \emph{r.h.s.} of \eqref{App_variation
parametrization} correspond to the variations of the endpoints
$y\equiv x(\sigma=1)$ and $x\equiv x(\sigma=0)$ respectively. The
last term involves all the points $x(\sigma)$ of $\CMcal{C}$ and
thus corresponds to the internal contribution associated with the
global deformation of $\CMcal{C}$. Let us perform the functional
derivation with respect to an internal point $x^{\alpha}(\sigma)$
along the contour $\CMcal{C}$ ($0<\sigma<1$ such that only the
last term in \eqref{App_variation parametrization} contributes):
\begin{equation}
\frac{\delta U(y,x;\CMcal{C})}{\delta
x^{\alpha}(\sigma)}=igU(1,\sigma)\dot{x}^{\beta}(\sigma)G_{\beta\alpha}(\sigma)U(\sigma,0)\;\;.
\end{equation}
A partial derivative with respect to $\dot{x}^{\beta}(\sigma)$
gives then Mandelstam's formula:
\begin{equation}\label{App_second derivation}
\frac{\de}{\de\dot{x}^{\beta}(\sigma)}\frac{\delta
U(y,x;\CMcal{C})}{\delta
x^{\alpha}(\sigma)}=-igU(1,\sigma)G_{\alpha\beta}(\sigma)U(\sigma,0)
\end{equation}
where $x(\sigma)$ and $\dot{x}(\sigma)$ have to be considered as
independent operators. Comparing \eqref{App_functional surface
derivation} with \eqref{App_second derivation}, we obtain the
following relations between the surface and functional derivatives
with respect to the point $x(\sigma)$:
\begin{eqnarray}
\frac{\de}{\de\dot{x}^{\beta}(\sigma)}\frac{\delta}{\delta
x^{\alpha}(\sigma)}&\equiv&\frac{\delta}{\delta\sigma^{\alpha\beta}(\sigma)}\;\;,\label{App_operator relation}\\
\frac{\delta}{\delta
x^{\alpha}(\sigma)}&\equiv&\dot{x}^{\beta}(\sigma)\frac{\delta}{\delta\sigma^{\alpha\beta}(\sigma)}\;\;.
\end{eqnarray}

\subsection{The equations of QCD}

It is worth going further in our study of the gauge line. Let us
then calculate the partial derivative (or path derivative
\cite{Migdal}) with respect to $x_{\lambda}(\sigma)$ of
Mandelstam's formula \eqref{App_second derivation}. We get:
\begin{eqnarray}
\frac{\de}{\de
x_{\lambda}(\sigma)}\frac{\de}{\de\dot{x}^{\beta}(\sigma)}\frac{\delta
U(y,x;\CMcal{C})}{\delta x^{\alpha}(\sigma)}&=&-ig\Big[\frac{\de
U(1,\sigma)}{\de
x_{\lambda}(\sigma)}G_{\alpha\beta}(\sigma)U(\sigma,0)+U(1,\sigma)G_{\alpha\beta}(\sigma)\frac{\de
U(\sigma,0)}{\de x_{\lambda}(\sigma)}\non\\
&&+U(1,\sigma)\frac{\de G_{\alpha\beta}(\sigma)}{\de
x_{\lambda}(\sigma)}U(\sigma,0)\Big]\;\;.\label{App_QCD equation}
\end{eqnarray}
On the other hand, from \eqref{App_variation parametrization}, one
can infer the actions of $\frac{\delta}{\delta
x_{\lambda}(\sigma)}$ on $U(1,\sigma)$ and $U(\sigma,0)$:
\begin{equation}
\left\{
\begin{array}{lll}
\frac{\de U(1,\sigma)}{\de x_{\lambda}(\sigma)}&=&igPU(1,\sigma)A^{\lambda}(\sigma)\\
\frac{\de U(\sigma,0)}{\de
x_{\lambda}(\sigma)}&=&-igPA^{\lambda}(\sigma)U(\sigma,0)
\end{array}
\right.
\end{equation}
since contributes, in every case, only one boundary term.
Eq.\eqref{App_QCD equation} becomes finally:
\begin{equation}
\frac{\de}{\de
x_{\lambda}(\sigma)}\frac{\de}{\de\dot{x}^{\beta}(\sigma)}\frac{\delta
U(y,x;\CMcal{C})}{\delta
x^{\alpha}(\sigma)}=-igU(1,\sigma)\Big(\frac{\de}{\de
x_{\lambda}(\sigma)}G_{\alpha\beta}(\sigma)+ig\big[A^{\lambda}(\sigma),G_{\alpha\beta}(\sigma)\big]\Big)U(\sigma,0)
\end{equation}
where one recognizes the expression of the covariant derivative:
\begin{equation}\label{App_covariant derivative definition}
\frac{\de}{\de
x_{\lambda}(\sigma)}\frac{\de}{\de\dot{x}^{\beta}(\sigma)}\frac{\delta
U(y,x;\CMcal{C})}{\delta
x^{\alpha}(\sigma)}=-igU(1,\sigma)D^{\lambda}G_{\alpha\beta}(\sigma)U(\sigma,0)\;\;.
\end{equation}
From this equation, it is possible to obtain two fundamental
results:

$\bullet$ by cyclic permutation over all the Lorentz indices, we
get with \eqref{App_operator relation}:
\begin{eqnarray}
\Big(\frac{\de}{\de x_{\lambda}(\sigma)}\frac{\delta
}{\delta\sigma^{\alpha\beta}(\sigma)}+\frac{\de}{\de
x_{\beta}(\sigma)}\frac{\delta
}{\delta\sigma^{\lambda\alpha}(\sigma)}+\frac{\de}{\de
x_{\alpha}(\sigma)}\frac{\delta
}{\delta\sigma^{\beta\lambda}(\sigma)}\Big)U(y,x;\CMcal{C})=\non\\
\;\;\;\;\;\;\;\;-igU(1,\sigma)\Big(D^{\lambda}G_{\alpha\beta}(\sigma)+D^{\beta}G_{\lambda\alpha}(\sigma)+D^{\alpha}G_{\beta\lambda}(\sigma)\Big)U(\sigma,0)
\end{eqnarray}
which is nothing else than the Bianchi identity of Yang-Mills
theory. The gauge line satisfies thus the following constraint:
\begin{equation}
\frac{\de}{\de x_{\lambda}(\sigma)}\frac{\delta
U(y,x;\CMcal{C})}{\delta\sigma^{\alpha\beta}(\sigma)}+(\textrm{cyclic
permutation})=0\;\;.
\end{equation}

$\bullet$ by contracting in \eqref{App_covariant derivative
definition} the Lorentz indices $\lambda$ and $\alpha$, we find:
\begin{equation}\label{App-equation of motion}
\frac{\de}{\de x_{\alpha}(\sigma)}\frac{\delta
U(y,x;\CMcal{C})}{\delta\sigma^{\alpha\beta}(\sigma)}=-igU(1,\sigma)D^{\alpha}G_{\alpha\beta}(\sigma)U(\sigma,0)\;\;.
\end{equation}
In this case, we recognize on the \emph{r.h.s} the first term
$D^{\alpha}G_{\alpha\beta}$ in the equation of motion of the
Yang-Mills gauge field.

\subsection{The Wilson loop: the colour trace of a closed gauge line}

When the gauge line endpoints of coordinates $x$ and $y$ coincide
in \eqref{App_gauge line}, $x(\sigma=0)=x(\sigma=1)$ and we obtain
a closed gauge line $U(x,x;\CMcal{C})$. Then, the trace on the
colour space gives the so-called Wilson loop $\phi(\CMcal{C})$:
\begin{equation}\label{App_wilson loop}
\phi(\CMcal{C})\equiv
Tr_c\,U(x,x;\CMcal{C})=Tr_c\,P\,e^{-ig\oint_{\CMcal{C}}
A_{\mu}(x)dx^{\mu}}\;\;.
\end{equation}
Let us pointing out that the Wilson loop does not depend in
\eqref{App_wilson loop} on the point $x$ from which is
parametrized the loop $\CMcal{C}$ because of the colour trace
$Tr_c$. Moreover, it is a manifestly gauge invariant functional.
Indeed, under a gauge transformation, we have:
\begin{equation}
U(x,x;\CMcal{C})\rightarrow
U'(x,x;\CMcal{C})=\Omega(x)U(x,x;\CMcal{C})\Omega^{\dagger}(x)
\end{equation}
so that
\begin{eqnarray}
\phi(\CMcal{C})\rightarrow\phi'(\CMcal{C})&=&Tr_c\,U'(x,x;\CMcal{C})\;\;,\non\\
&=&Tr_c\,\Big(\Omega(x)U(x,x;\CMcal{C})\Omega^{\dagger}(x)\Big)\;\;,\non\\
\phi'(\CMcal{C})&=&Tr_c\,U(x,x;\CMcal{C})=\phi(\CMcal{C})
\end{eqnarray}
by invoking the invariance of the trace under a cyclic
permutation.

\subsection{The Migdal-Makeenko equation and the loop equation}

When one reformulates a non-Abelian gauge theory as QCD into the
loop space \cite{Mandelstam,Polyakov 2,Nambu, Gervais}, all the
references to the gauge fields, gauge transformations,
gauge-fixing terms, ghosts and so on and so forth are discarded
and only remain gauge invariant functionals. The observables are
then expressed in terms of such functionals and the equations of
motion of the gauge fields describing QCD dynamics (without
fermions) are replaced by functional equations. The loop space
formalism being gauge invariant, the properties of the gauge group
become functional constraints. Examples of such constraints are
Mandelstam's constraints \cite{Mandelstam 2,Giles}, among which we
have for instance:

$\bullet$ the reparametrization invariance:
\begin{equation}
\phi(\CMcal{C})=\phi(\CMcal{C}')
\end{equation}
where $\CMcal{C}=\{\sigma\rightarrow x(\sigma)\}$ and
$\CMcal{C}'=\{\sigma'=f(\sigma)\rightarrow x(\sigma')\}$ with
$f(0)=0$ and $f(1)=1$.

$\bullet$ The reversal relation:
\begin{equation}
\phi(\CMcal{C})=\phi(\CMcal{C}')^{-1}
\end{equation}
with $x(\sigma')=x(1-\sigma)$.

Although this program gives rise to strong difficulties, a certain
number of issues have been resolved. For example, it has been
shown that the Yang-Mills equation of motion becomes, in loop
space, a functional equation satisfied by the Wilson loop, the
so-called Migdal-Makeenko equation \cite{Makeenko,Brandt}:
\begin{equation}\label{App_MM equation2}
\frac{\de}{\de
x^{\mu}}\frac{\delta}{\delta\sigma_{\mu\nu}(x)}\langle\phi(\CMcal{C})\rangle_A=-\frac{ig^2}{2}\oint_{\CMcal{C}}dy^{\nu}\delta^4(x-y)\langle\phi(\CMcal{C}_{yx})\phi(\CMcal{C}_{xy})-\frac{1}{N_c}\phi(\CMcal{C})\rangle_A\;\;,
\end{equation}
the averaging being defined in the path-integral formalism. Before
studying its most important properties, let us roughly check its
derivation. For this, we consider the equation of motion
\eqref{App-equation of motion} derived from Mandelstam's formula:
\begin{equation}
\frac{\de}{\de
x^{\mu}(\sigma)}\frac{\de}{\de\dot{x}_{\nu}(\sigma)}\frac{\delta\phi(\CMcal{C})}{\delta
x_{\mu}(\sigma)}=-igU(1,\sigma)D_{\mu}G^{\mu\nu}(\sigma)U(\sigma,0)
\end{equation}
or, in tensorial notation:
\begin{equation}\label{App_eq of motion from Mandelstam formula}
\frac{\de}{\de
x^{\mu}(\sigma)}\frac{\de}{\de\dot{x}_{\nu}(\sigma)}\frac{\delta\phi(\CMcal{C})}{\delta
x_{\mu}(\sigma)}=-\frac{ig}{\sqrt2}U^a_c(1,\sigma){D_{\mu}}^c_d{G^{\mu\nu}}^d_e(\sigma)U^e_a(\sigma,0)
\end{equation}
where we have defined
$(G_{\mu\nu})^a_b=(G_{\mu\nu}^i\frac{\lambda^i}{2})^a_b\equiv\frac{1}{\sqrt2}{G_{\mu\nu}}^a_b$.
In QCD, the equation of motion of the gluon field is:
\begin{equation}
D_{\mu}G^{\mu\nu}(x)=j^{\nu}_{F}(x)+j^{\nu}_{gf}(x)+j^{\nu}_{FP}(x)
\end{equation}
with $j^{\nu}_{F}(x)$ the fermionic current and $j^{\nu}_{gf}(x)$
and $j^{\nu}_{FP}(x)$ the currents associated with the
gauge-fixing term and the Faddeev-Popov ghosts respectively. If we
remark that the functional derivative of a path integral vanishes:
\begin{equation}
\frac{\delta}{\delta A}\langle Q[A]\rangle_A=0
\end{equation}
for any functional $Q[A]$ of the gauge field, then the equation of
motion reads in the path-integral formalism as:
\begin{equation}
\langle i\frac{\delta Q}{\delta A}\rangle_A=\langle\frac{\delta
S}{\delta A}Q\rangle_A\;\;.
\end{equation}
Without external fermionic source, $j^{\nu}_{F}(x)$ vanishes.
Moreover, since the Wilson loop $\phi(\CMcal{C})$ is gauge
invariant, the currents $j^{\nu}_{gf}(x)$ and $j^{\nu}_{FP}(x)$
cancel out each other (the Slanov-Taylor-Ward-Takahashi identity
shows that $j^{\nu}_{gf}(x)+j^{\nu}_{FP}(x)=0$ in the
path-integral formalism \cite{Brandt}). Therefore, we are led to
write out:
\begin{equation}
D_{\mu}G^{\mu\nu}(x)\equiv i\frac{\delta}{\delta A_{\nu}(x)}
\end{equation}
which should be understood in the weak sense, namely in terms of
mean values. As a result, the equation \eqref{App_eq of motion
from Mandelstam formula} becomes in the path-integral formalism:
\begin{equation}\label{App_funct derivative}
\frac{\de}{\de
x^{\mu}(\sigma)}\frac{\de}{\de\dot{x}_{\nu}(\sigma)}\frac{\delta\langle\phi(\CMcal{C})\rangle_A}{\delta
x_{\mu}(\sigma)}=-\frac{ig}{\sqrt2}\langle
P\Big(i\frac{\delta}{\delta
{A_{\nu}}^e_c(\sigma)}\Big)U^a_c(1,\sigma)U^e_a(\sigma,0)\rangle_A\;\;.
\end{equation}
Expanding as before \eqref{App_parametric heaviside expansion} the
gauge line $U^{a}_{c}(1,\sigma)$ as a power series, the
contribution of the order $O(g^n)$ involves the product of $n$
gauge fields:
\begin{equation}
U^a_c(1,\sigma)U^e_a(\sigma,0)\simeq
{A_{\mu_1}}^a_{c_1}(\sigma_1){A_{\mu_2}}^{c_1}_{c_2}(\sigma_2){A_{\mu_3}}^{c_2}_{c_3}(\sigma_3){A_{\mu_4}}^{c_3}_{c_4}(\sigma_4)\ldots
{A_{\mu_n}}^{c_{n-1}}_{c}(\sigma_n)U^e_a(\sigma,0)\;\;.
\end{equation}
For the sake of argument, let us consider the contribution of the
field ${A_{\mu_4}}^{c_3}_{c_4}(\sigma_4)$. The functional
derivative with respect to $\delta/\delta {A_{\nu}}^e_c(\sigma)$
gives:
\begin{equation}\label{App_example}
\begin{array}{c}
\delta^{\nu}_{\mu_4}\delta^{4}(x(\sigma_4)-x(\sigma))\Big\{{A_{\mu_1}}^a_{c_1}(\sigma_1){A_{\mu_2}}^{c_1}_{c_2}(\sigma_2){A_{\mu_3}}^{c_2}_e(\sigma_3)\Big[{A_{\mu_5}}^c_{c_5}(\sigma_5)\ldots
{A_{\mu_n}}^{c_{n-1}}_c(\sigma_n)\Big]\\
-\frac{1}{N_c}{A_{\mu_1}}^a_{c_1}(\sigma_1){A_{\mu_2}}^{c_1}_{c_2}(\sigma_2){A_{\mu_3}}^{c_2}_{c_3}(\sigma_3){A_{\mu_5}}^{c_3}_{c_5}(\sigma_5)\ldots{A_{\mu_n}}^{c_{n-1}}_e(\sigma_n)\Big\}U^e_a(\sigma,0)
\end{array}
\end{equation}
according to
\begin{equation}
\frac{\delta{A_{\mu}}^a_b(\sigma)}{\delta{A^{\nu}}^c_d(\sigma')}=\eta_{\mu\nu}\big(\delta^a_c\delta^d_b-\frac{1}{N_c}\delta^a_b\delta^c_d\big)\delta^{4}(x(\sigma)-x(\sigma'))\;\;
\end{equation}
where the tensorial structure on the colour indices comes from the
fact that the gluon fields are traceless. Because of the delta
function $\delta^4(x-y)$ in \eqref{App_example}, the points
$x\equiv x(\sigma)$ and $y\equiv x(\sigma_4)$ coincide in
space-time but can be different in loop space as they are
associated with different values of parameter. In this case, only
contributes the first term in \eqref{App_example} which
corresponds to the creation of an internal loop at the point $x$
of $\CMcal{C}$:
\begin{equation}
\big[{A_{\mu_5}}^{c}_{c_5}(\sigma_4)\ldots{A_{\mu_n}}^{c_{n-1}}_c(\sigma)\big]\;.
\end{equation}
Thus, this contribution gives rise in \eqref{App_MM equation2} to
the product of the two Wilson loops $\phi(\CMcal{C}_{yx})$ and
$\phi(\CMcal{C}_{xy})$.

On the other hand, when the parameters are equal
$\sigma=\sigma_4$, only contributes the second term in
\eqref{App_example} which corresponds to only one Wilson loop of
contour $\CMcal{C}$. Finally, we get the Migdal-Makeenko equation:
\begin{equation}
\frac{\de}{\de
x^{\mu}}\frac{\delta}{\delta\sigma_{\mu\nu}}\langle\phi(\CMcal{C})\rangle_A=-\frac{ig^2}{2}\oint_{\CMcal{C}}dy^{\nu}\delta^{4}(x-y)\langle\phi(\CMcal{C}_{yx})\phi(\CMcal{C}_{xy})-\frac{1}{N_c}\phi(\CMcal{C})\rangle_A
\end{equation}
where the current
\begin{equation}
j^{\nu}(x)\equiv-\frac{ig^2}{2}\oint_{\CMcal{C}}dy^{\nu}\delta^4(x-y)\langle\phi(\CMcal{C}_{yx})\phi(\CMcal{C}_{xy})-\frac{1}{N_c}\phi(\CMcal{C})\rangle_A
\end{equation}
is conserved, $\de_{\nu}j^{\nu}=0$, thanks to the antisymmetry
property of the surface derivative
$\delta/\delta\sigma_{\mu\nu}(x)$ in the exchange
$\mu\leftrightarrow\nu$.

Let us define the following gauge invariant one- and two-loop
functionals:
\begin{eqnarray}
W(\CMcal{C})&\equiv&\frac{1}{N_c}\langle\phi(\CMcal{C})\rangle_A\;,\label{App-one loop functional}\\
W_2(\CMcal{C}_1,\CMcal{C}_2)&\equiv&\langle\frac{1}{N_c}\phi(\CMcal{C}_1)\frac{1}{N_c}\phi(\CMcal{C}_2)\rangle_A
\end{eqnarray}
as the vacuum expectation values, respectively of the Wilson loop
$\phi(\CMcal{C})$ and of the product of two loops
$\phi(\CMcal{C}_1)\phi(\CMcal{C}_2)$ (up to normalization factors
in $1/N_c$). The Migdal-Makeenko equation takes then the following
form:
\begin{equation}\label{App_MM equation}
\frac{\de}{\de
x^{\mu}}\frac{\delta}{\delta\sigma_{\mu\nu}(x)}W(\CMcal{C})=-i\lambda\oint_{\CMcal{C}}dy^{\nu}\delta^{4}(x-y)\Big(W_2(\CMcal{C}_{yx},\CMcal{C}_{xy})-\frac{1}{N_c^2}W(\CMcal{C})\Big)\;\;,
\end{equation}
with $\lambda\equiv g^2 N_c/2$. Eq. \eqref{App_MM equation}
involves the three loops $\CMcal{C}$, $\CMcal{C}_{yx}$ (oriented
from $x$ to $y$) and $\CMcal{C}_{xy}$ (oriented from $y$ to $x$)
such that $\CMcal{C}_{yx}\cup\CMcal{C}_{cy}=\CMcal{C}$ and is,
consequently, a non-linear and non-closed functional equation: the
one-loop functional $W(\CMcal{C})$ (also called, roughly speaking,
Wilson loop) is related to the two-loop functional
$W_2(\CMcal{C}_{yx},\CMcal{C}_{yx})$. The Migdal-Makeenko equation
is the first equation of an infinite set of functional equations
relating the derivatives of $n-$loop functionals to the integrals
of $(n-1)-$, $n-$ and $(n+1)-$loop functionals.

Most of the time, one finds, in the literature, the
Migdal-Makeenko equation written in the Euclidean space-time $E^4$
with the metric tensor $\delta_{\mu\nu}=\textrm{diag}(1,1,1,1)$.
In this case, the Migdal-Makeenko equation \eqref{App_MM equation}
becomes:
\begin{equation}\label{App_euclidean MM equation}
\frac{\de}{\de
x_{\mu}}\frac{\delta}{\delta\sigma_{\mu\nu}(x)}W(\CMcal{C})=\lambda\oint_{\CMcal{C}}dy^{\nu}\delta^{4}(x-y)\Big(W_2(\CMcal{C}_{yx},\CMcal{C}_{xy})-\frac{1}{N_c^2}W(\CMcal{C})\Big)\;.
\end{equation}
There are two cases in which \eqref{App_euclidean MM equation} can
be simplified : in the Abelian case where $N_c=1$ and in the 't
Hooft limit where $N_c\rightarrow\infty$ with $\l$ finite.

$\bullet$ When $N_c=1$, one deals with an Abelian $U(1)$ gauge
theory with coupling constant $g'$ \cite{Brandt}. The gauge field
is no longer a matrix such that
\begin{equation}
\frac{\delta A_{\mu}(\sigma)}{\delta
A^{\nu}(\sigma')}=\delta_{\mu\nu}\delta^4(x(\sigma)-x(\sigma'))
\end{equation}
in \eqref{App_funct derivative}. As a result, only the first term
in \eqref{App_example} (without colour indices) contributes and
only remains on the \emph{r.h.s} of \eqref{App_euclidean MM
equation} the two-loop functional $W_2$. Moreover, the square
matrices $\phi(\CMcal{C}_{xy})$ and $\phi(\CMcal{C}_{yx})$ of
order $N_c$ become simple operators and
$W_2(\CMcal{C}_1,\CMcal{C}_2)$ reduces to the Wilson loop
$W(\CMcal{C})$. At the end of the day, the Abelian Migdal-Makeenko
equation takes the form:
\begin{equation}
\frac{\de}{\de
x_{\mu}}\frac{\delta}{\delta\sigma_{\mu\nu}(x)}W(\CMcal{C})={g'}^2\oint_{\CMcal{C}}dy^{\nu}\delta^{4}(x-y)W(\CMcal{C})
\end{equation}
which has the solution:
\begin{equation}\label{abelian solution}
W(\CMcal{C})=e^{-\frac{{g'}^2}{2}\oint_{\CMcal{C}}\oint_{\CMcal{C}}dx_{\mu}dy_{\nu}D_{\mu\nu}(x-y)}
\end{equation}
where $D_{\mu\nu}(x-y)$ is the coulombic propagator of the Abelian
gauge field $A_{\mu}$:
\begin{equation}\label{App_abelian propagator}
D_{\mu\nu}(x-y)=\frac{\delta_{\mu\nu}}{4\pi^2}\frac{1}{(x-y)^2}
\end{equation}
in the Feynman gauge.

$\bullet$ In the 't Hooft limit where $N_c\rightarrow\infty$ with
$\lambda$ finite \cite{'tHooft1,'tHooft2}, one can invoke the
following factorization property:
\begin{equation}
W_2(\CMcal{C}_1,\CMcal{C}_2)=W(\CMcal{C}_1)W(\CMcal{C}_2)+O(\frac{1}{N_c^2})
\end{equation}
in order to simplify \eqref{App_euclidean MM equation}. This
property has been demonstrated at any order of Perturbation
Theory: at each order, only remain, in the 't Hooft limit, the
planar diagrams in which every gluon is emitted and absorbed by
the same Wilson loop (in general, a gauge invariant operator). The
diagrams connected by gluon exchanges (namely, when a gluon
emitted by a Wilson loop is reabsorbed by another Wilson loop) are
suppressed in $1/N_c^2$. The factorization property has also been
proved in non-perturbative regime. The equation
\eqref{App_euclidean MM equation} thus becomes:
\begin{equation}
\frac{\de}{\de
x_{\mu}}\frac{\delta}{\delta\sigma_{\mu\nu}(x)}W(\CMcal{C})=\lambda\oint_{\CMcal{C}}dy^{\nu}\delta^{4}(x-y)W(\CMcal{C}_{yx})W(\CMcal{C}_{xy})\;\;.
\end{equation}
This is the so-called loop equation which is a non-linear but
closed functional equation since only one-loop functionals
contribute.

It is worth summarizing the theory in loop space when
$N_c\rightarrow\infty$ with $\lambda$ finite. We have the
following two equations:

$\bullet$ the loop equation:
\begin{equation}
\frac{\de}{\de
x_{\mu}}\frac{\delta}{\delta\sigma_{\mu\nu}(x)}W(\CMcal{C})=\lambda\oint_{\CMcal{C}}dy^{\nu}\delta^{4}(x-y)W(\CMcal{C}_{yx})W(\CMcal{C}_{xy})\;\;,
\end{equation}

$\bullet$ the Bianchi identity:
\begin{equation}
\frac{\de}{\de
x_{\alpha}}\frac{\delta}{\delta\sigma_{\mu\nu}(x)}W(\CMcal{C})+\Big(\textrm{cyclic
permutation}\Big)=0\;\;.
\end{equation}
The initial condition when the loop shrinks into a point is:
\begin{equation}
W(\mathds{I}_{\CMcal{C}})=1
\end{equation}
where $\mathds{I}_{\CMcal{C}}=\{\sigma\rightarrow x(\sigma)=x\}$
is the identity.

Finally, we conclude this section by recapitulating the
correspondence between the Yang-Mills theory and its formulation
in loop space:

$\bullet$ the equation of motion:
\begin{equation}
D_{\mu}G_{\mu\nu}(x)=0\Rightarrow \frac{\de}{\de
x_{\mu}}\frac{\delta}{\delta\sigma_{\mu\nu}(x)}W(\CMcal{C})=\lambda\oint_{\CMcal{C}}dy^{\nu}\delta^{4}(x-y)\Big(W_2-\frac{1}{N_c^2}W\Big)\;\;,
\end{equation}

$\bullet$ the Bianchi identity:
\begin{equation}
D_{\alpha}G_{\mu\nu}(x)+\big(\textrm{cyclic
permutation}\big)=0\Rightarrow \frac{\de}{\de
x_{\alpha}}\frac{\delta}{\delta\sigma_{\mu\nu}(x)}W(\CMcal{C})+\big(\textrm{cyclic
permutation}\big)=0\;\;.
\end{equation}

\subsection{The renormalization of the Wilson loop}

When one evaluates, in Perturbation Theory, line integrals in the
Wilson loop, one meets with singularities which have to be
regularized. And yet, the renormalization of the Wilson loop is
still far from being trivial since it behaves as a non-local
object.

$\bullet$ When the loop $\CMcal{C}$ is smooth (\emph{i.e.}
differentiable) and simple (\emph{i.e.} without nodes), the
leading perturbative contribution to $W(\CMcal{C})$ corresponds to
the one-gluon exchange (contribution of order $O(g^2))$. It
diverges linearly in $\frac{\pi}{a}L(\CMcal{C})$ where
$L(\CMcal{C})$ is the length of the contour $\CMcal{C}$ and $a$ is
a short distance cutoff. In 1980, Polyakov \cite{Polyakov},
Dotsenko and Vergeles \cite{Dotsenko} proved that the Wilson loop
is multiplicatively renormalizable for such a smooth and simple
contour: at any order, linear divergences appear which can be
gathered into a common factor $Z=e^{const.\frac{L(\CMcal{C})}{a}}$
where $const.$ is of order $O(1)$. Beyond the second order, it
also appears logarithmic divergences which can be then absorbed by
renormalization of the strong coupling constant $g$. At the end of
the day, we obtain:
\begin{equation}\label{App_multiplicative renormalization}
W(g,\CMcal{C})=Z\,W_R(g_R,\CMcal{C})
\end{equation}
where $W_R(g_R,\CMcal{C})$ is finite, provided that it is
expressed in terms of the renormalized strong coupling constant
$g_R$. The physical meaning of the linear divergence $Z$ is easily
explained if we consider $\ln(W(\CMcal{C}))$ as the effective
action of a test particle constrained to move along $\CMcal{C}$:
the factor $Z$ disappears with the renormalization of the mass.

$\bullet$ When the contour is no longer smooth but has a cusp of
angle $\gamma$, the renormalization still remains multiplicative
\cite{Brandt,Brandt2}:
\begin{equation}
W(g,\CMcal{C})=Z(\gamma)\tilde{W}(g,\CMcal{C})
\end{equation}
where $\tilde{W}(g,\CMcal{C})$ refers to the Wilson loop on the
\emph{l.h.s} of \eqref{App_multiplicative renormalization}. The
factor $Z(\gamma)$ is an additional logarithmic divergence which
depends locally on the loop $\CMcal{C}$ (at the vicinity of the
cusp). Thus, this anomalous divergence cannot be absorbed by
renormalization of $g$.

$\bullet$ Let us consider a loop $\CMcal{C}$ which intersects one
or several times at the same space-time point and which is smooth
everywhere else. At the beginning of the eighties, Brandt,
Gocksch, Sato and Neri \cite{Brandt,Brandt2} showed that the
Wilson loop cannot be renormalized alone in this case. It must be
renormalized by mixing with all the other loops
$W_{n_i}(\{\CMcal{C}^i_j\})\equiv W^i(\{\CMcal{C}^i_j\})$ where
$i=1,2,\ldots,I$ ($I$ is the number of sets) and
$j=1,2,\ldots,n_i$ ($n_i$ is the number of loops $\CMcal{C}^i_j$
in the set $i$). Given a set $i$, the $n_i$ loops
$\CMcal{C}^i_j$'s must be identical to the corresponding sections
of $\CMcal{C}$, both in space-time (every $\CMcal{C}^i_j$ draws
the same path than $\CMcal{C}$) and in direction (every
$\CMcal{C}^i_j$ is oriented as $\CMcal{C}$).

In order to illustrate quickly the mechanism, let us consider the
simplest case of a loop with only one node (there is thus only one
independent angle $\g$). To the contour $\CMcal{C}$ is associated
two sets: $\{\CMcal{C}_1^1\}$ which is nothing else than
$\CMcal{C}$ when the path ordering prescription $P$ is taken into
account and $\{\CMcal{C}_1^2,\CMcal{C}_2^2\}$ corresponding to the
two sections of the loop on either side of the crossing point. The
set of the Wilson loops $W^1(\CMcal{C}_1^1)=W(\CMcal{C})$ and
$W^2(\CMcal{C}^2_1,\CMcal{C}^2_2)$ mix then together under
renormalization through the square matrix of order 2 $Z(\gamma)$.

The generalization to more complicated loops gives:
\begin{equation}
W^i(g,\{\CMcal{C}^i_j\})=\sum_{k=1}^{I}Z^{ik}(\gamma)\tilde{W}^k(g,\{\CMcal{C}^k_l\})
\end{equation}
where $Z(\gamma)$ is a matrix of order $I$ depending on all the
independent angles at the considered node.

\subsection{The Wilson loop in QCD$_2$}

Since 't Hooft's seminal paper in 1974 \cite{'tHooft2}, numerous
physicists have sought to understand the properties of QCD in
$(1+1)-$dimensional space-time (see, \emph{e.g.}
\cite{Frishman,Abdalla}). As a matter of fact, the Migdal-Makeenko
equation turns out to be, in this case, exactly solvable in the
large $N_c$ limit. In the sequel, I will essentially refer to the
review \cite{Makeenko}; the interested reader could fruitfully
read the pioneering works \cite{Kostov,Kazakov} where is detailed
the resolution.

Within the loop space formalism, the theory is manifestly gauge
invariant: let us choose the axial gauge $n\cdot A=0$. In
$(1+1)-$dimensional space-time, $n^{\mu}=(0,1)$, thus $A_1$
vanishes and only remains $A_0$. The interest of the axial gauge
is twofold: first of all, it discards the gluon self-interactions
(the commutator $\big[ A_{\mu},A_{\nu}\big]$ in the strength field
tensor being then equal to zero). Secondly, the ghosts decouple
from the theory and can be ignored. As a result, QCD$_2$ in the
axial gauge looks like, at first sight, an Abelian theory.

From the diagrammatic point of view, the Wilson loop
$W(\CMcal{C})$ sums, in the 't Hooft limit, the disconnected
planar diagrams, \emph{i.e.} those which are only made up of free
propagators (in the axial gauge, there are not three- or
four-gluon interactions). In a first time, we consider the
simplest case of a smooth and simple loop $\CMcal{C}$. One finds
($\lambda=g^2 N_c$):
\begin{equation}\label{App_solution 2 dim QCD}
W(\CMcal{C})=e^{-\frac{\lambda}{2}\oint_{\CMcal{C}}\oint_{\CMcal{C}}dx_{\mu}dy_{\nu}D_{\mu\nu}(x-y)}
\end{equation}
which strongly mimics the Abelian solution \eqref{abelian
solution}. The gluon propagator
\begin{equation}
D_{\mu\nu}(x-y)\equiv\frac{1}{N_c}Tr_c\langle0|
A_{\mu}(x)A_{\nu}(y)|0\rangle
\end{equation}
reads in the axial gauge as
\begin{equation}
D_{\mu\nu}(x-y)=\frac{1}{2}\delta_{\mu0}\delta_{\nu0}|x_1-y_1|\delta(x_0-y_0)\;\;.
\end{equation}
Because of the delta function $\delta(x_0-y_0)$ involving the time
coordinates, the interaction is instantaneous. Although this
result is valid in general, it is easier to demonstrate it in
Perturbation Theory. For this, we expand the (non-renormalized)
Wilson loop $W(\CMcal{C})$ in powers of $g$
\cite{Makeenko,Brandt2} (we are unaware here of the existence of
regularization schemes required in order to deal with well-defined
calculations):
\begin{equation}\label{App_expansion}
W(\CMcal{C})=1+\sum_{n=2}^{\infty}(-ig)^n\oint_{\CMcal{C}}dx_1^{\mu_1}\times\ldots\times\oint_{\CMcal{C}}dx_n^{\mu_n}\theta(\CMcal{C};1,\ldots,n)G^{(n)}_{\mu_1\ldots\mu_n}(x_1,\ldots,x_n)\;\;.
\end{equation}
The prescription $\theta(\CMcal{C};1,\ldots,n)$ puts in order the
points $x_1,\ldots,x_n$ along the loop $\CMcal{C}$ and $G^{(n)}$
is the Green function with $n$ external legs attached to
$\CMcal{C}$:
\begin{equation}\label{App-green function}
G^{(n)}_{\mu_1\ldots\mu_n}(x_1,\ldots,x_n)\equiv\frac{1}{N_c}Tr_c\,\langle0|A_{\mu_1}(x_1)\ldots
A_{\mu_n}(x_n)|0\rangle\;\;,
\end{equation}
satisfying the normalization condition $G^{(0)}(0)=1$. Since the
theory behaves, in the axial gauge, as an Abelian theory, the
gauge fields do not interact each other such that $n$ must be even
$n=2k$. The expansion \eqref{App_expansion} then becomes:
\begin{equation}
\begin{array}{lll}
W(\CMcal{C})&=&1+\displaystyle\sum_{k=1}^{\infty}(-ig)^{2k}\oint_{\CMcal{C}}dx_1^{\mu_1}\times\ldots\times\oint_{\CMcal{C}}dx_{2k}^{\mu_{2k}}\theta(\CMcal{C};1,\ldots,2k)\\
&&\times N_c^{k-1}D_{\mu_1\mu_2}(x_1-x_2)\ldots
D_{\mu_{2k-1}\mu_{2k}}(x_{2k-1}-x_{2k})\;\;.
\end{array}
\end{equation}
At the lowest order, we have:
\begin{equation}
W(\CMcal{C})=1+(-ig)^{2}\oint_{\CMcal{C}}dx_1^{\mu_1}\oint_{\CMcal{C}}dx_{2}^{\mu_{2}}\theta(\CMcal{C};1,2)N_cD_{\mu_1\mu_2}(x_1-x_2)+O(g^4)\;\;.
\end{equation}
Moreover, $D_{\mu_1\mu_2}(x_1-x_2)=D_{\mu_2\mu_1}(x_2-x_1)$ and
$\theta(\CMcal{C};1,2)+\theta(\CMcal{C};2,1)=1$ such that
\begin{equation}
W(\CMcal{C})=1-\frac{g^2N_c}{2}\oint_{\CMcal{C}}dx_1^{\mu_1}\oint_{\CMcal{C}}dx_{2}^{\mu_{2}}D_{\mu_1\mu_2}(x_1-x_2)+O(g^4)\;\;.
\end{equation}
One can generalize this mechanism at any order to obtain the
solution \eqref{App_solution 2 dim QCD} by taking the exponential
of the expansion of $W(\CMcal{C})$.

For a smooth and simple loop $\CMcal{C}$, the exponential factor
in \eqref{App_solution 2 dim QCD} is easily evaluated:
\begin{equation}
\oint_{\CMcal{C}}\oint_{\CMcal{C}}dx_{\mu}dy_{\nu}D_{\mu\nu}(x-y)=A(\CMcal{C})
\end{equation}
where $A(\CMcal{C})$ represents the area of the (plane) surface
with $\CMcal{C}$ as boundary. The Wilson loop takes then the form:
\begin{equation}\label{App_area law dim 2 QCd}
W(\CMcal{C})=e^{-\frac{\lambda}{2}A(\CMcal{C})}
\end{equation}
which is the area law in QCD$_2$. It is worth pointing out that
this behaviour \eqref{App_area law dim 2 QCd} is valid in the
non-Abelian case as well, the difference appearing only for more
complicated loops. So, let us focus on a contour having one node.
We have then two configurations: the first configuration gives
similar results in the Abelian and non-Abelian theories:
$A(\CMcal{C})=A_1+A_2$ and
$W(\CMcal{C})=e^{-\frac{\lambda}{2}(A_1+A_2)}$. The second
configuration gives $A(\CMcal{C})=A_1+4A_2$ but
$W(\CMcal{C})=e^{-\frac{\lambda}{2}(A_1+4A_2)}$ in the Abelian
case whereas $W(\CMcal{C})=(1-\lambda
A_2)e^{-\frac{\lambda}{2}(A_1+2A_2)}$ in the non-Abelian case
where contribute only the planar diagrams in the 't Hooft limit.
The generalization for any loop gives in QCD$_2$:
\begin{equation}
W(\CMcal{C})=\sum_{i}P_i(A_1,\ldots,A_n)
\end{equation}
where the $P_i$'s are exponential functions of the $n$ surfaces of
areas $A_i$ ($i=1,\ldots,n$) which make up $A(\CMcal{C})$.

\end{document}